\documentclass{ws-rv9x6}
\usepackage{setspace,bm}

\usepackage{ws-rv-thm}   
\usepackage[square]{ws-rv-van}   
\usepackage{nicefrac}
\usepackage{shuffle}
\usepackage{amssymb}
\usepackage{url}

\makeindex

\def\tsplit#1#2#3{$\begin{array}{c}\text{#1} \\[-#3pt] \text{#2}\end{array}$}
\def\tsplit#1#2{$\begin{array}{c}\text{#1} \\[-5pt] \text{#2}\end{array}$}

\newif\ifdraft
\drafttrue
\newif\ifpreprint
\preprinttrue

\def\half{ \nicefrac{1}{2} }

\def\sect#1{section~{\ref{#1}}}
\def\fig#1{fig.~{\ref{#1}}}

\def\eqn#1{eq.~(\ref{#1})}

\def\eqns#1#2{eqs.~(\ref{#1}) and~(\ref{#2})}

\def\Tab#1{Table~{\ref{#1}}}

\def\nn{\nonumber}

\newenvironment{polynomial}
  {\par\vspace{\abovedisplayskip}%
   \setlength{\leftskip}{\parindent}%
   \setlength{\rightskip}{\leftskip}%
   \medmuskip=4mu plus 2mu minus 2mu
   \binoppenalty=0
   \noindent$\displaystyle}
  {$\par\vspace{\belowdisplayskip}}

\def\be{\begin{equation}}
\def\ee{\end{equation}}

\def\bea{\begin{align}}
\def\eea{\end{align}}

\newcommand{\cT}[1]{ {\rm T}^{a_{#1}} }
\newcommand{\cTi}[1]{ {\rm T}^{{#1}} }

\newcommand\cN{ {\cal N} }

\newcommand\NeqFour{{\cN=4}}

\newcommand{\mtrx}[1]{\bm{#1}}

\newcommand\Mtree[1]{{\cal M}^{\rm tree}_#1}
\newcommand\Atree[1]{{\cal A}^{\rm tree}_#1}
\newcommand\AtreeCO[1]{{ A}^{\rm tree}_#1}
\newcommand\AtreeVec[1]{{\mtrx{A}}^{\rm tree}_#1}
\newcommand\MtreeVec[1]{\mtrx{ {\cal M} }^{\rm tree}_#1}

\newcommand\Mloop[2]{{\cal M}^{#1\text{-loop}}_{#2}}
\newcommand\Aloop[2]{{\cal A}^{#1\text{-loop}}_{#2}}

\newcommand{\ECO}{ {\rm ECO} }

\newcommand{\GraphsTree}[1]{\Gamma^{\rm tree}_{#1, {\rm UO}}}
\newcommand{\GraphsTreeCO}[1]{\Gamma^{\rm tree}_{#1,  \sigma}}

\newbox\charbox
\newbox\slabox
\def\s#1{{      
        \setbox\charbox=\hbox{$#1$}
        \setbox\slabox=\hbox{$/$}
        \dimen\charbox=\ht\slabox
        \advance\dimen\charbox by -\dp\slabox
        \advance\dimen\charbox by -\ht\charbox
        \advance\dimen\charbox by \dp\charbox
        \divide\dimen\charbox by 2
        \raise-\dimen\charbox\hbox to \wd\charbox{\hss/\hss}
        \llap{$#1$} }}

\begin{document}

\chapter*{TASI 2014\\
Lectures on Gauge and Gravity Amplitude Relations}

\author[J.~J.~M.~Carrasco]{ John Joseph M. Carrasco}

\address{
Stanford Institute for Theoretical Physics and  Department of Physics,\\ 
Stanford University, Stanford, CA 94305, USA\\
}

\begin{abstract}
In these lectures I talk about  simplifications and universalities found in scattering amplitudes for gauge and gravity theories.  In contrast to Ward identities, which are understood to arise from familiar symmetries of the classical action, these structures are currently only understood in terms of graphical organizational principles, such as the gauge-theoretic color-kinematics duality and the gravitational double-copy structure, for local representations of multi-loop $S$-matrix elements.  These graphical principles make manifest new relationships in and between gauge and gravity scattering amplitudes.  My lectures will focus on arriving at such graphical organizations for generic theories with examples presented from maximal supersymmetry, and their use in unitarity-based multi-loop integrand construction.
\end{abstract}


\body

\section*{Outline}
\begin{enumerate}
\item The first section will give a sense for the type of structures arising from the color-kinematics duality, with an example 
of the loop-momentum integrand for a two-loop amplitude in the maximally supersymmetric theory.  We will look closely at the three-point Feynman rule for gluons and gravitons.
\item Since all of our understanding of loop-momentum integrands at loop level can be encoded in trees, the second section lays out some 
very special properties of tree amplitudes and their description in terms of cubic graphs.
\item The third section presents a state of the art technique for extracting generic integrands of multi-loop scattering amplitudes from tree-level data. First I  discuss  how one can use tree-level properties to {\em verify} a candidate (graph-organized) amplitude at the integrand level. I  introduce the notion 
of spanning cuts, and discuss how the existence of hierarchical approaches to verification leads to a natural method of integrand construction, called the {\it method of maximal cuts}.  
\item The fourth section discusses the exploitation of the color-kinematics duality to make calculations of multi-loop amplitudes in gauge theory simpler, and then those of gravity amplitudes trivial, making clear the functional nature of the loop-level duality.
\item The fifth section concludes with a brief discussion of exciting open questions and touches on some progress in the literature.
\item I include a bonus appendix that offers ready formulae for tree-level calculation and unitarity sums in 4-dimensions, so as to allow quick access to actual expressions at tree and cut-loop level, with references to far more in-depth treatments.  
\end{enumerate}

\section{Introduction}

The early 1990s marked the first wave of a new approach to perturbative scattering amplitudes, one that drew inspiration from the idealistic analytic $S$-matrix program of the 1960's, invoked pragmatic computational acumen and book-keeping from recently developed quantum field theory calculations, and applied broad insight developed in the first string theory revolution.  Many of the tools of that time, including spinor-helicity notation, color-decomposition and unitarity methods, were introduced to students nearly twenty years ago at the 1995 TASI~\cite{LanceTasi1996}. In the subsequent decades, the tool-chest for calculating perturbative scattering amplitudes has only grown larger.

\begin{figure}
\centerline{\includegraphics[width=4.5in]{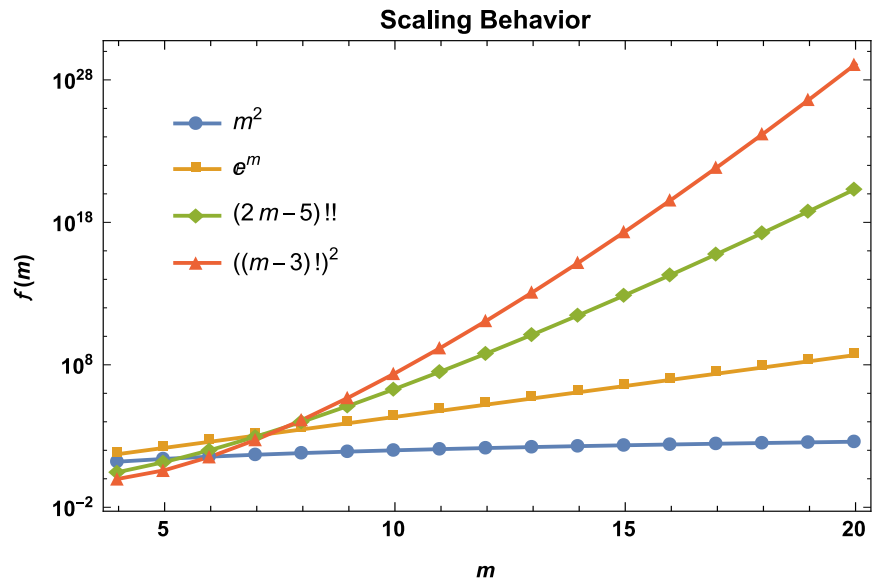}}
\caption{Comparison of various scaling behaviors.  Quadratic scaling, $m^2$, is associated with the number of interaction terms for classical $m$-body systems with two-body interactions.   Note that the far sharper exponential scaling, $e^m$, associated with the number of cubic graphs contributing to a color-ordered tree-level amplitude, is still subdominant for $m>7$ to the factorial scaling associated with the number of cubic graphs associated with $m$-particle tree-level interaction, $(2m-5)!!$, which itself is subdominant to the number of terms in the smallest known Kawai-Lewellen-Tye (KLT) representation of generic tree-level gravity amplitudes $((m-3)!)^2$.  } \label{fig:scalingBehavior}
\end{figure}

Why aren't we done?  Why are there still important open perturbative questions in standard model calculations relevant to current collider experiments, in formal quantum field theory,  in condensed matter, in cosmology, in astrophysics? For any given theory, in general, there is a factorial  increase in complexity with the number of particles interacting (the {\em multiplicity}) as well as with the order of quantum effect (the {\em loop order}).   Factorial increase is incredibly steep, see~\fig{fig:scalingBehavior}.  Contrast this with the costs of a classical $N$-body simulation (naively $N^2$).   The numbers can get ridiculous quickly --- the expense of a universe simulation involving 100-billion galaxies is the same order of magnitude as the size of a 21 particle tree-level scattering amplitude.  Typical brute force advance in automation or computational resources does not really advance the field --- it just allows the saturation of previously developed ideas.  In contrast great ideas do indeed open new avenues --- they leap-shot us ahead --- but until we find the factorial-beating idea, there is always a threshold we can challenge ourselves with.  This field not only allows room for new ideas and new insights, but demands it to progress.

There are many reasons to get excited about  scattering amplitudes. One my favorites is that they make actual observable predictions. This is no small thing even when talking about imaginary theories.  Consequently, amplitude theory is an area where the symmetries of relativistic quantum field theories 
 come to life.  As we probe the scattering amplitudes of a theory, in a completely unambiguous way, free of any gauge dependence, we can start hearing what the symmetries of the theory have to say about how the world evolves.     This may teach us about important symmetries nobody had ever noticed a theory possessed\footnote{See, e.g., the symmetry I discuss in Problem~\ref{probMethodInt} which launched a revolution in understanding the planar maximal supersymmetric gauge theory in four dimensions.}, but we can hope it leads us further --- into stories making manifest a primacy of algorithmic ideas capturing the truly fundamental.

In these lectures I tell an aspirational story.  Our goal is to find the best ways of extracting and manipulating the predictive information from the theories we are considering: the best ways to represent it, encode it, and compress it. We will wonder: what are other questions this information is answering? In the end  we will have built a web of understanding between initially very different questions --- those involving amplitudes for gauge theories containing particles like photons and gluons, and those for theories containing gravitons.  

Also, here is an important question to keep in mind: what information do we {\em not} care about?  What information are we completely happy to discard?  Do we care about the integrands of amplitudes, or only about their predictions after integrating over all the loop momenta of the virtual particles?  After carrying out this integration, we will have certainly thrown out some information --- but what we keep is all that is necessary to be relevant to measurement.  Furthermore, there are relations and structures that are only manifest post-integration.  This  year's TASI has included beautiful lectures by Claude Duhr concerning some of the language developing around the results of integrating: multiple polylogarithms and the fascinating relations between them.  As spectacular as this important subject is, there is still much to be understood at the integrand level --- structures that are universal even when talking about effective field theories (theories that only hold for certain energy scales and which must be completed either in the infrared or the ultraviolet). These lectures will solely concentrate on structure of the integrands of the $S$-matrix.

Because they are simple enough to write on a blackboard and to really get one's head around, I will use examples from the maximally supersymmetric gauge theory, called $\cN=4$ super Yang-Mills~\cite{NeqFourDefinition}  in four dimensions.  Even though I will mainly provide supersymmetric examples, the approaches I discuss and the ways of organizing the amplitudes are quite generic.

\subsection{Gluons for (almost) nothing and gravitons for free: an example}
\label{ex2l}

Let me give you one very provocative example~\cite{twoLoopNeq4YM} at two loops in the maximally supersymmetric gauge theory.
There are two distinct integrals that contribute to the ($2\rightarrow2$) scattering of two gluons into two gluons at this loop order.  We can represent them in terms of graphs:
\be 
\label{planarDef}
{\rm planar} = \vcenter{ \includegraphics[width=3in]{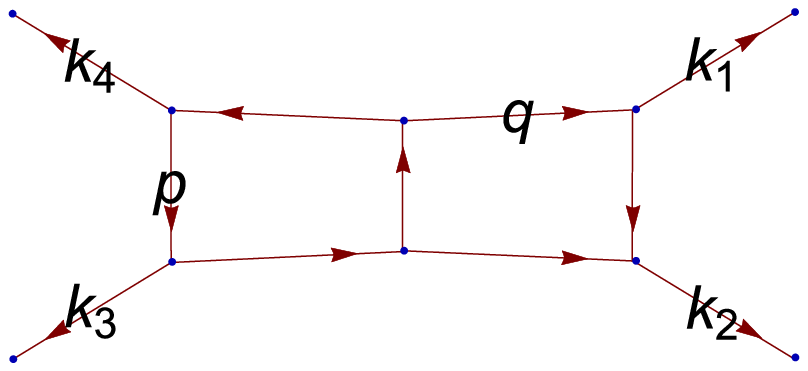}} \, , 
\ee
\be
\label{nonplanarDef}
{\rm nonplanar} = \vcenter{ \includegraphics[width=3in]{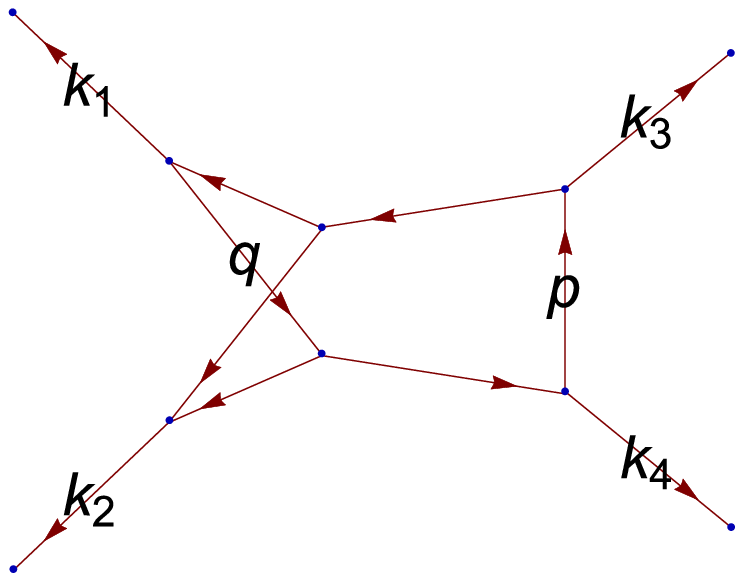}} \, .
\ee
The labeled momenta $p$ and $q$ are independent loop momenta to be integrated over in each graph, and the external momenta are on-shell (i.e.~$k_i^2=0$).  Every vertex maintains conservation of momenta, so the momenta along unlabeled edges can be determined by the momenta labeled in each graph.  Since both graphs are being integrated over, we are free to take their integrands under the same integral.

Every graph has an integrand, which we can break into parts, ignoring the ubiquitous measure:
\begin{equation}
{\rm integrand}({\rm graph}) = \frac{n(g) c(g)}{d(g)} \,.
\end{equation} 
Each graph has a propagator structure. We call this structure the denominator $d(g)$ associated with each graph since it always appears downstairs. It is simply the product of the Lorentz square of the momentum flowing through each internal edge of the graph.  E.g., for the planar graph: 
\begin{eqnarray}
d({\rm planar}) &=& p^2 q^2 \left(p-k_3\right){}^2 \left(k_4+p\right){}^2 \nonumber\\
 &&\null\times \left(q-k_1\right){}^2 \left(k_1+k_2-q\right){}^2 \left(k_4+p+q\right){}^2. 
\end{eqnarray}

There are color weights $c(g)$ associated with each graph, which come from dressing each vertex with the $f^{abc}$ structure constants associated with the gauge group being considered, e.g.
\be
\label{colorFactorPlanar}
c({\rm planar}) = f^{a_1 b_1 b_2} f^{b_1 a_2 b_3} f^{b_3 b_4 b_5} f^{b_4 a_3 b_6} f^{b_6 a_4 b_7}  f^{b_7 b_2 b_5}.
\ee
I give each external momentum $k_i$ the color index $a_i$. For the internal lines, I just assign some numbers $1$ through $7$, they get the color indices $b_i$; each $b_i$ shows up in two vertices. The repeated color indices $b_i$ are to be summed over. I will talk a little more about color factors in a little bit, but it will not be so important for this first point.  

Last but absolutely not least, there will also be a kinematic numerator $n(g)$ associated with each graph. The kinematic weight in the numerator associated with both graphs will be:  $s^2 t \, \AtreeCO{4}(1234)$ as drawn, where $\AtreeCO{4}(1234)$ is simply the four-point tree, and $s$ and $t$ are the four-point Mandelstam invariants.


\begin{convention}
{\em Some four-point notation}:
Conservation of momenta in a massless four-point process is so prevalent that the kinematic invariants are given handy annotations.  There are three kinematic invariants between the four (outgoing) momenta $k_1, k_2 ,k_3, k_4$, typically called {\it Mandelstam invariants}, with the following definitions for massless $k_i$:
\begin{align}
s&\equiv(k_1+k_2)^2=(k_3+k_4)^2 \,, \\
t&\equiv(k_1+k_4)^2=(k_2+k_3)^2 \,, \\
u&\equiv -s-t = (k_1+k_3)^2=(k_2+k_4)^2 \,.
\end{align}
\end{convention}

\begin{problem} 
Work out the denominator structure associated with the nonplanar graph.
\end{problem}

\begin{problem} 
Label the graph in \eqn{planarDef} with the same internal color-indices $b_i$  to reproduce \eqn{colorFactorPlanar}. Label the color-indices in \eqn{nonplanarDef} and write down the associated color factor. 
\end{problem}

Here is the crucial point:  The quantities $n(g)$, $c(g)$ and $d(g)$ for two graphs, $g\,=\,{\rm planar,\, nonplanar}$, are all the information we need to calculate the correction to $2\rightarrow2$ scattering in the maximally supersymmetric theory:
\begin{eqnarray}
\Aloop{2}{4} &=&  \int \frac{d^D p}{(4 \pi)^D} \frac{d^D q}{(4 \pi)^D}
\Biggl( \frac{n({\rm planar}) c({\rm planar})}{d({\rm planar})} \nonumber\\
&&\null\hskip0.3cm
  + \frac{n({\rm nonplanar}) c({\rm nonplanar})}{d({\rm nonplanar})}\ +\ ({\text{ext. perms}}) \Biggr),
\label{ym2Loop}
\end{eqnarray}
where the external permutations are simply permuting all external legs.  I have elided an overall symmetry factor that takes care of over-counting, but this compact expression in \eqn{ym2Loop} is really all the information associated with the two-loop correction to $2\rightarrow2$ scattering in the maximally supersymmetric theory. 
We will see how seamlessly this expression arises from consideration of on-shell quantities.

Here is something else that is interesting.  If we replace each color factor with another copy of a numerator factor, leaving the propagator structure the same, we get~\cite{twoLoopNeq8GR} the scattering amplitude in the maximally supersymmetric (${\cal N}=8$) supergravity theory:
\begin{eqnarray}
\Mloop{2}{4} &=&  \int \frac{d^D p}{(4 \pi)^D} \frac{d^D q}{(4 \pi)^D}
\Biggl( \frac{n({\rm planar}) n({\rm planar})}{d({\rm planar})} \nonumber\\
&&\null\hskip0.3cm
  + \frac{n({\rm nonplanar}) n({\rm nonplanar})}{d({\rm nonplanar})}\ +\ ({\text{ext. perms}}) \Biggr).
\label{gr2Loop}
\end{eqnarray}

We should notice a few things in the gauge theory example. Both the denominator factors and the color factors arise from simple algorithms applied to the graph structure. The algorithms themselves have nothing to do with the theory. The values of the structure constants change when we change gauge groups, but they do not care how much supersymmetry is around. The denominator is the same for any gauge or gravity theory. The heart of the theory can be mapped to understanding what the kinematic numerator factors are. The differences between gauge theories (having the same gauge group) can be understood as the differences between these kinematic factors. We will see that the difference between gauge theory integrands and gravity theory integrands involves replacing the color-weight with another kinematic weight, when the kinematic weight is written in the correct gauge~\cite{BCJ,BCJLoop}.  

Are there simple generic approaches to understand what these kinematic numerators should be? To quote a famous philosopher, ``It depends upon what the meaning of the word `is' is.'' I will take us to the cutting edge of current methods applicable to generic quantum field theories. We will engage with a special kinematic structure that dances so closely with the  color structure definitive of gauge theories, one can believe we're starting to glimpse the predictive core of what it means to be a gauge theory. At the end hopefully you'll be in a position to judge for yourself how simple and beautiful this understanding and associated methods are, and how far we have yet to go.

What is the best way forward? My favorite approach is to ask questions, so let us look at this simple integrand and ask some questions.

Here are two: 
\begin{enumerate}
\item  Why are the numerator kinematic weights for the two different graph topologies identical?  
\item Why, when we replace the color factor, graph by graph, with another copy of the kinematic numerator factor, do we get a supergravity scattering amplitude?   
\end{enumerate}
It turns out the answers to these two questions are intimately related.  But to discuss it we first need some background.

\subsection{Antisymmetry, Jacobi Relations, and the Color-Kinematic Duality}
To answer these questions we need to learn a little more about color factors, the $c(g)$.  They do two things for a living.  The first thing is that the $f^{abc}$ have a cyclic symmetry, but pick up a minus sign under odd permutations, i.e.~they are totally antisymmetric:
\be
f^{abc} = f^{bca} = f^{cab}= - f^{bac} = -f^{acb} = -f^{cba} \,.
\label{ftotantisym}
\ee
It turns out that propagator structures do not care about odd permutations of vertices, but color-factors and kinematic weights will --- but only up to a sign. If one vertex is flipped (permuted by one of the three odd permutations in \eqn{ftotantisym}) then we care, but if two vertices are flipped, then neither numerator nor color factor will distinguish between the graphs. Once we're at the point of writing down numerators and color factors associated with graphs we might as well identify any two graphs that have the same color factors and same numerator factors.  When two graphs, $g_a$ and $g_b$ have the same topology, but differ by an odd number of odd vertex permutations, I will annotate the relation as follows:  $g_b = \overline{g_a}$.  The antisymmetry of the corresponding color factors can be written as:
\be
\label{antiCol}
c(g_a) = - c(\overline{g_a}) \,.
\ee

 The second thing our color factors do is obey Jacobi relations, because they're composed of the structure constants of a given Lie algebra. Here is a true fact for any color-indices $b_i$
\be
f^{b_1 b_2 b_0} f^{b_0 b_3 b_4}  = f^{b_4 b_1 b_0} f^{b_0 b_2 b_3} +   f^{b_3 b_1 b_0 } f^{b_0 b_4 b_2} \,,
\ee
where we sum over the repeated index $b_0$.  This is a Jacobi relation.  I will annotate the three graphs as $g_s$, $g_t$ and $g_u$. The Jacobi relation can be written:
\be
\label{jacCol}
c(g_s) = c(g_t) + c(g_u) \,.
\ee
It does not matter whether this product of $f$'s in a color factor is embedded in a longer stream of $f$'s as long as the rest of the stream is shared between all three graphs. What this means is that the color factors of graphs obey Jacobi relations around individual edges.  Each edge defines a relation between three graphs (see problem~\ref{opProblem}). The labels $g_s$, $g_t$, and $g_u$ can be seen as relative to a particular edge in a graph.  As an example let us consider three four-loop graphs, whose color factors are related as follows:
\begin{multline}
c\left(  \!\!\!\!\! \!\!\!\!\!   \vcenter{ \includegraphics[width=2in, bb=0 0 260 115]{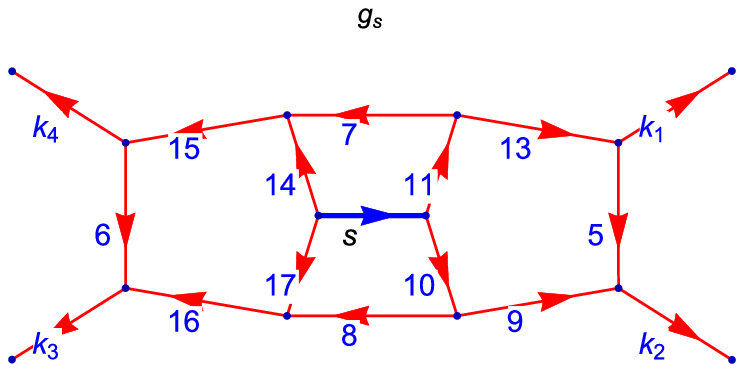}} \!\!\!\!\! \!\!\!\!\! \!\!\!\!\! \!\!\!\!\!  \!\!\!\!\! \!\!\!\!\! \!\!\!\!\! \!\!\!\!\!  \!\!\!\!\! \!\!\!\!\! \!\!\!\!\! \!\!\!\!\!  \!\!\!\!\! \!\!\!\!\!  \!\!\!\!\! \!\!\!\!\! \!\!\!\!\!  \!\!\!\!\! \!\!\!\!\! \!\!\!\!\! \!\!\!\!\! \right) = 
c\left(  \!\!\!\!\! \!\!\!\!\! \vcenter{ \includegraphics[width=2in, bb=0 0 260 115]{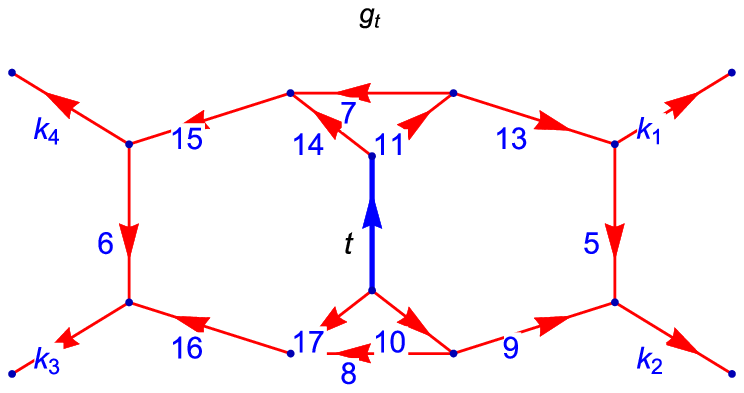}}  \!\!\!\!\! \!\!\!\!\! \!\!\!\!\! \!\!\!\!\!  \!\!\!\!\! \!\!\!\!\! \!\!\!\!\! \!\!\!\!\!  \!\!\!\!\! \!\!\!\!\! \!\!\!\!\! \!\!\!\!\!  \!\!\!\!\! \!\!\!\!\!  \!\!\!\!\! \!\!\!\!\! \!\!\!\!\!  \!\!\!\!\! \!\!\!\!\! \!\!\!\!\! \!\!\!\!\! \right) \\
 + c\left( \!\!\!\!\! \!\!\!\!\! \vcenter{ \includegraphics[width=2in, bb=0 0 260 115]{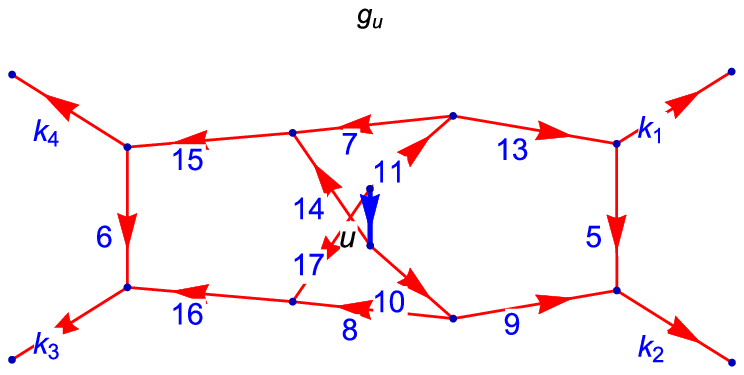}}  \!\!\!\!\! \!\!\!\!\! \!\!\!\!\! \!\!\!\!\!  \!\!\!\!\! \!\!\!\!\! \!\!\!\!\! \!\!\!\!\!  \!\!\!\!\! \!\!\!\!\! \!\!\!\!\! \!\!\!\!\!  \!\!\!\!\! \!\!\!\!\!  \!\!\!\!\! \!\!\!\!\! \!\!\!\!\!  \!\!\!\!\! \!\!\!\!\! \!\!\!\!\! \!\!\!\!\!  \right),
\end{multline}
where we see that $g_s$, $g_t$, and $g_u$ are relative to the edge $s$ of the first graph.  The edge we ``Jacobi'' around is labeled $s$, $t$, and $u$ in each graph respectively.  Note that each of the three graphs is identical, except for how the $s$, $t$, or $u$ edge is glued into the rest of the graph.

The next thing to realize is something called  {\em the color-kinematic duality}~\cite{BCJ,BCJLoop} that holds in our two-loop representation.  This duality is now known to be true~\cite{LagrangianSquare} for gauge theories at tree level, and it is conjectured to hold to all loop orders. (It has been demonstrated to hold through four loops for four-point scattering in the maximally supersymmetric theory~\cite{BCJLoop,  BCJ4Loop}.)  The idea is that there is a representation that allows the kinematic weights to obey antisymmetry and Jacobi identities for each edge of every graph, in the same way that color factors do:
\begin{align}
\label{antiKin}
n(g) &= -n(\overline{g}) \,,\\
\label{jacKin}
n(g_s) &= n(g_t)+n(g_u) \,.
\end{align}
The momentum running through all the edges of the three graphs in \eqn{jacKin}, other than $s$,
$t$, $u$, is taken to be the same; the momentum flowing through $s$, $t$, and $u$ is
determined by momentum conservation.  As we will see, the set of numerator factors for an amplitude integrand can be altered without changing the full integrand, a property known as generalized gauge invariance.  Choices for the numerator factors that satisfy \eqns{antiKin}{jacKin} are referred to as color-dual representations or choosing a color-dual gauge.  Our two-loop representation is color-dual because there are only two graphs contributing and $n({\rm planar}) = n({\rm nonplanar})$.  It is not hard to see that this satisfies Jacobi (\eqn{jacKin}) on every edge. Each such kinematical Jacobi relation, if it contains any nonzero numerators, contains exactly two nonzero numerators (with the correct relative sign) and one that is zero.

The final thing to realize is something called {\em the double-copy property}~\cite{BCJ,BCJLoop} of gravity theories.  For scattering amplitudes in many gravity theories, it is possible to write their kinematic numerators, graph by graph, as a product of two gauge-theory (YM) numerators that are in a color-dual gauge:
\begin{align}
n_{\rm GR}(g) = n_{\rm YM}(g) \tilde{n}_{\rm YM}(g).
\end{align}
For special gravity theories (like $\cN=8$ supergravity) both gauge numerators come from the same gauge theory (although only one needs to be in a color-dual gauge).  A more general class of gravity theories can be described using numerators from two distinct gauge theories.  I will go into more detail a little later, once we have a better grasp of numerators at tree level.  But first let us close this lecture by considering the absolute simplest case, in order to get some comfort with these ideas.

\subsection{Coming to grips with gravity as a double copy}

I claimed that, once gauge-theory graph numerators are written in the correct way, one can express numerators for a related gravity theory as a product of gauge-theory numerators.  This result can be proven recursively at tree level~\cite{LagrangianSquare}. Loop-level relations are then expected to follow via unitarity.  But where does this notion come from?

First let us travel back in time to 1967 when Bryce DeWitt wrote his seminal papers on the perturbative quantization of gravity, of which his third~\cite{DeWitt:1967uc} will be the most relevant here.  Here we find the three-point Feynman rule for Yang-Mills theory (eq.~(2.1) of ref.~\cite{DeWitt:1967uc}),
\be
\frac{\delta S^3}{
\delta A^{a}_\mu  \delta A^{b}_\sigma  \delta A^{c}_\rho} \to   i f^{abc} \left(\left(k_1{}^{\rho }-k_2{}^{\rho }\right) \eta ^{\mu 
   \sigma }+\left(k_2{}^{\mu }-k_3{}^{\mu }\right) \eta ^{\sigma  \rho
   }+\left(k_3{}^{\sigma }-k_1{}^{\sigma }\right) \eta ^{\rho  \mu
   }\right)
\ee
where particle 1 comes in with momenta $k_1$ and color index $a$,  particle 2 comes in with momenta $k_2$ and color-index $b$ and particle 3 comes in with momenta $k_3$ and color-index $c$.   Also, and more to the point, DeWitt gives us the 171 term\footnote{DeWitt actually introduces shorthand for symmetrization and permutation so the expression takes up less space on the page --- for pedagogy I have expanded it out so that  students can play with the full expression without fear of misinterpreting which symbols should be permuted over.}  three-point Feynman rule for three gravitons (eq.~(2.3) of ref.~\cite{DeWitt:1967uc}),
\begin{tiny}
\begin{polynomial}
\frac{\delta S^3}{
\delta \varphi_{\mu\nu}  \delta \varphi_{\sigma\tau}  \delta \varphi_{\rho\lambda}} \to
2 \eta ^{\mu  \tau } \eta ^{\nu  \sigma } k_1{}^{\lambda } k_1{}^{\rho
   }+2 \eta ^{\mu  \sigma } \eta ^{\nu  \tau } k_1{}^{\lambda }
   k_1{}^{\rho }-2 \eta ^{\mu  \nu } \eta ^{\sigma  \tau } k_1{}^{\lambda
   } k_1{}^{\rho }+   2 \eta ^{\lambda  \tau } \eta ^{\mu  \nu }
   k_1{}^{\sigma } k_1{}^{\rho }+2 \eta ^{\lambda  \sigma } \eta ^{\mu 
   \nu } k_1{}^{\tau } k_1{}^{\rho }+   \eta ^{\mu  \tau } \eta ^{\nu 
   \sigma } k_2{}^{\lambda } k_1{}^{\rho }+   \eta ^{\mu  \sigma } \eta
   ^{\nu  \tau } k_2{}^{\lambda } k_1{}^{\rho }+\eta ^{\lambda  \tau }
   \eta ^{\nu  \sigma } k_2{}^{\mu } k_1{}^{\rho }+\eta ^{\lambda  \sigma
   } \eta ^{\nu  \tau } k_2{}^{\mu } k_1{}^{\rho }+\eta ^{\lambda  \tau }
   \eta ^{\mu  \sigma } k_2{}^{\nu } k_1{}^{\rho }+\eta ^{\lambda  \sigma
   } \eta ^{\mu  \tau } k_2{}^{\nu } k_1{}^{\rho }+\eta ^{\lambda  \tau }
   \eta ^{\nu  \sigma } k_3{}^{\mu } k_1{}^{\rho }+\eta ^{\lambda  \sigma
   } \eta ^{\nu  \tau } k_3{}^{\mu } k_1{}^{\rho }-\eta ^{\lambda  \nu }
   \eta ^{\sigma  \tau } k_3{}^{\mu } k_1{}^{\rho }+\eta ^{\lambda  \tau
   } \eta ^{\mu  \sigma } k_3{}^{\nu } k_1{}^{\rho }+\eta ^{\lambda 
   \sigma } \eta ^{\mu  \tau } k_3{}^{\nu } k_1{}^{\rho }-\eta ^{\lambda 
   \mu } \eta ^{\sigma  \tau } k_3{}^{\nu } k_1{}^{\rho }+\eta ^{\lambda 
   \nu } \eta ^{\mu  \tau } k_3{}^{\sigma } k_1{}^{\rho }+\eta ^{\lambda 
   \mu } \eta ^{\nu  \tau } k_3{}^{\sigma } k_1{}^{\rho }+\eta ^{\lambda 
   \nu } \eta ^{\mu  \sigma } k_3{}^{\tau } k_1{}^{\rho }+\eta ^{\lambda 
   \mu } \eta ^{\nu  \sigma } k_3{}^{\tau } k_1{}^{\rho }+2 \eta ^{\mu 
   \nu } \eta ^{\rho  \tau } k_1{}^{\lambda } k_1{}^{\sigma }+2 \eta
   ^{\mu  \nu } \eta ^{\rho  \sigma } k_1{}^{\lambda } k_1{}^{\tau }-2
   \eta ^{\lambda  \rho } \eta ^{\mu  \nu } k_1{}^{\sigma } k_1{}^{\tau
   }+2 \eta ^{\lambda  \nu } \eta ^{\mu  \rho } k_1{}^{\sigma }
   k_1{}^{\tau }+2 \eta ^{\lambda  \mu } \eta ^{\nu  \rho } k_1{}^{\sigma
   } k_1{}^{\tau }+\eta ^{\mu  \tau } \eta ^{\nu  \rho } k_1{}^{\sigma }
   k_2{}^{\lambda }+\eta ^{\mu  \rho } \eta ^{\nu  \tau } k_1{}^{\sigma }
   k_2{}^{\lambda }+\eta ^{\mu  \sigma } \eta ^{\nu  \rho } k_1{}^{\tau }
   k_2{}^{\lambda }+\eta ^{\mu  \rho } \eta ^{\nu  \sigma } k_1{}^{\tau }
   k_2{}^{\lambda }+\eta ^{\nu  \tau } \eta ^{\rho  \sigma }
   k_1{}^{\lambda } k_2{}^{\mu }+\eta ^{\nu  \sigma } \eta ^{\rho  \tau }
   k_1{}^{\lambda } k_2{}^{\mu }+\eta ^{\lambda  \tau } \eta ^{\nu  \rho
   } k_1{}^{\sigma } k_2{}^{\mu }-\eta ^{\lambda  \rho } \eta ^{\nu  \tau
   } k_1{}^{\sigma } k_2{}^{\mu }+\eta ^{\lambda  \nu } \eta ^{\rho  \tau
   } k_1{}^{\sigma } k_2{}^{\mu }+\eta ^{\lambda  \sigma } \eta ^{\nu 
   \rho } k_1{}^{\tau } k_2{}^{\mu }-\eta ^{\lambda  \rho } \eta ^{\nu 
   \sigma } k_1{}^{\tau } k_2{}^{\mu }+\eta ^{\lambda  \nu } \eta ^{\rho 
   \sigma } k_1{}^{\tau } k_2{}^{\mu }+2 \eta ^{\nu  \rho } \eta ^{\sigma
    \tau } k_2{}^{\lambda } k_2{}^{\mu }+\eta ^{\mu  \tau } \eta ^{\rho 
   \sigma } k_1{}^{\lambda } k_2{}^{\nu }+\eta ^{\mu  \sigma } \eta
   ^{\rho  \tau } k_1{}^{\lambda } k_2{}^{\nu }+\eta ^{\lambda  \tau }
   \eta ^{\mu  \rho } k_1{}^{\sigma } k_2{}^{\nu }-\eta ^{\lambda  \rho }
   \eta ^{\mu  \tau } k_1{}^{\sigma } k_2{}^{\nu }+\eta ^{\lambda  \mu }
   \eta ^{\rho  \tau } k_1{}^{\sigma } k_2{}^{\nu }+\eta ^{\lambda 
   \sigma } \eta ^{\mu  \rho } k_1{}^{\tau } k_2{}^{\nu }-\eta ^{\lambda 
   \rho } \eta ^{\mu  \sigma } k_1{}^{\tau } k_2{}^{\nu }+\eta ^{\lambda 
   \mu } \eta ^{\rho  \sigma } k_1{}^{\tau } k_2{}^{\nu }+2 \eta ^{\mu 
   \rho } \eta ^{\sigma  \tau } k_2{}^{\lambda } k_2{}^{\nu }+2 \eta
   ^{\lambda  \tau } \eta ^{\rho  \sigma } k_2{}^{\mu } k_2{}^{\nu }+2
   \eta ^{\lambda  \sigma } \eta ^{\rho  \tau } k_2{}^{\mu } k_2{}^{\nu
   }-2 \eta ^{\lambda  \rho } \eta ^{\sigma  \tau } k_2{}^{\mu }
   k_2{}^{\nu }+\eta ^{\mu  \tau } \eta ^{\nu  \sigma } k_1{}^{\lambda }
   k_2{}^{\rho }+\eta ^{\mu  \sigma } \eta ^{\nu  \tau } k_1{}^{\lambda }
   k_2{}^{\rho }+\eta ^{\lambda  \nu } \eta ^{\mu  \tau } k_1{}^{\sigma }
   k_2{}^{\rho }+\eta ^{\lambda  \mu } \eta ^{\nu  \tau } k_1{}^{\sigma }
   k_2{}^{\rho }+\eta ^{\lambda  \nu } \eta ^{\mu  \sigma } k_1{}^{\tau }
   k_2{}^{\rho }+\eta ^{\lambda  \mu } \eta ^{\nu  \sigma } k_1{}^{\tau }
   k_2{}^{\rho }+2 \eta ^{\mu  \tau } \eta ^{\nu  \sigma } k_2{}^{\lambda
   } k_2{}^{\rho }+2 \eta ^{\mu  \sigma } \eta ^{\nu  \tau }
   k_2{}^{\lambda } k_2{}^{\rho }-2 \eta ^{\mu  \nu } \eta ^{\sigma  \tau
   } k_2{}^{\lambda } k_2{}^{\rho }+2 \eta ^{\lambda  \nu } \eta ^{\sigma
    \tau } k_2{}^{\mu } k_2{}^{\rho }+2 \eta ^{\lambda  \mu } \eta
   ^{\sigma  \tau } k_2{}^{\nu } k_2{}^{\rho }+\eta ^{\nu  \tau } \eta
   ^{\rho  \sigma } k_1{}^{\lambda } k_3{}^{\mu }+\eta ^{\nu  \sigma }
   \eta ^{\rho  \tau } k_1{}^{\lambda } k_3{}^{\mu }-\eta ^{\nu  \rho }
   \eta ^{\sigma  \tau } k_1{}^{\lambda } k_3{}^{\mu }+\eta ^{\lambda 
   \tau } \eta ^{\nu  \rho } k_1{}^{\sigma } k_3{}^{\mu }+\eta ^{\lambda 
   \nu } \eta ^{\rho  \tau } k_1{}^{\sigma } k_3{}^{\mu }+\eta ^{\lambda 
   \sigma } \eta ^{\nu  \rho } k_1{}^{\tau } k_3{}^{\mu }+\eta ^{\lambda 
   \nu } \eta ^{\rho  \sigma } k_1{}^{\tau } k_3{}^{\mu }+\eta ^{\nu 
   \tau } \eta ^{\rho  \sigma } k_2{}^{\lambda } k_3{}^{\mu }+\eta ^{\nu 
   \sigma } \eta ^{\rho  \tau } k_2{}^{\lambda } k_3{}^{\mu }+\eta
   ^{\lambda  \tau } \eta ^{\rho  \sigma } k_2{}^{\nu } k_3{}^{\mu }+\eta
   ^{\lambda  \sigma } \eta ^{\rho  \tau } k_2{}^{\nu } k_3{}^{\mu }+\eta
   ^{\lambda  \tau } \eta ^{\nu  \sigma } k_2{}^{\rho } k_3{}^{\mu }+\eta
   ^{\lambda  \sigma } \eta ^{\nu  \tau } k_2{}^{\rho } k_3{}^{\mu }+\eta
   ^{\mu  \tau } \eta ^{\rho  \sigma } k_1{}^{\lambda } k_3{}^{\nu }+\eta
   ^{\mu  \sigma } \eta ^{\rho  \tau } k_1{}^{\lambda } k_3{}^{\nu }-\eta
   ^{\mu  \rho } \eta ^{\sigma  \tau } k_1{}^{\lambda } k_3{}^{\nu }+\eta
   ^{\lambda  \tau } \eta ^{\mu  \rho } k_1{}^{\sigma } k_3{}^{\nu }+\eta
   ^{\lambda  \mu } \eta ^{\rho  \tau } k_1{}^{\sigma } k_3{}^{\nu }+\eta
   ^{\lambda  \sigma } \eta ^{\mu  \rho } k_1{}^{\tau } k_3{}^{\nu }+\eta
   ^{\lambda  \mu } \eta ^{\rho  \sigma } k_1{}^{\tau } k_3{}^{\nu }+\eta
   ^{\mu  \tau } \eta ^{\rho  \sigma } k_2{}^{\lambda } k_3{}^{\nu }+\eta
   ^{\mu  \sigma } \eta ^{\rho  \tau } k_2{}^{\lambda } k_3{}^{\nu }+\eta
   ^{\lambda  \tau } \eta ^{\rho  \sigma } k_2{}^{\mu } k_3{}^{\nu }+\eta
   ^{\lambda  \sigma } \eta ^{\rho  \tau } k_2{}^{\mu } k_3{}^{\nu }+\eta
   ^{\lambda  \tau } \eta ^{\mu  \sigma } k_2{}^{\rho } k_3{}^{\nu }+\eta
   ^{\lambda  \sigma } \eta ^{\mu  \tau } k_2{}^{\rho } k_3{}^{\nu }+2
   \eta ^{\lambda  \tau } \eta ^{\rho  \sigma } k_3{}^{\mu } k_3{}^{\nu
   }+2 \eta ^{\lambda  \sigma } \eta ^{\rho  \tau } k_3{}^{\mu }
   k_3{}^{\nu }-2 \eta ^{\lambda  \rho } \eta ^{\sigma  \tau } k_3{}^{\mu
   } k_3{}^{\nu }+\eta ^{\mu  \tau } \eta ^{\nu  \rho } k_1{}^{\lambda }
   k_3{}^{\sigma }+\eta ^{\mu  \rho } \eta ^{\nu  \tau } k_1{}^{\lambda }
   k_3{}^{\sigma }+\eta ^{\lambda  \nu } \eta ^{\mu  \rho } k_1{}^{\tau }
   k_3{}^{\sigma }+\eta ^{\lambda  \mu } \eta ^{\nu  \rho } k_1{}^{\tau }
   k_3{}^{\sigma }+\eta ^{\mu  \tau } \eta ^{\nu  \rho } k_2{}^{\lambda }
   k_3{}^{\sigma }+\eta ^{\mu  \rho } \eta ^{\nu  \tau } k_2{}^{\lambda }
   k_3{}^{\sigma }-\eta ^{\mu  \nu } \eta ^{\rho  \tau } k_2{}^{\lambda }
   k_3{}^{\sigma }+\eta ^{\lambda  \tau } \eta ^{\nu  \rho } k_2{}^{\mu }
   k_3{}^{\sigma }+\eta ^{\lambda  \nu } \eta ^{\rho  \tau } k_2{}^{\mu }
   k_3{}^{\sigma }+\eta ^{\lambda  \tau } \eta ^{\mu  \rho } k_2{}^{\nu }
   k_3{}^{\sigma }+\eta ^{\lambda  \mu } \eta ^{\rho  \tau } k_2{}^{\nu }
   k_3{}^{\sigma }-\eta ^{\lambda  \tau } \eta ^{\mu  \nu } k_2{}^{\rho }
   k_3{}^{\sigma }+\eta ^{\lambda  \nu } \eta ^{\mu  \tau } k_2{}^{\rho }
   k_3{}^{\sigma }+\eta ^{\lambda  \mu } \eta ^{\nu  \tau } k_2{}^{\rho }
   k_3{}^{\sigma }+2 \eta ^{\lambda  \rho } \eta ^{\nu  \tau } k_3{}^{\mu
   } k_3{}^{\sigma }+2 \eta ^{\lambda  \rho } \eta ^{\mu  \tau }
   k_3{}^{\nu } k_3{}^{\sigma }+\eta ^{\mu  \sigma } \eta ^{\nu  \rho }
   k_1{}^{\lambda } k_3{}^{\tau }+\eta ^{\mu  \rho } \eta ^{\nu  \sigma }
   k_1{}^{\lambda } k_3{}^{\tau }+\eta ^{\lambda  \nu } \eta ^{\mu  \rho
   } k_1{}^{\sigma } k_3{}^{\tau }+\eta ^{\lambda  \mu } \eta ^{\nu  \rho
   } k_1{}^{\sigma } k_3{}^{\tau }+\eta ^{\mu  \sigma } \eta ^{\nu  \rho
   } k_2{}^{\lambda } k_3{}^{\tau }+\eta ^{\mu  \rho } \eta ^{\nu  \sigma
   } k_2{}^{\lambda } k_3{}^{\tau }-\eta ^{\mu  \nu } \eta ^{\rho  \sigma
   } k_2{}^{\lambda } k_3{}^{\tau }+\eta ^{\lambda  \sigma } \eta ^{\nu 
   \rho } k_2{}^{\mu } k_3{}^{\tau }+\eta ^{\lambda  \nu } \eta ^{\rho 
   \sigma } k_2{}^{\mu } k_3{}^{\tau }+\eta ^{\lambda  \sigma } \eta
   ^{\mu  \rho } k_2{}^{\nu } k_3{}^{\tau }+\eta ^{\lambda  \mu } \eta
   ^{\rho  \sigma } k_2{}^{\nu } k_3{}^{\tau }-\eta ^{\lambda  \sigma }
   \eta ^{\mu  \nu } k_2{}^{\rho } k_3{}^{\tau }+\eta ^{\lambda  \nu }
   \eta ^{\mu  \sigma } k_2{}^{\rho } k_3{}^{\tau }+\eta ^{\lambda  \mu }
   \eta ^{\nu  \sigma } k_2{}^{\rho } k_3{}^{\tau }+2 \eta ^{\lambda 
   \rho } \eta ^{\nu  \sigma } k_3{}^{\mu } k_3{}^{\tau }+2 \eta
   ^{\lambda  \rho } \eta ^{\mu  \sigma } k_3{}^{\nu } k_3{}^{\tau }-2
   \eta ^{\lambda  \rho } \eta ^{\mu  \nu } k_3{}^{\sigma } k_3{}^{\tau
   }+2 \eta ^{\lambda  \nu } \eta ^{\mu  \rho } k_3{}^{\sigma }
   k_3{}^{\tau }+2 \eta ^{\lambda  \mu } \eta ^{\nu  \rho } k_3{}^{\sigma
   } k_3{}^{\tau }-\eta ^{\lambda  \tau } \eta ^{\mu  \sigma } \eta ^{\nu
    \rho } k_1\cdot k_2-\eta ^{\lambda  \sigma } \eta ^{\mu  \tau } \eta
   ^{\nu  \rho } k_1\cdot k_2-\eta ^{\lambda  \tau } \eta ^{\mu  \rho }
   \eta ^{\nu  \sigma } k_1\cdot k_2+\eta ^{\lambda  \rho } \eta ^{\mu 
   \tau } \eta ^{\nu  \sigma } k_1\cdot k_2-\eta ^{\lambda  \sigma } \eta
   ^{\mu  \rho } \eta ^{\nu  \tau } k_1\cdot k_2+\eta ^{\lambda  \rho }
   \eta ^{\mu  \sigma } \eta ^{\nu  \tau } k_1\cdot k_2+2 \eta ^{\lambda 
   \tau } \eta ^{\mu  \nu } \eta ^{\rho  \sigma } k_1\cdot k_2-\eta
   ^{\lambda  \nu } \eta ^{\mu  \tau } \eta ^{\rho  \sigma } k_1\cdot
   k_2-\eta ^{\lambda  \mu } \eta ^{\nu  \tau } \eta ^{\rho  \sigma }
   k_1\cdot k_2+2 \eta ^{\lambda  \sigma } \eta ^{\mu  \nu } \eta ^{\rho 
   \tau } k_1\cdot k_2-\eta ^{\lambda  \nu } \eta ^{\mu  \sigma } \eta
   ^{\rho  \tau } k_1\cdot k_2-\eta ^{\lambda  \mu } \eta ^{\nu  \sigma }
   \eta ^{\rho  \tau } k_1\cdot k_2-2 \eta ^{\lambda  \rho } \eta ^{\mu 
   \nu } \eta ^{\sigma  \tau } k_1\cdot k_2+2 \eta ^{\lambda  \nu } \eta
   ^{\mu  \rho } \eta ^{\sigma  \tau } k_1\cdot k_2+2 \eta ^{\lambda  \mu
   } \eta ^{\nu  \rho } \eta ^{\sigma  \tau } k_1\cdot k_2-\eta ^{\lambda
    \tau } \eta ^{\mu  \sigma } \eta ^{\nu  \rho } k_1\cdot k_3-\eta
   ^{\lambda  \sigma } \eta ^{\mu  \tau } \eta ^{\nu  \rho } k_1\cdot
   k_3-\eta ^{\lambda  \tau } \eta ^{\mu  \rho } \eta ^{\nu  \sigma }
   k_1\cdot k_3+2 \eta ^{\lambda  \rho } \eta ^{\mu  \tau } \eta ^{\nu 
   \sigma } k_1\cdot k_3-\eta ^{\lambda  \sigma } \eta ^{\mu  \rho } \eta
   ^{\nu  \tau } k_1\cdot k_3+2 \eta ^{\lambda  \rho } \eta ^{\mu  \sigma
   } \eta ^{\nu  \tau } k_1\cdot k_3+2 \eta ^{\lambda  \tau } \eta ^{\mu 
   \nu } \eta ^{\rho  \sigma } k_1\cdot k_3-\eta ^{\lambda  \nu } \eta
   ^{\mu  \tau } \eta ^{\rho  \sigma } k_1\cdot k_3-\eta ^{\lambda  \mu }
   \eta ^{\nu  \tau } \eta ^{\rho  \sigma } k_1\cdot k_3+2 \eta ^{\lambda
    \sigma } \eta ^{\mu  \nu } \eta ^{\rho  \tau } k_1\cdot k_3-\eta
   ^{\lambda  \nu } \eta ^{\mu  \sigma } \eta ^{\rho  \tau } k_1\cdot
   k_3-\eta ^{\lambda  \mu } \eta ^{\nu  \sigma } \eta ^{\rho  \tau }
   k_1\cdot k_3-2 \eta ^{\lambda  \rho } \eta ^{\mu  \nu } \eta ^{\sigma 
   \tau } k_1\cdot k_3+\eta ^{\lambda  \nu } \eta ^{\mu  \rho } \eta
   ^{\sigma  \tau } k_1\cdot k_3+\eta ^{\lambda  \mu } \eta ^{\nu  \rho }
   \eta ^{\sigma  \tau } k_1\cdot k_3-\eta ^{\lambda  \tau } \eta ^{\mu 
   \sigma } \eta ^{\nu  \rho } k_2\cdot k_3-\eta ^{\lambda  \sigma } \eta
   ^{\mu  \tau } \eta ^{\nu  \rho } k_2\cdot k_3-\eta ^{\lambda  \tau }
   \eta ^{\mu  \rho } \eta ^{\nu  \sigma } k_2\cdot k_3+2 \eta ^{\lambda 
   \rho } \eta ^{\mu  \tau } \eta ^{\nu  \sigma } k_2\cdot k_3-\eta
   ^{\lambda  \sigma } \eta ^{\mu  \rho } \eta ^{\nu  \tau } k_2\cdot
   k_3+2 \eta ^{\lambda  \rho } \eta ^{\mu  \sigma } \eta ^{\nu  \tau }
   k_2\cdot k_3+\eta ^{\lambda  \tau } \eta ^{\mu  \nu } \eta ^{\rho 
   \sigma } k_2\cdot k_3-\eta ^{\lambda  \nu } \eta ^{\mu  \tau } \eta
   ^{\rho  \sigma } k_2\cdot k_3-\eta ^{\lambda  \mu } \eta ^{\nu  \tau }
   \eta ^{\rho  \sigma } k_2\cdot k_3+\eta ^{\lambda  \sigma } \eta ^{\mu
    \nu } \eta ^{\rho  \tau } k_2\cdot k_3-\eta ^{\lambda  \nu } \eta
   ^{\mu  \sigma } \eta ^{\rho  \tau } k_2\cdot k_3-\eta ^{\lambda  \mu }
   \eta ^{\nu  \sigma } \eta ^{\rho  \tau } k_2\cdot k_3-2 \eta ^{\lambda
    \rho } \eta ^{\mu  \nu } \eta ^{\sigma  \tau } k_2\cdot k_3+2 \eta
   ^{\lambda  \nu } \eta ^{\mu  \rho } \eta ^{\sigma  \tau } k_2\cdot
   k_3+2 \eta ^{\lambda  \mu } \eta ^{\nu  \rho } \eta ^{\sigma  \tau }
   k_2\cdot k_3
\end{polynomial}
\end{tiny}

\begin{problem} Using symbolic manipulation software, derive these Feynman rules starting from their respective actions.    \end{problem}

Recall that Feynman rules for off-shell Green's functions depend on the gauge, and hence are unphysical. To talk about physical observables we need to take the external particles to be physical --- ``on shell".  So this means particles $k_1$, $k_2$, and $k_3$ have to be massless if we want to talk about the three-particle scattering amplitudes of either gluons or gravitons. Note that this means that $k_i \cdot k_j=0$.   This also means, for 3-point amplitudes, that $k_i$ have to be complex.  This will not be a bother to us, but it is something to pay attention to if you're going to plug in numbers.  To get a physical amplitude for external gluons we must contract \eqn{ym3ptOnShell} with three transverse polarization vectors.  Gluons can have either positive or negative helicity polarization states in four dimensions, satisfying $\varepsilon^{\pm}_i \cdot k_i=0$.   In pure Yang-Mills at three-points, in four dimensions, we can only have two negative-helicity gluons scattering into a positive helicity gluon or vice-versa,  either choice will restrict some terms in our expressions.   We are free\footnote{See, e.g., eqs.~(31)--(35) of ref.~\cite{LanceTasi1996}.} to choose our polarization vectors such that $\varepsilon^{\pm}_i \cdot \varepsilon^{\pm}_j =0$.  Considering, without loss of generality, gluons 1 and 2 to be negative helicity, and gluon 3 to be positive helicity we have our on-shell Feynman rule for Yang-Mills theory (YM):
\begin{equation}
\label{ym3ptOnShell}
\left \langle\frac{\delta S^3}{
 \delta A^{-a}_\mu  \delta A^{-b}_\sigma  \delta A^{+c}_\rho} \right \rangle_{\text{on-shell}}\to -2   i f^{abc}  \left(k_1{}^{\sigma } \eta ^{\mu  \rho }-k_2{}^{\mu }
   \eta ^{\rho  \sigma }\right)\,.
\end{equation}
Here I also used conservation of momentum to express $k_3=-k_1-k_2$.

Now gravitons are also massless and also must be in one of two states in four dimensions, which can be taken to have helicity $\pm2$.   Their polarization tensors then factorize into a product of spin-1 polarization vectors: $\varepsilon^{\pm\pm}_i{}^{\mu\nu} = \varepsilon^{\pm}_i{}^{\mu}  \varepsilon^{\pm}_i{}^{\nu}$.  The constraints from dotting the graviton Feynman rule into polarizations are much more dramatic!  Taking into account all constraints of masslessness and properties of polarization vectors we find immediately that,
\begin{align*}
\varepsilon^{-}_1{}_{\mu\nu} \varepsilon^{-}_2{}_{\sigma\tau} \varepsilon^{+}_3{}_{\rho\lambda}  \frac{\delta S^3}{
\delta \varphi_{\mu\nu}  \delta \varphi_{\sigma\tau}  \delta \varphi_{\rho\lambda}}  &=
4 \left(\left(\varepsilon^{-} _1\cdot \varepsilon _3^{+}\right) \left(k_1\cdot
   \varepsilon _2^{-}\right)-\left(\varepsilon _2^{-}\cdot \varepsilon _3^{+}\right)
   \left(k_2\cdot \varepsilon _1^{-}\right)\right){}^2\,
\end{align*}
so the on-shell graviton Feynman rule at 3-points factorizes:
\begin{equation}
\label{grav3ptOnShell}
\left \langle
 \frac{\delta S^3}{
\delta \varphi^-_{\mu\nu}  \delta \varphi^-_{\sigma\tau}  \delta \varphi^+_{\rho\lambda}}
\right \rangle_{\text{on-shell}}\to 4 \left(k_1{}^{\sigma } \eta ^{\mu  \rho }-k_2{}^{\mu }
   \eta ^{\rho  \sigma }\right) \left(k_1{}^{\tau } \eta ^{\nu  \lambda }-k_2{}^{\nu }
   \eta ^{\lambda  \tau }\right)\,.
\end{equation}

\begin{problem}
Find non-trivial on-shell four-dimensional complex momenta $k_i$ that satisfy:  $k_i^2=0$ and $k_1+k_2+k_3=0$.    Either the spinor helicity\footnote{See references given in the bonus appendix for these lecture notes, or consider the expressions given in \eqn{helicityProducts}.} angle-products $\langle i , j\rangle$  between these momenta will all vanish or the square products $[ i, j]$ will all vanish.   This will tell you whether you've got the appropriate momenta to either consider a $({-}{-}{+})$ scattering amplitude or a $({+}{+}{-})$ scattering amplitude.  Find the appropriate polarization tensors (see \cite{LanceTasi1996} for help), and dot them into the above off-shell Feynman rules.  Using  your momenta,  polarization tensors, and the off-shell Feynman rules, verify \eqns{ym3ptOnShell}{grav3ptOnShell}. 
\end{problem}

We can take away two immediate lessons.  First, that Feynman rules are a horribly complicated way to calculate in these theories --- intermediate expressions are huge, but end-result on-shell physical quantities are much more tractable.  More to the point, even back in 1967, if we had looked at physical quantities, we would have discovered that perturbatively, at least, gravitons want to behave as gluons, but with a dynamical ``color group'':  the ``$f^{abc}$" of the gravity theory is another copy of the kinematic factor.   

This double-copy behavior between on-shell tree-level scattering amplitudes for gauge theory and gravity was first realized for all multiplicity at tree level in string theory by Kawai, Lewellen, and Tye~\cite{KLT} when they observed that closed-string tree-level scattering amplitudes can be written as the sum over permutations of products of (ordered) open tree-level scattering amplitudes.   This squaring of the 3-point Yang-Mills scattering amplitude to get the 3-point gravity scattering amplitude can be seen as the first example of the low-energy limit of the relations that they found.  I will talk more about the field-theory version of the KLT relations towards the end of the next lecture.  

Conversely, from a graph perspective, this squaring can be seen as the simplest possible example of the graph-based double-copy of individual numerators of ref.~\cite{BCJ}.  Since the three-point amplitude only has one graph, and no propagator, one can think of the $f^{abc}$ as the $c(g)$ of the graph, being replaced by another $n(g)$ to get the gravity theory amplitude.   I will spend the majority of the next lecture describing how we can describe scattering amplitudes in terms of graphs.


\section{Learning about trees, and learning about graphs}
 
\subsection{Color-dressed and color-ordered tree-level amplitudes}
Already you have had lectures at this TASI from Zvi Bern and Mark Spradlin that discussed the approach of on-shell recursion~\cite{BCFW} to build higher-point color-ordered (color-stripped) tree-level amplitudes from lower-order tree-level amplitudes.  Mark also discussed the Grassmann-encoding of external states using on-shell superspace~\cite{Nair, GGK, FreedmanGenerating}.  I  take it as given that you can write down a color-ordered tree-level amplitude in some representation.  If not please consult some of the other lectures in this program, and the excellent introductions in refs.~\cite{LanceTasi1996, ElvangHuangBook,LanceTasi2013}.  I have  included a bonus appendix with some quick tree-level data just so that you can have  data ready at hand --- but this appendix will be no substitute for understanding where these expressions originate.

In Zvi Bern's first lecture of this year's TASI, he taught us that we can express a color-dressed tree-level amplitude in terms of a trace over the generating functions of color and color-ordered (color-stripped) partial amplitudes:
\be
\Atree{m}(12\ldots m) = \sum_{\sigma\in {\rm perm}(2\ldots m)}
{\rm Tr} \left( \cT{1} \cT{\sigma(2)} \cT{\sigma(3)}\cdots \cT{\sigma(m)} \right) 
\AtreeCO{m}\left( 1,  \sigma  \right) \,.
\ee
In this expression the full color-dressed amplitude  $\Atree{m}$ on the left is given by a  sum over all permutations $\sigma$ of labels $2,3,\ldots,m$ on the right hand side, and $\AtreeCO{m}$ are the color ordered tree amplitudes associated with such permutations.  We can rewrite the full color-dressed scattering amplitude in terms of purely cubic graphs (graphs with only trivalent vertices):
\be
\Atree{m}(12\ldots m) = \sum_{g \in \GraphsTree{m}} \frac{c(g) n(g)}{d(g)} \,,
\label{CubicExp}
\ee
where $\GraphsTree{m}$ is the set of cubic tree graphs with distinct momentum routings that have $0$ loops (or closed cycles) and $m$ external edges,  $c(g)$ is obtained by dressing every vertex with $f^{abc}$ structure constants, $d(g)$ is the product of the square of the momentum running along each propagator, and $n(g)$ is the theory-specific kinematic numerator weight (everything else). In terms of cubic graphs, a given color-ordered tree --- i.e., one associated with a particular permutation of legs $2,\ldots,m$, is given by:
\be
\AtreeCO{m}(1, \sigma ) = \sum_{g \in \GraphsTreeCO{m} } \frac{ n(g)}{d(g)} \,.
\ee
where $\GraphsTreeCO{m}$ are the tree graphs that can contribute to the given color-ordering $\sigma$.  

I will tell you how to write down color-ordered graphs, and color-dressed ``unordered" graphs a little later.  But let me get a point about cubic graphs out of the way.  I have claimed we can write Yang-Mills amplitudes, and consequently also gravity amplitudes, solely in terms of cubic graphs. However, if you're familiar with Feynman rules you might be suspicious.  For gluons, yes we have 3-point interactions, but we also have Feynman rules for four-point interactions. For gravity, not only do we have 3- and 4-point graviton interactions, but we have a Feynman rule 
for 5-point graviton interactions, a Feynman rule for 6-point graviton interactions; in fact, for each multiplicity there is an additional Feynman rule. But here I am claiming we can write all of this --- encode it --- in terms of just cubic graphs.

\subsubsection{We only have to worry about graphs with trivalent vertices.}
Trivalent vertices are a fancy way of saying vertices with only 3 edges.  We call graphs with all trivalent vertices cubic graphs.

There are two points to realize here.  One is that our graphs are {\em not} Feynman graphs.  They are something different --- they are abstract graphs involving only cubic vertices, independent (unlike Feynman graphs) of the state of particles on internal lines.  They will track conservation of color and conservation of momenta at vertices.  We can think of each of these graphs as representing a sum over all internal states of Feynman graphs that have the same routing of color and momenta.  

The second point is that assigning the information associated with higher-point graphs to the graphs associated with cubic vertices is nothing more than book-keeping. What are the graphs doing? They're tracking the contributions associated with certain routings of momentum and certain routings of color for gauge theories.  

First let us ignore color, and just talk about kinematics.  So consider a gravity theory, or the color-ordered contribution to gluon scattering.  We know that Yang-Mills theory require four-point Feynman rules, and gravity theories require additional higher-point Feynman rules.   Any contributions involving higher than three-point interactions are called {\em contact terms}.   What is the difference between the book-keeping associated with a four-point contact term and just the cubic Feynman rules? Let us talk about a generic cubic graph $g_3$ and a related graph $g_4$ that is identical to $g_3$ except that one of $g_3$'s edges, say with momentum $p$, has been replaced with a four-point vertex. The graph $g_4$ has all the same propagators as $g_3$ except that it is missing a $p^2$.  So 
\be  
d(g_3)=p^2 d(g_4),
\ee
 recalling that we use $d(g)$ to denote the mapping of a graph $g$ to its propagator contributions. Let us say that the original kinematic numerator contribution for the cubic graph is $n(g_3)$, and the one for the contact graph is $n(g_4)$.  Highlighting the contribution of these two graphs to some scattering amplitude $\Aloop{L}{m}$ we see the following:
\begin{align}
\Aloop{L}{m}  &=  \ldots  +\frac{n(g_3)}{d(g_3)} + \frac{n(g_4)}{d(g_4)}+ \ldots \\  
    &=  \ldots  +\frac{ n(g_3)}{d(g_3)} + \frac{n(g_4) }{ d(g_4)  } \times \frac{p^2}{p^2} + \ldots\\
    &=  \ldots+  \frac{ n(g_3)}{d(g_3)} + \frac{n(g_4) p^2}{ d(g_3)} + \ldots\\
    &=  \ldots + \frac{{\widehat n}(g_3)}{d(g_3)} + \dots\,,
\end{align}
where, in the last line, we have simply defined a new mapping from $g_3$ to its kinematic numerator contribution:  ${\widehat n}(g_3) \equiv n(g_3) +p^2 n(g_4)$.  In doing so we have absorbed the contact contribution associated with the 4-point graph $g_4$ into $g_3$.  The absorption of higher-point contact terms follows in exactly the same manner.

Now let us discuss color.   Let us restrict to scattering processes in which all particles transform in the adjoint representation --- which involves dressing cubic vertices simply with $f^{abc}$.   What does color do differently with four-point vertices?   Nothing at all, except the four-point contact term will have some color factors associated with it --- but fortunately just the color factors associated with the three different ways of expanding a four-point contact term into underlying cubic vertices connected by an edge.  Let us say that the four edges of the four-point contact term are labeled with outgoing momenta: $k_1, k_2, k_3, k_4$.  The Feynman rule associated with such a contact term is generically: 
\begin{align}
F_{4pt} &\equiv f^{a_1 a_2 b}f^{b a_3 a_4} n_{12} +  f^{a_1 a_3 b}f^{b a_2 a_4} n_{13} +   f^{a_4 a_1 b}f^{b a_2 a_3} n_{14} \\
            &= c_{12} n_{12} +  c_{13} n_{13} +  c_{14} n_{14} \\
            &= c_{12} n_{12} \frac{(k_1+k_2)^2}{(k_1+k_2)^2} +  c_{13} n_{13} \frac{(k_1+k_3)^2}{(k_1+k_3)^2} +  c_{14} n_{14} \frac{(k_1+k_4)^2}{(k_1+k_4)^2} \,.
\end{align} 
The color factors $c_{1i}$ are products of structure constants $f^{abc}$.  The $n_{1i}$ stand for the kinematic parts of the gluonic Feynman rule (which in general just involves tensor contractions); their precise form is irrelevant to the point here.

One should recognize that the three color factors $c_{1i}$ are exactly those associated with each of the 3 ways of expanding out a four-point vertex into two three-point vertices connected by an edge. So any quartic color-dressed contact term has three graphs it could be shared between.  As the color-factors obey Jacobi identities,  we can rewrite any of the $c_{1i}$ in terms of the other two: i.e.~$c_{12} = c_{13}+c_{14}$.  As this works both with color and kinematic weights of graphs, there is no barrier to assigning higher-point contact terms to cubic graphs in either adjoint gauge theory or gravity theory scattering amplitudes.  These different choices, our book-keeping regarding how to assign contact information to cubic graphs, we will refer to as a choice between different {\em generalized} gauges.

\subsubsection{Our graphs want structure.}

Recall from \eqns{antiCol}{antiKin} that both the color factors and the kinematic weights associated with the Yang-Mills graphs care about antisymmetry around vertices:
\begin{align}
c(g)&=-c(\overline{g}) \,,\\
n(g)&=-n(\overline{g}) \,.
\end{align}
Our graphs need to be able to encode this antisymmetry.  Typically one thinks of graphs as a set of vertices, or nodes, connected by legs, or edges. The edges may have a direction associated with them, in which case we have a ``directed graph''.  We want our vertices to have structure too --- to care about order.  

A handy way to represent trivalent graphs is just to list each vertex as a triplet of entries labeling the three edges that end at that vertex, and give the entries signs to indicate direction.  For a four-point graph:
\begin{equation}
g_s(a,b,c,d)  =  \left\{  \begin{matrix} (a,b,i)\\
(-i,c,d) 
\end{matrix} \right\},
\end{equation}
where every edge is labeled outgoing to its vertex, and $i$ extends from vertex $(a,b,i)$ to vertex $(-i,c,d)$.  We will employ {\em necklaces}, ordered lists whose cyclic permutations are identified, to represent our vertices.  For cubic necklaces this means they obey
\begin{equation}
(i,j,k) = (j,k,i) = (k,i,j) \ne (i,k,j).
\end{equation}
Then our graph representation will encode all the structure we need.  
\begin{equation}
 \left\{  \begin{matrix} (a,b,i)\\
(-i,c,d) 
\end{matrix} \right\}  = \left\{  \begin{matrix} (b,i,a)\\
(c,d,-i) 
\end{matrix} \right\}  =   \left\{  \begin{matrix} 
(c,d,-i) \\
(a,b,i)
\end{matrix} \right\} \ne    \left\{  \begin{matrix} 
(a,i,b) \\
(c,d,-i)
\end{matrix} \right\} \,.
\end{equation}
We will label an external edge, i.e.~an edge that connects to an external leg, simply with that leg's label.

Notice that we do not care about the order in which we list our vertices, only the order in which we attach our edges to cubic vertices, up to cyclic permutations. In fact when worrying about numerator and color factors we'll go further, and identify graphs up to even number of odd (flip) permutations of vertices. So as far as numerators and color factors are concerned, we will equate any two graphs that differ only by an even number of vertex flips. 

So:
\begin{equation}
n(g_s(a,b,c,d) ) \equiv n( \left\{  \begin{matrix} (a,b,i)\\
(-i,c,d) 
\end{matrix} \right\}  )=  n( \left\{  \begin{matrix} (a,i,b)\\
(c,-i,d) 
\end{matrix} \right\} ) 
\end{equation}
and
\begin{equation}
n(\overline{g}_s(a,b,c,d) ) \equiv n( \left\{  \begin{matrix} (a,i,b)\\
(-i,c,d) 
\end{matrix} \right\}  )=  n(\left\{  \begin{matrix} (a,b,i)\\
(c,-i,d) 
\end{matrix} \right\} ).
\end{equation}

Our map $c(g)$ from this graph representation to the color factor is straightforward,
\begin{align}
c\left(  \left\{ \begin{matrix} v_1\\ v_2 \\ \vdots \\ v_n  \end{matrix}\right\}  \right)  &=  \prod_{i=1}^n c(v_i) \\
&{\text{with}} \nonumber\\
   c( v_a \equiv (i,j,k) ) &= f^{ijk},
\end{align}
 and it is an easy exercise to see that $c(g) = - c(\overline{g})$.

\begin{problem}[From a graph's edge to a Jacobi relation!]
\label{opProblem} 
Start with the edge $e_g$ of some cubic graph $g$.  We can ask what four-point subgraph of $g$ contains $e_g$.   Let us refer to its four bounding edges as $\{a,b,c,d\}$. Then the vertex necklaces containing $e_g$ are ${\cal V}_g \equiv \{ (a,b,e_g), (-e_g,c,d) \}$.  We will refer to all vertices that do not touch edge $e_g$ as $\left({\cal V}(g) \setminus {\cal V}(e_g)\right)$.

 Now we will create three operators  $\widehat{s}$, $\widehat{t}$, and $\widehat{u}$ that each take an edge, and give back a graph.   Consider first the trivial operator $\widehat{s}$,
\be
\label{shat}
{\widehat{s}} ( e_g) = \left\{ \bigg({\cal V}(g) \setminus {\cal V}(e_g)\bigg) \cup   \left\{  \begin{matrix}
 (a,b, e) \\
  (-e,d,f)  \end{matrix} \right\} \right \} = g\,.
\ee
Note that this graph has all the same vertices as $g$, so all this operator did was give back the graph that belongs to the edge $e_g$.  This is fine; it's handy to have an operator that does that.

Now consider a non-trivial operator $\widehat{t}$,
\be
\label{that}
{\widehat{t}} ( e_g) = \left\{  \bigg({\cal V}(g) \setminus {\cal V}(e_g)\bigg)\cup    \left\{  \begin{matrix}
 (d, a, t) \\
  (-t,b,c)  \end{matrix} \right\} \right \}
\ee
This new graph, $g_t \equiv \widehat{t} ( e_g) $  is a different graph than $g$.  It has all the same vertices except for two, but these two have a different connectivity --- now edges $a$ and $d$ meet in a vertex, as do edges $c$ and $b$.   Additionally there is no edge $e$, rather $\widehat{t}(e_g)$ has a new edge $t$.  The edge $e$ is nowhere to be found in this graph.

Finally ready for the problem:  Define an operator $\widehat{u}$ such that in all cases:
\be
c\circ \widehat{s}(  e_g) =c\circ\widehat{t} ( e_g) + c\circ\widehat{u}(e_g),
\ee and for color-dual representations  
\be
n\circ \widehat{s} ( e_g) =n\circ \widehat{t} (e_g) + n\circ \widehat{u} (e_g),
\ee
where $c\circ\widehat{t}(e_g) \equiv c(\widehat{t}(e_g))$, etc.
\end{problem}

\subsubsection{External color-order of a tree graph.}

Consider the collapse operator,  ${\cal{C}}_e(g)$, which takes an internal edge $e$ of graph $g$ and collapses it. In terms of our vertex-based representation above, this means taking the two necklaces which contain the reference to edge $e$:  $v(e)$ and $v(-e)$ and merging them into a single necklace which maintains the relative order between the necklaces.

\be
\label{collapseOperator}
{\cal{C}}_e \left(   \left\{  \begin{matrix} \vdots \\
 (a,b, e) \\
  (-e,d,f)\\ \vdots \end{matrix} \right\} \right)  =    \left\{  \begin{matrix} \vdots \\
 (a,b, d,f) \\ \vdots \end{matrix} \right\} \,.
\ee

Note that the two vertices being merged can always be put in the above form where the outgoing label of the to-be-collapsed edge is rotated to the rightmost slot of the necklace, and the ingoing label of the to-be-collapsed edge is rotated to the leftmost slot of the necklace. One can see that the operation proceeds straightforwardly to collapse higher-point vertices by realizing that the $b$ and $d$ in \eqn{collapseOperator} can stand in for lists of ordered edges connecting to that vertex.  This operator is only undefined when collapsing cycles (i.e.~when both $e$ and $-e$ belong in the same vertex), which we will not consider.

Restricting ourselves to $m$-point trees, we can use ${\cal C}_e$ repeatedly to collapse all internal lines. The resulting necklace, a single $m$-tuple, is defined to be the {\em external color-order} ${\text{ECO}}(g)$ of the graph.  As it is a necklace, all cyclic permutations are identified.  For four points,
\be
(a,b,c,d) = (b,c,d,a) = (c,d,a,b) = (d,a,b,c)\,.
\ee

Note that two graphs that have the same color factor, and whose numerators have been identified, can have different external color-orders! Consider:
\begin{align*}
 \ECO\left( g_{ab} =  \left\{  \begin{matrix} (a,b,s)\\
(-s,c,d)  \end{matrix}\right\} \right)  &\equiv (a,b,c,d)   \\
&\ne\\
 \ECO\left( g_{ba} = \left\{  \begin{matrix} (b,a,s)\\
(-s,d,c)  \end{matrix}\right\} \right)  &\equiv (b,a,d,c) 
\end{align*}
even though $c(g_{ab})=c(g_{ba})$.

\begin{problem}
How many cubic graphs can one write down at the four-point level?  What are they?  How many unique external orders; how many naively distinct color-ordered amplitudes? Contrast with the number of graphs that are distinct under color-factors.  What relations can you immediately write down between color-ordered amplitudes after imposing antisymmetry of numerator factors? What relations can you write down after imposing the Jacobi relations?
\end{problem}
\begin{solution}
There are four ways of choosing the first external leg, three ways of choosing the second external leg, and  two ways of choosing the third external leg, so one might be tempted to say $4!$, but this over-counts by two as the two internal vertices can be swapped freely: there should be 12 distinct graphs.  Let us see how this comes about. Without loss of generality choose the first vertex to be the one that has $k_1$. So we consider first the 6 four-point graphs defined
\begin{align}
g_{12} &=  \left\{  \begin{matrix} (k_1,k_2,i)\\
(-i,k_3,k_4)  \end{matrix}\right\} & g_{21} &=  \left\{  \begin{matrix} (k_2,k_1,i)\\
(-i,k_4,k_3)  \end{matrix}\right\} \\
g_{41} &=  \left\{  \begin{matrix} (k_4,k_1,i)\\
(-i,k_2,k_3)  \end{matrix}\right\}  &
g_{14} &=  \left\{  \begin{matrix} (k_1,k_4,i)\\
(-i,k_3,k_2)  \end{matrix}\right\} \\
g_{31} &=  \left\{  \begin{matrix} (k_3,k_1,i)\\
(-i,k_4,k_2)  \end{matrix}\right\} &
g_{13} &=  \left\{  \begin{matrix} (k_1,k_3,i)\\
(-i,k_2,k_4)  \end{matrix}\right\} 
\end{align}
and then get their odd-permutation conjugates $\overline{g}_{12},\overline{g}_{21},\overline{g}_{41},\overline{g}_{14}, \overline{g}_{31}, \overline{g}_{13}$ (let us say odd in their second vertex permutation to be specific).  They have the following external orders:
\begin{align}
\ECO(g_{12}) &= (1234)&
\ECO(g_{21}) &=  (1432) &
\ECO(g_{41} )&=  (1234) \\
\ECO(g_{14} )&=  (1432) &
\ECO(g_{31}) &= (1423) &
\ECO(g_{13} )&=  (1324) \nn \\
\ECO(\overline{g}_{12}) &= (1243)&
\ECO(\overline{g}_{21}) &=  (1342) &
\ECO(\overline{g}_{41} )&=  (1324) \nn \\
\ECO(\overline{g}_{14} )&=  (1423) &
\ECO(\overline{g}_{31}) &= (1243) &
\ECO(\overline{g}_{13} )&=  (1342) \nn 
\end{align}
so in fact there are only $3!=6$ distinct external orders. So we can write down the 6 color-ordered tree-amplitudes associated with distinct color-orders.  (Any cyclic permutation would of course be the same.)
\begin{align}
\AtreeCO{4}(1234) &= \frac{n}{d} \circ  g_{12} +\frac{n}{d} \circ g_{41}  &
\AtreeCO{4}(1432) &= \frac{n}{d} \circ  g_{21} +\frac{n}{d} \circ g_{14}   \\
\AtreeCO{4} (1243)&= \frac{n}{d} \circ  \overline{g}_{12} +\frac{n}{d} \circ \overline{g}_{13}  &
\AtreeCO{4} (1342) &= \frac{n}{d} \circ  \overline{g}_{13} +\frac{n}{d} \circ \overline{g}_{21} \nn \\
\AtreeCO{4} (1324)&= \frac{n}{d} \circ  g_{13} +\frac{n}{d} \circ \overline{g}_{41}  &
\AtreeCO{4}(1423)&= \frac{n}{d} \circ  g_{31} + \frac{n}{d} \circ \overline{g}_{14} \nn
\end{align}
These 6 expressions are not all distinct, as we can see by using symmetric and anti-symmetric properties of the numerator function, and equivalence under even permutations of necklace orders.  By considering such symmetries, we find that:
\begin{align}
n_s\equiv n(g_{12})&=  n(g_{21}) = -n(\overline{g}_{12}) = -n(\overline{g}_{21}),  \\
n_u\equiv n(g_{13})&=  n(g_{31}) = -n(\overline{g}_{13})  = -n(\overline{g}_{31}),  \nn \\
n_t\equiv n(g_{14})&=  n(g_{41}) = -n(\overline{g}_{14}) = -n(\overline{g}_{41}). \nn
\end{align}
One should note that the color factors $c(g)$ also obey these relations.  Additionally, if one removes the negative signs, then they hold for the denominators $d(g)$:
\begin{align}
s\equiv d(g_{12})&=  d(g_{21}) = d(\overline{g}_{12}) = d(\overline{g}_{21})  \,, \\
u\equiv d(g_{13})&=  d(g_{31}) = d(\overline{g}_{13})  = d(\overline{g}_{31}) \,, \nn \\
t\equiv d(g_{14})&=  d(g_{41}) = d(\overline{g}_{14}) = d(\overline{g}_{41})  \,. \nn
\end{align}
Immediately we see that
\begin{align}
\AtreeCO{4}(1234) &=
\AtreeCO{4}(1432)=  \frac{n_s}{s} + \frac{n_t}{t} \,,\\
\AtreeCO{4} (1243)&= 
\AtreeCO{4} (1342)= - \frac{n_s}{s} -\frac{n_u}{u} \,,  \nn \\
\AtreeCO{4} (1324)&= \AtreeCO{4}(1423)= -\frac{n_t}{t}+\frac{n_u}{u} \,, \nn
\end{align}
giving us, as a byproduct, the relation:
\begin{equation}
\AtreeCO{4}(1423)= -(\AtreeCO{4}(1234) + \AtreeCO{4} (1342)).
\end{equation}
What happens if we impose Jacobi, $n_s = n_t + n_u$, between the numerators?  We find that all the color-ordered amplitudes can in fact be expressed in terms of just one color order: 
\begin{equation}
\AtreeCO{4} (1342) =  \frac{t}{u} \AtreeCO{4}(1234).
\end{equation}
\end{solution}

\begin{problem}
Verify that the $\widehat{t}$ operator  worked out in problem \ref{opProblem} does not change the color-order of any of the above graphs when applied to any edges.  What about the $\widehat{u}$ operator?
\end{problem}

\subsubsection{Relations between Color-Ordered Trees}
After working through that four-point problem, we can read off the first examples of a number of relations.
 
\begin{enumerate}
\item Cyclic Symmetry~\cite{LanceTasi1996}.  \be
\AtreeCO{m}(1,2,\ldots,m)= \AtreeCO{m}(2,\ldots,m,1)\,.
 \ee
This we get for free, simply from the fact that there is a graph-based decomposition and the definition of color-ordered amplitudes in terms of graphs.
\item Reflection (anti)-Symmetry~\cite{LanceTasi1996}.
\be
\AtreeCO{m}(1,2,3,\ldots,m-1,m) = (-1)^m\AtreeCO{m}(1,m,m-1,\ldots,3,2)\,.
\ee
This we get from the antisymmetry of kinematic numerators.
\item Photon Decoupling~\cite{LanceTasi1996}.
\be
 \sum_{\sigma\in {\rm cyclic~perm}(2,\ldots,m)}\AtreeCO{m}(1,\sigma) =0\,.
\ee 
This we also get from the antisymmetry of kinematic numerators.
\item Kleiss-Kuijf (KK) Relations: There is an $(m-2)!$ dimensional basis of $m$-point color-ordered scattering amplitudes under $\{\pm 1,0\}$.  This means we can express every $m$-point color-ordered amplitude as a linear combination of the $(m-2)!$ color-ordered scattering amplitudes arising from fixing the position of two of the leg labels (say $1$ and $m$), and with coefficients only of $\{-1,0,1\}$.
\begin{equation}
\label{KK}
\AtreeCO{m}(1,\{\alpha\},m,\{ \beta \}) = (-1)^{|\beta|} \!\!\!\! \sum_{ \sigma \in {\rm OP}(\{\alpha\},\{\beta^R\})} \!\!\!\! \AtreeCO{m}(1,\sigma,m)
 \end{equation}
where the sum is over the {\em ordered permutations} (OP): all permutations merging the sets
$\{\alpha\}$ and $ \{\beta^R\}$ that maintain the order of the
individual elements belonging to each set within the merged set.
I introduce the notation $\{\beta^R\}$ to represent the set $\{\beta\}$ with inverted order, and $|\beta|$ is the number of elements of $\{\beta\}$. 
These relations were first conjectured in ref.~\cite{KleissKuijf} and later proven in
ref.~\cite{LanceColor}, using the antisymmetry of kinematic numerators and the fact that color factors satisfy Jacobi identities.
\item Bern-Carrasco-Johansson (BCJ) Relations: There is an $(m-3)!$ dimensional basis of $m$-point color ordered scattering amplitudes under functions of external momentum Lorentz invariants.  This means we can express every $m$-point color-ordered amplitude as a linear combination of the $(m-3)!$ color-ordered scattering amplitudes arising from fixing the position of three external legs (say 1,2,3), where each tree has a coefficient that is a function of momentum invariants.  This we get from imposing Jacobi relations on kinematic numerators.  As you might expect the formula is a little more complicated:
 
\begin{equation}
\AtreeCO{m}(1,2,\{\alpha\},3,\{\beta\})=
\!\!\!\!\!\!\!\!\! \sum_{ \sigma  \in {\rm POP}(\{\alpha\},\{\beta\})}  \!\!\!\!\!\!\!\!\! 
\AtreeCO{m}(1,2,3,\sigma )  \prod_{k=4}^n  
{\frac{{\cal F}_k(3,\sigma ,1)}{ s_{2,4,\ldots,k} }} \,,
\label{allnBCJ}
\end{equation}
where $n =|\{\alpha\}|+3$ is the position in the list $\{1,2,\{\alpha\},3,\{\beta\}\}$ of $k_3$, and the sum runs over {\em partially ordered permutations} (POP) of
the merged $\{\alpha\}$ and $\{\beta\}$ sets. This gives all
permutations of $\{\alpha\} \bigcup \{\beta\}$ consistent with the
order of the $\{\beta\}$ elements. Either $\alpha$ or $\beta$ may be empty, trivially so for the $\alpha$ case.
The  function ${\cal F}_k$ associated with leg $k$ is given by,
\begin{align}
{\cal F}_k(\{ \rho\} ) &=
\left\{ 
\begin{array}{ll}
          \sum_{l=t_k}^{m-1} {\mathbf S}_{k,\rho_l} & \mbox{if $t_{k-1} < t_{k}$}\\
        - \sum_{l=1}^{t_k} {\mathbf S}_{k,\rho_l} & \mbox{if $t_{k-1} > t_{k}$}
\end{array} \right\} \nn \\&
 \null +
\left\{ 
\begin{array}{ll}
         s_{2,4,\ldots,k} & \mbox{if $t_{k-1} < t_{k}< t_{k+1}$}\\
        -s_{2,4,\ldots,k} & \mbox{if $t_{k-1} > t_{k} > t_{k+1}$}\\
         0 & \mbox{else}
\end{array} \right\} \,,
\hskip 1 cm 
\end{align}
where $t_k$ is the position of leg $k$ in the set $\{\rho\}$,
except for $t_3$ and $t_{n+1}$ which are always defined to
be,
\begin{equation}
t_3\equiv t_5 \,, \hskip 2cm t_{n+1} \equiv 0 \,.
\end{equation}
(Yes, for $|\{\alpha\}|=1$ we have $n=4$, and this means that $t_3=t_5=t_{n+1}=0$.) 
The expression ${\mathbf S}_{i,j}$ is given by,
\begin{equation}
\label{bcjFinal}
{\mathbf S}_{i,j}=\left\{ 
\begin{array}{ll}
        s_{i,j}  & \mbox{if $i< j$ or $j=1$ or $j=3$}\\
        0& \mbox{else}
\end{array} \right\} \,.
\end{equation}
These relations were first conjectured in ref.~\cite{BCJ}, and then proven, first as a low-energy limit of string-theory relations~\cite{Monodromy}, and then directly using the BCFW relations in field theory~\cite{amplituderelationProof,CachazoBCJProof}.
\begin{convention}
I invoked here a common shorthand for momentum invariants:
\begin{equation}
s_{i,j}=(k_i+k_j)^2, \hskip 1.3cm s_{i,j,\ldots,l}=(k_i+k_j+k_{j+1} +\cdots+k_l)^2,
\end{equation}
\end{convention}
\begin{problem}  The above can look prohibitively complicated until you get the hang of it.  Work out a couple of examples at the 5 and 6 point level and verify by comparing with ref.~\cite{BCJ}. Bonus points for verifying numerically by evaluating maximally-helicity-violating (MHV) scattering amplitudes at some kinematic points.  Extra bonus points for verifying the relations for the non-MHV 6-point configuration.
\end{problem}

\end{enumerate}

\subsubsection{Which cubic graphs contribute to the full color-dressed tree-amplitude?}
\label{cubicGraphsFull}
If we want to understand what information a theory tells us contributes to graph-organized expressions,  we first need to know which graphs contribute to amplitudes.  The place to begin for both gravity and gauge theory graphs is the set of unordered tree-level  graphs with $m$ external legs: $\GraphsTree{m}$. This is the set of graphs that is relevant for both the $m$-point color-dressed tree-level gauge theory amplitude, recall \eqn{CubicExp}, 
\be
\Atree{m} = \sum_{g\in\GraphsTree{m}} \frac{c(g) n(g)}{d(g)},
\ee
as well as the $m$-point tree-level gravity amplitude,
\be
\Mtree{m}  = \sum_{g\in\GraphsTree{m}} \frac{n_{\rm GR}(g)}{d(g)}.
\ee
Note that the use of the word {\it unordered} refers to the set, not to the graphs.  We will use our same ordered graphs, whose vertices can be represented as necklaces.  The necklaces distinguish flips in the order in which edges enter a vertex.  However, any graphs that are identical up to antisymmetry of necklaces need be represented by only one exemplar in the set of unordered graphs. How do we arrive at this set?

\begin{figure}
\begin{center}
\includegraphics[scale=.45]{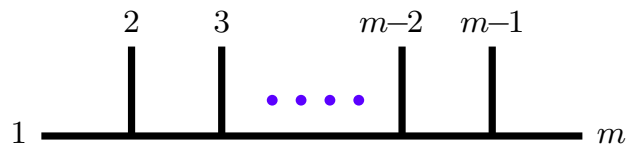}
\end{center}
\caption{Half-ladder tree-level graph. }
\label{halfLadder}
\end{figure}

We could start by writing down an $m$-point graph labeled according to any permutation. (For example, it could be the ``half-ladder'' graph with long edge running from 1 to $m$, and rungs $2,3,\ldots,m-1$, c.f. \fig{halfLadder}.)  Call this the first element of our set $\GraphsTree{m}$. We can add to our set $\GraphsTree{m}$ each graph arising from applying the $\widehat{t}$ operator (see problem \ref{opProblem}) to each internal edge, as well as every graph arising from applying the $\widehat{u}$ operator to each internal edge. We can now do this to every graph in our $\GraphsTree{m}$ that we have not already exhausted with our $\widehat{t}$ and $\widehat{u}$ operations, discarding any duplicates we land on, and ignoring any antisymmetry distinctions due to vertex order.  We can do this until $\GraphsTree{m}$ is closed under $\widehat{t}$ and $\widehat{u}$. This procedure is certainly doable, and there are aspects of it that I find aesthetically appealing.  Unfortunately, there is a certain amount of graph-isomorphism exhaustion\footnote{To be fair, the graph isomorphism at this level is simply comparing the propagator structure, so perhaps is not so expensive.  That being said, I find it offensive to have to continuously check, ``Have I found this graph already?''.  I'd very much like to believe there is a labeling and an algorithm that lets us bypass any such nonsense.  I suspect there is a very real benefit to spending time thinking about the geometric object generated by all applications of $\widehat{t}$ and $\widehat{u}$ in combination with the constraints of the numerator being able to satisfy Jacobi relations, and the adjunct completely algorithmic scalar propagator structure.}  that one must go through to handle the ``discarding any duplicates.'' There is quite possibly a nice graph-theoretic algorithm that has a controlled application of $\widehat{t}$ and $\widehat{u}$ that would never generate duplicates, and spans the space, but I do not know it.  At least not as a manifest application of $\widehat{u}$ and $\widehat{t}$. 
 
 Rather I will provide now an efficient constructive algorithm\footnote{I am told this is the graphical equivalent of Schwinger-Dyson recursion~\cite{SchI,SchII,Dys}.} as follows.  I The approach is straightforward and simply requires knowing: $\GraphsTree{m-1}$, with the understanding that \(\GraphsTree{3}\) consists simply of the single cubic vertex. For every edge $e$ of every graph of \(\GraphsTree{m-1}\), generate a new graph simply by attaching to $e$ the edge $m$. Done.  
 
  The cubic graph contributing to the three-point color-dressed amplitude ${\cal A}(123)$ is 
\begin{equation} 
g_3 =  \left\{  \begin{matrix} (k_1,k_2,k_3) \end{matrix}\right\}.
\end{equation}
 
To get the graphs contributing to ${\cal A}(1234)$ we  add an edge $k_4$ to each of the three edges: $\{k_1, k_2,k_3\}$.
Here are the resulting graphs:
 \begin{align}
g_s &=  \left\{  \begin{matrix} (k_1,k_2,i)\\
(-i,k_3,k_4)  \end{matrix}\right\} \,, \\
g_t &=  \left\{  \begin{matrix} (k_4,k_1,i)\\
(-i,k_2,k_3)  \end{matrix}\right\} \,, \\
g_u &=  \left\{  \begin{matrix} (k_3,k_1,i)\\
(-i,k_4,k_2)  \end{matrix}\right\} \,. 
\end{align}
 
\begin{problem} These are the only graphs that we need for the unordered set of four-point graphs.  Note that we could have chosen different exemplars (graphs with identical color and numerator factors, but having different color-orders) by adding $k_4$ to the three-point graph in different vertex orders. Find some different versions. Verify that it does not matter what exemplars we have as long as we get each distinct kinematic channel $s$, $t$, and $u$.
\end{problem}

To continue to the 5-point case we again add an edge that connects $k_5$ to each edge, internal and external, of each four-point graph.  In general the size of the unordered set will be:
\begin{equation}
| \GraphsTree{m} | = (2m-5)!!
\end{equation}

\begin{problem} Work out the 5-point example.  Prove the counting of this set goes as $(2m-5)!!$ generically, using the fact that every cubic $m$-point tree graph has $m$ external edges and $(m-3)$ internal edges.
\end{problem}

\subsubsection{Which cubic graphs contribute to a color-ordered tree-amplitude?}
\label{cubicGraphsCD}
There is a tremendous advantage to using color-ordered tree-amplitudes, as we will discover when we consider generalized unitarity cuts of gauge theories.  To use them we will need to know what set of graphs contributes to a color-ordered tree-amplitude.  As already addressed it is the set of all the distinct graphs that have the same color-order.  So a perfectly cromulent strategy would be to write down all cubic graphs, and then cluster them by color-order.  One might still want a more efficient algorithm. 

In a manner similar to the first suggestion above for generating unordered graphs, we could write down a  color-ordered graph (say the half-ladder), labeled according to that color-order, and close the set under $\widehat{t}$. Unfortunately, there is still a large amount of graph-isomorphism exhaustion to handle.  I also hope there is a nice graph-theoretic algorithm that has a controlled application of $\widehat{t}$ that would never generate duplicates, and spans the color-ordered space, but I do not know this algorithm either.  
 
 Here is the most efficient algorithm\footnote{I have heard it referred to as the graphical equivalent of Berends-Giele recursion~\cite{BerendsGiele}.} I know, and it comes from knowing all the color-ordered cubic graphs at the $(m-1)$-point level.  Because of color-ordering it is possible to canonically talk of which edges are to the {\em left-of} and which edges are to the {\em right-of} other edges within a graph.  
 For every graph in $\GraphsTreeCO{m-1}$, consider every edge (either internal or external) to the {\it inclusive-right-of} edge $m-1$ and to the {\it inclusive-left-of} edge $1$. For each of those edges generate a graph by inserting an edge $m$ such that it is to the {\it right-of} edge $m-1$ and to the {\it left-of} edge 1. These directions left-of and right-of can be made precise from the necklace-considerations of the definition of graphs, but this would be a little pedantic. Let me show an example:
 
 The graph contributing to the color-ordered amplitude $A(123)$ is simply
\begin{equation} 
g_3 =  \left\{  \begin{matrix} (k_1,k_2,k_3) \end{matrix}\right\} \,.
\end{equation}
To get the graphs contributing to $A(1234)$ we must find all edges to the left of $k_1$ and to the right of $k_3$ inclusive.  These are in fact edges $k_1$ and $k_3$.  Adding $k_4$ to each of these edges in such a way that it remains to the left of $k_1$ and to the right of $k_3$ results in the following graphs contributing to $A(1234)$:
 \begin{align}
g_{41} &=  \left\{  \begin{matrix} (k_4,k_1,i)\\
(-i,k_2,k_3)  \end{matrix}\right\} \,, \\
g_{12} &=  \left\{  \begin{matrix} (k_1,k_2,i)\\
(-i,k_3,k_4)  \end{matrix}\right\} \,.
\end{align}
 
For $g_{12}$, the edges to the left of $k_1$ and to the right of $k_4$, inclusive, are $\{k_1, i, k_4\}$.
 Here are the three graphs that result from adding $k_5$ to each of those edges, such that $k_5$ is to the left of $k_1$ and to the right of $k_4$:
 \begin{align}
g_{51,2,34} &=  \left\{  \begin{matrix} (k_5,k_1,j)\\
(-j, k_2,i)\\
(-i,k_3,k_4)  \end{matrix}\right\} \,, \\
g_{12,\overline{5},34} &=  \left\{  \begin{matrix} (k_1,k_2,i)\\
(-i,j,k_5)\\
(-j,k_3,k_4)  \end{matrix}\right\} \,, \\
g_{12,3,45} &=  \left\{  \begin{matrix} (k_1,k_2,i)\\
(-i,k_3, j )\\
(-j, k_4,k_5)  \end{matrix}\right\} \,.
\end{align}

For $g_{41}$, the edges to the left of $k_1$ and to the right of $k_4$ inclusive are simply $\{k_1, k_4\}$.  Here are the two graphs that result from adding $k_5$ to each of those edges such that $k_5$ is to the left of $k_1$ and to the right of $k_4$:
\begin{align}
g_{51,\overline{4},23} &=  \left\{  \begin{matrix} (j,k_5,k_1)\\
(k_4,-j,i)\\
(-i,k_2,k_3)  \end{matrix}\right\} \,, \\
g_{45,1,23} &=  \left\{  \begin{matrix} (k_4, k_5, j)\\
(-j,k_1,i)\\
(-i,k_2,k_3)  \end{matrix}\right\} \,.
\end{align}

One should see immediately that all five of these graphs have the color order $(12345)$, and in fact there are no (distinct) additional graphs that have this color order.  
\begin{problem}
Verify that applying $\widehat{t}$ to internal edge $i$ or $j$ of any of these graphs results in one of these graphs.
\end{problem}

 A little poking around should convince one that the number of color-ordered graphs that contribute at each multiplicity at tree level follows the Catalan number $C_{m-2}$:
\begin{equation}
|\GraphsTreeCO{m}|  = C_{m-2} =  \frac{2^{(m-2)} (2m - 5)!!}{(m-1)!} \,.
\end{equation}

\begin{problem}
Prove this counting using either the counting of the unordered set, or by relating the problem to Euler's Polygon Division problem. (Hint: What is the dual graph of an $m$-point cubic tree graph?)
\end{problem}

Just to catch our breath for a second, let us look at a graphical relation between color dressed and color-ordered amplitudes, 
\begin{align}
\Atree{m} &= \sum_{g \in  \GraphsTree{m} }  \frac{c(g) n(g)}{d(g)} \label{uo1}\\
    &= \sum_{\sigma \in {\rm perm}(2,\ldots,m)} 
{\rm Tr} \left( \cT{1} \cT{\sigma(2)} \cT{\sigma(3)}\cdots \cT{\sigma(m)} \right) 
  \AtreeCO{m}(1,\sigma)\\
    &= \sum_{\sigma \in {\rm perm}(2,\ldots,m)}
{\rm Tr} \left( \cT{1} \cT{\sigma(2)} \cT{\sigma(3)}\cdots \cT{\sigma(m)} \right) 
  \sum_{ g \in \GraphsTreeCO{m} } \frac{n(g)}{d(g)} \label{uo2} \,.
\end{align}
\begin{problem} The equality between \eqn{uo1} and \eqn{uo2} can be easily established on a case by case basis.  Show this at 5 points.
\end{problem}

\begin{problem} The equality between  \eqn{uo1} and \eqn{uo2} can be realized generically by expressing $c(g)$ in a trace basis, and realizing that all color-factors can be expressed in terms of color factors of half-ladders via Jacobi relations.  How does this relate to the counting of the number of graphs contributing to color-ordered sets?
\end{problem}

\subsection{Finding a color-kinematic dual representation at tree level}

We can look at \eqn{uo1} as a denominator-weighted inner product between a $(2m - 5)!!$ vector of color factors and a  $(2m-5)!!$ vector of kinematic numerator weights,
\begin{equation}
\Atree{m} = \mtrx{c}{}^{\intercal} \cdot \mtrx{ {\rm D}^{\rm UO}} \cdot \mtrx{n}
\label{cDnmatrixeq}
\end{equation}
where $\mtrx{ {\rm D}^{\rm UO}}$ is the diagonal $(2m - 5)!! \times (2m-5)!!$ matrix containing the relevant denominator of each graph, $D^{\rm UO}_{i,j} = \delta_{i,j}/d(g_j)$.

The $(2m - 5)!!$ color factors in \eqn{cDnmatrixeq} are not all independent; they are related by the color Jacobi relations.  There is a basis of $(m-2)!$ master graphs; for example, the half-ladder graphs with legs 1 and 2 along the long edge, and rungs labeled by arbitrary permutations of $\{3,4,\ldots,m\}$.
The $\mtrx{c}$ can be expressed linearly in terms of the $(m-2)!$ master graphs by simply solving the Jacobi relations, which can be obtained by applying the $\widehat{s} = \widehat{t} + \widehat{u}$ operator to each edge of every graph belonging to the unordered set.  This is a vastly redundant set of equations between all $(2m-5)!!$ graphs, but it can always be reduced to $(2m-5)!!$ relations relating $\mtrx{c} = \mtrx{{\rm J}} \cdot \mtrx{c}_{\rm master}$, where $\mtrx{{\rm J}}$ is some sparse $(m-2)! \times (2m - 5 )!!$ integer matrix representing the solution of the color Jacobi relations. 

Such a $\mtrx{{\rm J}}$ is by no means unique; there is a freedom of choice in terms of the $(m-2)!$ master graphs.   
Ref.~\cite{LanceColor} used a choice of all half-ladder graphs for the color-masters in order to prove the Kleiss-Kujif relations discussed earlier in~\eqn{KK}.  Any such $\mtrx{{\rm J}}$ applies equally well to numerator factors, assuming that they are in a representation that obeys the color-kinematic Jacobi identities.  A necessary condition for having a color-kinematics dual representation is to say that
$\mtrx{n} =  \mtrx{{\rm J}} \cdot \mtrx{n}_{\rm master}$.   How do we find such a representation?  
For theories that admit a color-dual representation at tree level, we can solve for the master numerators in terms of the color-ordered tree-amplitudes (calculated e.g.~through recursion relations).

In somewhat more detail, we can first use the Kleiss-Kuijf relations to reduce the number of 
distinct color-ordered amplitudes to $(m-2)!$.  We consider the $(m-2)!$ dimensional vector,
\be
\AtreeVec{m}\equiv 
\{  \AtreeCO{m}(12\sigma_i) \} \,,
\ee  
where $i$ runs over the $(m-2)!$ permutations $\sigma_i$ of labels $\{3,\ldots,m\}$. Anti-symmetry of numerator functions allows us to express this vector of tree amplitudes in terms of the vector of graphs contributing to the unordered set, using some matrix of denominators weighted by the relevant signs:
\begin{equation}
\AtreeVec{m}  =   \mtrx{{\rm D}^{\rm O} } \cdot  \mtrx{n},
\end{equation}
where $ \mtrx{{\rm D}^{\rm O}}$ is not diagonal.  Rather it is a  sparse $(m-2)! \times (2m-5)!!$ matrix, where $({\rm D}^{\rm O})_{ij}=  {\rm sgn}_i(g_j)/d(g_j)$.  Here ${\rm sgn}_i (g_j)$  is the relative sign of the relevant term in the color-factor of graph $g_j$ with respect to the color-order associated with permutation $\sigma_i$, and it vanishes when the color-factor of the graph is incompatible with the color-order for $\sigma_i$.

\begin{problem}
What is the relationship between $ \mtrx{{\rm J}}$, $\mtrx{ {\rm D}^{\rm UO}}$, and $\mtrx{ {\rm D}^{\rm O}}$?
\end{problem}

Here now we can invoke the relation $\mtrx{{\rm J}}$ putatively expressing $\mtrx{n}$ in terms of $\mtrx{n}_{\rm master}$:
\begin{equation}
\label{treeJacSoln}
\AtreeVec{m} =  (\mtrx{ {\rm D}^{\rm O}}\cdot \mtrx{  {\rm J} } ) \cdot  \mtrx{n}_{\rm master} \,.
\end{equation}

The whole question of whether or not we can find a color-kinematic dual representation is simply the question of whether or not we can invert the matrix: $(\mtrx{ {\rm D}^{\rm O}}\cdot \mtrx{  {\rm J} } ) $.  Since there is no unique solution this is actually a pseudo-inverse operation, but the point is, for gauge theories, we can always solve \eqn{treeJacSoln} systematically (through, e.g.~Gaussian elimination)  to find $\mtrx{n}_{\rm master}$ in terms of $\AtreeVec{m} $.  It turns out however that we will only constrain $(m-3)!$ of the masters, and these will be functions of the remaining $(m-3)!(m-3)$ master numerators. So in general only $(m-3)!$ need to be non-vanishing, and the rest can be anything.  Nothing physical will ever depend on the value of that extraneous {\it post-Jacobi} gauge freedom.

Let us see how this plays out at the four-point level, where we already worked out all color-ordered tree-amplitudes in terms of graphs in the first problem:
\begin{equation}
\left(\begin{matrix}
\AtreeCO{4}(1234) \\
\AtreeCO{4}(1342)\\
\AtreeCO{4}(1423)\\
\end{matrix}\right) =  \left( 
\begin{matrix} 
n(g_s)/s + n(g_t)/t\\
-n(g_s)/s - n(g_u)/u\\
-n(g_t)/t + n(g_u)/u\\
\end{matrix}
\right) \,.
\end{equation}
But recall from the solution to the first problem (and obvious by inspection), the Kleiss-Kuijf relations give the third amplitude in terms of a linear superposition of the first two, so we need only the following:
\begin{align}
\left(
\begin{matrix}
\AtreeCO{4}(1234) \\
\AtreeCO{4}(1342)\\
\end{matrix}
\right)
&= \left( 
\begin{matrix} 
1/s & 1/t& 0\\
-1/s & 0 & -1/u
\end{matrix}
\right) \, \left(
\begin{matrix}
n(g_s)\\
n(g_t)\\
n(g_u)
\end{matrix}
\right)\\
&=
 \left( 
\begin{matrix} 
1/s & 1/t& 0\\
-1/s & 0 & -1/u
\end{matrix}
\right) \, \left(
\begin{matrix}
1 & 0 \\
0 & 1 \\
1 & -1 
\end{matrix}
\right)
\, \left(
\begin{matrix}
n(g_s)\\
n(g_t)
\end{matrix}
\right) \\
&=
 \left( 
\begin{matrix} 
1/s & 1/t \\
t/ (s u) & 1/u 
\end{matrix}
\right) 
\, \left(
\begin{matrix}
n(g_s)\\
n(g_t)
\end{matrix}
\right) \,.
\end{align}

In the second line we used the relation $n(g_u)=n(g_s)-n(g_t)$ to construct $\mtrx{{\rm J}}$ and found the compact form of $({\rm D}^{\rm O} \, {\rm J})$ using the kinematic relation between Mandelstam variables, $s+t+u=0$.  This matrix is singular.  However, we can use the first relation to define $n(g_s)$ in terms of $\AtreeCO{4}(1234)$ and $n(g_t)$:
\begin{align}
n(g_s) &\equiv s \AtreeCO{4}(1234) - \frac{s}{t} n(g_t) ,\\
\AtreeCO{4}(1342) &= \frac{t}{ u} \AtreeCO{4}(1234) - \frac{1}{u} n(g_t) + \frac{1}{u} n(g_t) = \frac{t}{u} 
\AtreeCO{4}(1234).
\end{align}
We see in the second line that, upon using the first relation to define $n(g_s)$, we have completely removed any constraints from $n(g_t)$. Instead we find that imposing the kinematic-Jacobi relations has reduced the dimension of the basis of color-ordered amplitudes from $(m-2)!$ to $(m-3)!$ (for $m=4$).  As $n(g_t)$ is entirely unconstrained by observables, it can be an arbitrary function of external momenta, without affecting any of the values of the scattering amplitudes --- i.e.~without affecting anything physical.

\begin{problem} Do this for the 5-point case. 
\end{problem}

If we carry out the same procedure for the 6-point case, we find that out of the $(6\times2-5)!!=105$ graphs contributing to the unordered set of graphs,  only $(6-2)!\,=24$ are masters, of which only $(6-3)! \,=6$  need be non-vanishing.  The rest of the 18 graphs are unconstrained by any physical observables, nor are they constrained by Jacobi.

In general the $(m-3)! \times (m-3)$ unconstrained master numerators can be taken as a measure of the remaining gauge-freedom or redundancy.  What could this redundancy be used for?  A fine guess would be that it is somehow related to a freedom to find similar representations at loop level. Before we get to loop level, I will emphasize a few points relating to physical observables, and double-copying to gravity.

\subsubsection{Physical Observables}

Why should one care about the ability to put tree-level amplitudes in terms of a color-dual representation around local graphs?  Most important is that it generalizes naturally to loop level, which we will get to in due course.  Staying at tree level, the first physical result is that it leads directly to observable relations between gauge-invariant objects. The possibility of a color-dual representation ---  kinematic numerators satisfying all Jacobi relations --- is sufficient to find a reduction to the BCJ ($(m-3)!$ dimensional) basis of the gauge invariant color-ordered tree-level scattering amplitudes mentioned in \eqn{allnBCJ}.  Recall this is the basis where any color-ordered scattering amplitude can be represented as a sum over products between functions of momentum invariants and the set of $(m-3)!$ color-ordered amplitudes with the order of three legs fixed.  So just as kinematic anti-symmetry and color-factor Jacobi identities leads to an $(m-2)!$ dimensional Kleiss-Kuijf basis (see \eqn{KK}), the ability to impose the kinematic-Jacobi relations leads to the BCJ basis.

  All of this discussion holds for $\cN=4$ super Yang-Mills tree-level super amplitudes. All of this holds for non-supersymmetric pure Yang-Mills at tree level.  All of this holds for every theory ``in-between.''  What does ``in-between'' mean? Any theory that can be arrived at by deforming $\cN=4$ by either projecting out states, or otherwise breaking symmetries.  All of these satisfy the $(m-3)!$ relations, as do natural lifting\footnote{Lifting theories consistently to higher dimensions is often referred to as {\em oxidation}.} to equivalent theories in higher dimensions.

Is there any matter we can add to gauge theories where this does not hold?  Absolutely!  We  have counterexamples. There are straightforward  theories which do not satisfy the $(m-3)!$ relations:  theories with multiple flavors of fermions, or even scalars, if these flavors are not related by supersymmetry.  For such theories, kinematic-Jacobi relations on edges between distinct flavors are precluded, as they would generate explicit flavor violation which is not present. Of course kinematic-Jacobi relations can still be imposed on edges where any two of the four attached lines are gluonic.  It is an open research problem as to whether there is a good generalization of global color-kinematic duality for these theories.

\begin{problem}  Consider 2-flavor scalar QCD.  How many color-ordered amplitudes are there with 4 external scalars, where two scalars have one flavor and the other two have another?
\end{problem}

Staying at tree level, the second physical result is that gravity amplitudes can be expressed simply in terms of color-ordered tree-amplitudes, which is the subject of the next subsection.

\subsubsection{Gravity tree-level amplitudes for free}
\label{gravTreeAmplitudes}
We discussed that, for some gauge theories, it was possible to find color-kinematic dual representations, i.e.~(kinematic) Jacobi-satisfying numerators.  For such theories, Jacobi-satisfying representations can be given as functions of color-ordered partial amplitudes expressed in terms of the BCJ $(m-3)!$-dimensional basis.  The ``double-copy construction''~\cite{BCJ,LagrangianSquare} means that when one has a Jacobi-satisfying  $m$-point gauge representation given in $\mtrx{n}$, then an associated scattering amplitude in a related gravity theory can be written in the form,
\begin{equation}
{\cal M}^{\rm tree}_{m} = \sum_{g \in  \GraphsTree{m} } \frac{n(g) \widetilde{n}(g)}{d(g)},
\label{gDCT}
\end{equation}
where the gauge-theory kinematic numerator $\widetilde{n}$ can be from the same gauge theory or a completely different theory.  Also, while the external states of $n$ and $\widetilde{n}$ can be the same, they are not required to be.  How does this work?

It is easiest for me to introduce the states in the gravity theory by talking about the states of the associated gauge theory in four-dimensions.  Of course many of the theories also have an interpretation in higher dimensions (``oxidize'' to higher dimensions), as do their states.  But for the lectures, let us restrict ourselves to looking at the four-dimensional states, which can be classified according to their helicity:  plus-helicity gluon $A_+$,  plus-helicity fermion $f_+$, scalar $s$, negative-helicity fermion $f_-$, and negative-helicity gluon $A_-$. 

The states in the gravity theory are straightforward to identify with double-copies of Yang-Mills states. The sum of the helicities of the two gauge-theory states should equal the helicity of the gravity state.  The plus-helicity graviton is given by
\begin{equation}
h_{++} = A_+ \otimes A_+.
\end{equation}
The plus-helicity gravitini can come in a couple of ways:
\begin{align}
\psi_{+}^{(1,\half)} &= A_+ \otimes f_+ \,,\\ 
\psi_{+}^{(\half,1)} &=  f_+ \otimes A_+ \,.
\end{align}
The two gravitini can be distinguished by the gauge theories from which the gluon came.  In general, the two theories may be very different, so these gravitini may behave very differently.
The gravitating plus-helicity vector can come in three different ways: 
\begin{align}
A_{+}^{(1,0)} &= A_+ \otimes s \,,\\
A_{+}^{(\half,\half)} &= f_+ \otimes f_+ \,,\\
A_{+}^{(0,1)} &=  s  \otimes A_+ \,.
\end{align}
The gravitating plus-helicity spin one-half can come in 4 different ways: 
 \begin{align}
\xi_{+}^{(1,-\half)} &= A_+ \otimes f_- \,,\\
\xi_{+}^{(\half,0)} &= f_+ \otimes s\,,\\
\xi_{+}^{(0,\half)} &=  s  \otimes f_+\,,\\
\xi_{+}^{(-\half,1)} &=  f_- \otimes A_+ \,.
\end{align}
And finally the gravitating scalars can come in 5 different ways:
\begin{align}
\tau^{(1,-1)} &= A_+ \otimes A_- \,,\\
\tau^{(\half,-\half)} &= f_+ \otimes f_-\,,\\
\tau^{(0,0)} &=  s  \otimes s\,,\\
\tau^{(-\half,\half)} &= f_- \otimes f_+ \,,\\
\tau^{(-1,1)} &=  A_- \otimes A_+\,. 
\end{align}
All the possible negative helicity states are simply the CP conjugates of the above.

For factorizable supergravity theories, in which the entire Fock space can be factored as above, there is not a lot of subtlety to identifying the gravity theory we get by taking the double copy of various gauge theories. I have listed, for example, the factorizable~\cite{factorizableSG} supergravity theories in \Tab{tabFactorizable}.  

\begin{problem}
Using the fact that
 the $\cN=4$ SYM vector multiplet is $\{ 1 A_+,  4 f_+, 6 s, 4 f_-, 1 A_- \}$,
  the $\cN=2$ SYM vector multiplet is $\{ 1 A_+,  2 f_+, 2 s, 2 f_-, 1 A_- \}$,
 and the $\cN=1$ SYM vector multiplet is $\{ 1 A_+,  1 f_+, 0 s, 1 f_-, 1 A_-\}$,
 work out the supergravity states in the factorized theories in \Tab{tabFactorizable}.
 \end{problem}

{
\renewcommand{\arraystretch}{1.25}
\begin{table}[ht]
\tbl{\small Factorizable four-dimensional $\cN\ge 1$  supergravity theories arising from straightforward double-copy~\cite{factorizableSG}.%
}
{\begin{tabular}{|c|c|c|c|}
\hline
\# & {\bf $\cal N$}& {\bf Factors} &{\bf Supergravity} \\ 
\hline
1 & $8$ &  $\cN=4  \text{ SYM}    \otimes  \cN=4  \text{ SYM}    $ & {\small pure $\cN=8$ SG} \\[2pt]
\hline
2 & $6$ &  $\cN=4  \text{ SYM}    \otimes  \cN= 2  \text{ SYM}   $ & {\small pure $\cN=6$ SG} \\[2pt]
\hline
3 & $5$ &  $ \cN=4  \text{ SYM}  \otimes  \cN=1 \text{ SYM}  $ & {\small pure $\cN=5$ SG} \\[2pt]
\hline
4 &  $4$ &  $\cN=4  \text{ SYM} \otimes (\cN=0 \text{ YM} + n_v \text{ scalars} )$ & \small{\tsplit{$\cN=4$ SG,}{$n_v$ vector multiplets}} \\[0pt]
5& $4$ &  $\cN=2 \text{ SYM}  \otimes \cN=2 \text{ SYM}  $ & \small{\tsplit{$\cN=4$ SG,}{$2$ vector multiplets}} \\[0pt]
\hline
6& $3$ &  $\cN=2  \text{ SYM} \otimes \cN=1 \text{ SYM}  $ & \small{\tsplit{$\cN=3$ SG,}{$1$ vector multiplet}} \\[0pt]
\hline
7 & $2$ &  $\cN=2 \text{ SYM} \otimes (\cN=0 \text{ YM} + n_v \text{ scalars} )$ & \small{\tsplit{$\cN=2$ SG,}{$n_v +1$ vector multiplets}} \\[0pt]
8 & $2$ &  $\cN=1  \text{ SYM}  \otimes \cN=1  \text{ SYM}   $ & \small{\tsplit{$\cN=2$ SG,}{$1$ hypermultiplet}} \\[0pt]
\hline
9 & $1$ &  $\cN=1 \text{ SYM} \otimes (\cN=0 \text{ YM} + n_v \text{ scalars} )$ & \small{\tsplit{$\cN =1$ SG, $n_v$ vector}{ and $1$ chiral multiplets}} \\[0pt]
\hline
\end{tabular}}

\label{tabFactorizable}
\end{table}
}

The vast majority of supergravity theories are not factorizable.  Consider pure non-supersymmetric gravity --- it only has gravitons.  Yet when we talk about the double copy of pure Yang-Mills we find 2 additional scalars:
\begin{equation}
\{A_+,A_-\}\otimes\{A_+,A_-\} = \{ h_{++}, \tau^{(1,-1)},\tau^{(-1,1)}, h_{--}\}.
\end{equation}
The double copy of pure Yang-Mills results in a gravitational theory with two scalars (which can be associated with an axion and dilaton).  At tree level, the scattering amplitudes for all external gravitons decouple from the axion and dilaton: these amplitudes are the same in pure gravity as well as in generic $\cN$ supergravity theories. At loop level, however, we do indeed have scalars running around the loops, so the loop-level integrands are different in these various theories, even for amplitudes with all external gravitons. To handle these more general supergravity theories at the level of multi-loop integrands, we require additional book-keeping that simply lets us project out unwanted states in a coordinated manner. Within the past couple of years there has been fast progress in developing the book-keeping for double-copy construction in non-factorizable supergravities, as well as the associated structural insight~\cite{Chiodaroli:2013upa, Johansson:2014zca}, including generalizations~\cite{Chiodaroli:2014xia} to gauged (super)-gravities like Einstein-Yang-Mills and supersymmetric generalizations~(see also ref.~\cite{Bern:1999bx} for an early double-copy understanding).

\subsection{Emergence of Tree-level Invariant Relations Between Gauge and Gravity Theories}

Now consider that we can write our gravity tree-level expression using the  unordered-set propagator matrix $\mtrx{{\rm D}^{\rm UO}}$, and let us take both copies to be color-kinematics dual representations, so we have, rewriting \eqn{gDCT} in matrix form:
\begin{equation}
\label{gravDoubleCopyTree}
\MtreeVec{m}= \mtrx{n}^{\intercal} \cdot \mtrx{ D^{\rm UO} } \cdot \mtrx{\widetilde{n}} \,.
\end{equation}

A property that has been exploited to great success in understanding the structure and ultraviolet behavior of lower-supersymmetry supergravity theories~\cite{Bern:2012cd, Bern:2012gh,Bern:2012efa, Carrasco:2013ypa, Bern:2013qca,Bern:2013uka,Bern:2014sna,Bern:2014lha}, is that only one  copy of the gauge numerator factors need be in a color-dual form to generate the associated gravity amplitude at tree level.   How can we see this easily?  In addition to the gauge theory numerators $\mtrx{\widetilde{n}}$, let us consider a different, non-Jacobi satisfying representations $\mtrx{\widetilde{n}'}$.  This representation must produce the same physical gauge theory amplitudes as does $\mtrx{\widetilde{n}}$.  Hence $\mtrx{c}^\intercal \cdot \mtrx{D^{UO}} \cdot \mtrx{\Delta} = 0$, where $\mtrx{\Delta} \equiv \mtrx{\widetilde{n}} - \mtrx{\widetilde{n}'}$.  But this vanishing cannot depend on any special properties of $\mtrx{c}^\intercal \cdot \mtrx{D^{UO}}$ other than the fact that $\mtrx{c}$ satisfies Jacobi relations and anti-symmetry --- since numerator representations are independent of the gauge group.  This means that if we replace $\mtrx{c}$ with a Jacobi-satisfying $\mtrx{{n}}$, we will also see that  $\mtrx{n}^\intercal  \cdot  \mtrx{D^{UO} } \cdot  \mtrx{\Delta } = 0$.  The relation allows us to shift \eqn{gravDoubleCopyTree} away from a Jacobi-satisfying $\mtrx{\widetilde{n}}$ to a non-Jacobi-satisfying $\mtrx{\widetilde{n}'}$,
i.e.~$\MtreeVec{m}= \mtrx{n}^{\intercal} \cdot \mtrx{ D^{\rm UO} } \cdot \mtrx{\widetilde{n}'}$.

That said, we can take \eqn{gravDoubleCopyTree} and express $\mtrx{n}$ and $\mtrx{\widetilde{n}}$ in terms of their color-ordered tree-amplitudes as described above.  We arrive at an expression that involves only gauge-invariant expressions:
\begin{equation}
\MtreeVec{m} = (\AtreeVec{m}){}^{\intercal}\cdot \mtrx{ {\cal K}}\cdot \widetilde{\AtreeVec{m}}
\label{eqKLT}
\end{equation}
where $\mtrx{{\cal K}}$ is in general a non-diagonal, sparse matrix of  rational functions of momentum invariants.  These momentum invariants come from the propagators of $\mtrx{D^{\rm UO}}$, but also from the pseudo-inverses of $\mtrx{D^{O}}$, $\mtrx{J}$, $\mtrx{\widetilde{D}^{O}}$, and $\mtrx{\widetilde{J}}$.  This $\mtrx{{\cal K}}$ has come to be called a momentum kernel~\cite{KLTproofs}. It can be derived from the existence of the BCJ relations at tree level, or directly from the vanishing $\alpha'$ limit of stringy generalizations of the BCJ relations.  The momentum kernel is not unique --- it depends on the basis of $\mtrx{A}$ and $\mtrx{\widetilde{A}}$.  There exist many forms of $\mtrx{{\cal K}}$, due to the freedom in taking the pseudo-inverses, or, if you like, the choice of which $(m-3)!$ dimensional basis is selected for each copy.

The set of all-multiplicity tree-level equations of the form~(\ref{eqKLT}) are known as the Kawai, Lewellen, and Tye (or KLT) relations~\cite{KLT, KLTfield,otherKLT,KLTproofs}.  Let me emphasize how special these are. They represent a gauge-invariant relationship between gravity and gauge theory at the level of tree-level scattering amplitudes.  (Actually, \eqn{eqKLT} is the low-energy limit of the full, stringy KLT relations.) Now in some sense these relationships are not as aesthetically pleasing as the graph by graph double-copy, because explicit forms for $\mtrx{{\cal K}}$ get incredibly convoluted as the multiplicity increases. Take a look at the closed form for the BCJ relations, given in eqs.~(\ref{allnBCJ})--(\ref{bcjFinal}), and mentally square it. That said, there is something very powerful in the realization that there is a choice-independent relationship between gauge and gravity classically.  This was an important early hint that there is something very genuine connecting the two --- not a coincidence of similar notations. The most compact field theory version of ${\cal K}$, and first all-multiplicity expression, was written down in ref.~\cite{KLTfield}, and proven much more recently~\cite{KLTproofs} after understanding the $(m-3)!$ relations between color-ordered scattering amplitudes~\cite{BCJ}.  This is a $(m-3)! \times (m-3)!$ matrix in terms of a specific color-ordered BCJ basis of color-ordered tree-amplitudes.

\begin{problem}
Show that $ \mtrx{{\cal K}} = ((\mtrx{ D^{\rm O}})^{\intercal}){}^{-1} \cdot (\mtrx{J^{\intercal}}){}^{-1} \cdot \mtrx{ D^{\rm UO}} \cdot \widetilde{\mtrx{J}}{}^{-1}  \, (\mtrx{ \widetilde{D}{}^{\rm O}}){}^{-1}$. 
\end{problem}

\begin{problem}
Find $\mtrx{{\cal K}}$ for 4-point tree amplitudes, assuming a basis of $\Atree{4}(1234)$ and  $\widetilde{\Atree{4}}(1234)$.  Now find a basis with $\widetilde{\Atree{4}}(1423)$ instead. Do you like one better?
\end{problem}

Now is as good a time as any to mention the following fact.  The type of algebraic (pseudo)-inversion between kinematic numerators and gauge-invariant quantities, which was exploited to establish gauge-invariant relations between gravity and gauge amplitudes at tree level, does not yet exist at loop level.  As we will discuss later, in section~\ref{ckexploit}, the dependence of loop-level numerator factors on internal loop momenta means that the kinematic-Jacobi identities are {\it functional} constraints, rather than the algebraic constraints found at tree level.  These functional constraints cannot be simply inverted to solve for the numerators. Correspondingly, there is no equivalent of the closed-form gauge-invariant KLT relations at the loop level.  Instead it is the generalized-gauge-dependent numerator double-copy which does in fact generalize.  Before we discuss this generalization, we will take the time to understand functional graph-organized integrands, leading to natural verification and construction.

\section{Building Loop-Level Amplitudes}

I'm going to spend this lecture talking about on-shell ideas and methods for verifying and constructing multi-loop scattering amplitudes.  These approaches~\cite{UnitarityMethod,GeneralizedUnitarity,GeneralizedUnitarity2,BCF} have influenced much of modern multiloop calculation in gauge and gravity theories.  I will try to give a pedagogic introduction to the spirit behind them, or at least how I've found them most useful.  

Generalizing from graph representations of tree-level amplitudes to graph representations of loop-level amplitudes is not only straightforward, but incredibly useful. The biggest distinction, and challenge, is that now one has to track each graph's automorphisms, both external and internal. Rather than thinking of the integrand as simply some algebraic expression under an integral sign that one has to poke and prod with carefully chosen delta-functions (corresponding to putting legs on shell in generalized unitarity cuts), we can exploit the ability of graphs to simultaneously make manifest both the conservation of momentum and color, in order to target individual poles.  So we will keep this abstract structure of graphs and functional mappings to numerator weights, in addition to the algorithmic mapping to color factors and scalar denominators. 

We will see that the value of maintaining functional graph-structured integrands is two-fold:
\begin{enumerate}
\item When organized graphically, a functional integrand allows for the easy {\it verification} of full non-planar scattering amplitude integrands by consideration of their behavior along a finite number of kinematic channels.
 \item One can invert the verification process to generate a handy method of {\it construction}. 
\end{enumerate}

\subsection{Verification}
When I discuss unitarity methods I like to discuss verification first, then construction.  Why? For multi-loop integrands, as for many things, checking whether an expression is correct is easier than constructing one from scratch --- but knowing efficient methods of verification leads to natural methods of construction that are provably bounded in complexity.
 
How do we know if an expression is a valid (correct) integrand for a scattering amplitude? Sometimes the most basic answers are the best places to start. If we integrate the candidate integrand, then we must get the same expression as if we had calculated by integrating all Feynman graphs.
\be
\int \text{candidate integrand} =  \sum \int \text{Feynman graphs}.
\ee
This condition is absolutely necessary and obviously sufficient --- certainly from the action/QFT perspective this is how scattering 
amplitudes are defined. We'll move under the integrand sign shortly, but we want to always be talking about gauge-invariant objects, so it  is convenient to consider various kinematic limits that take internal legs on-shell, always maintaining gauge invariance.  Any given 
kinematic limit throws away information, but we will consider verifying the integrand on a set of complementary kinematic limits of internal (and external) momenta, so as to ensure that our integrand has all the information contained in the Feynman graphs.  The gauge-invariant limits are called {\it generalized unitarity cuts}.  A sufficient condition for verifiability is that all generalized unitarity cuts must be satisfied:
\(\)
\begin{multline}
{\rm cut} \left( \text{candidate integrand} \right) = {\rm cut} \left( \text{Feynman graphs} \right),\\
 \forall~{\rm cut} \in \{\text{Unitarity Cuts}\}.
\end{multline}

Think of each cut on the Feynman graph side as organizing the Feynman graphs into a product of tree-level scattering amplitudes joined by shared graph-internal edges.  These internal shared edges will be on-shell  ``external'' states of the tree-level scattering amplitudes.  Each of the momenta of these shared edges will emerge from one amplitude, $A(\ldots,e,\ldots)$, and enter another one, $A(\ldots,-e,\ldots)$.  Since in general there will be multiple particle types that can have the same momenta, a cut will involve the sum over the product of scattering amplitudes, with the cut-legs taking on all states available in the theory.  We can consider for gauge theories 
either {\em color-dressed cuts} --- cuts sewing full color-dressed tree scattering amplitudes --- or {\em color-ordered cuts}, cuts sewing only color-ordered scattering amplitudes.  As the number of graphs contributing to color-ordered trees is far fewer, these cuts can be much more manageable.

Generalized unitarity cuts at $L$-loops have no total-derivative ambiguity as long as all unintegrated loop-momenta are frozen by cuts.  So if one wishes to avoid integration one should cut down to trees --- cutting at least $L$-independent loop momenta for an $L$-loop scattering amplitude. One can of course cut more legs; this would involve sewing additional trees, and some of the cut loop-momenta would depend on other cut-loop momenta. So we consider a generic unitarity cut involving $I$ trees:
\be
{\rm cut}(\text{Feynman graphs}) \equiv \sum_{\rm states} {\rm A}^{(1)}_{\rm tree}  {\rm A}^{(2)}_{\rm tree }\cdots  {\rm A}^{(I)}_{\rm tree } \,.
\label{itrees}
\ee

\subsubsection{Extracting cut information from state sums}

How do we carry out the state-sums in \eqn{itrees}? A thorough description of state-sum book-keeping would take us pretty far afield from gauge and gravity theory relations, and there are nice references available which I will point you towards.  In four dimensions it is incredibly efficient to use the handy Grassmann superspace you learned in other TASI  lectures for tree-level recursion to trivially take care of the 4D supersums.  Naturally I recommend looking at ref.~\cite{SuperSum} where we provide a  discussion with two different algorithms to calculate in maximal supersymmetry, with one approach that trivially generalizes to less supersymmetry.  This latter one I quote in the bonus appendix,~\sect{appendix}, to these lecture notes.  This however is only in four-dimensions --- a great place to start, but if we are regulating with dimensional regularization, strictly calculating in four dimensions can cause us to miss important data --- pieces that are only visible when the cut loop momenta lives outside of four dimensions.  Fortunately there's a superspace in higher dimensions for us too.  Check out refs.~\cite{Donal, SixDimSusy,SixDim, Donal10} after you get comfortable manipulating the graphs as described here, when you want to ensure that your 4D calculations are not missing anything critical.

\subsubsection{Extracting cut information from a graph-organized integrand}
To compare with the state-sum, we must be able to extract the information relevant to this cut from our candidate integrand. Here is where a graph-organized integrand shines. If we have described our candidate integrand in terms of mappings from labeled graph topologies to numerator functions, denominators, and color factors, we can exploit the structure of our graphs to directly isolate the contribution to a given cut. We simply consider what tree graphs contribute to each of the tree scattering amplitudes, label them according to the momentum labels of each tree graph, and take the outer product.  The set of graphs contributing to a cut with $I$ trees would then be:
\be
{\cal G}_{\rm cut} = {\cal G}_{\rm tree~(1)}\otimes {\cal G}_{\rm tree~(2)}\otimes \cdots \otimes {\cal G}_{{\rm tree}~(I)}
\ee
Note that the graphs contributing from each tree amplitude  ${\cal G}_{{\rm tree}~(I)}$ will be either  $\GraphsTree{m}$ from \sect{cubicGraphsFull}, or $\GraphsTreeCO{m}$ from  \sect{cubicGraphsCD}, depending on whether we are considering gravity/color-dressed or color-ordered (color-stripped) gauge theory cuts, respectively. 
For color-dressed cuts, color-ordered, and gravity cuts, we extract our relevant cut information as:
\begin{align}
{\rm cut_{\rm CD} } (  \text{candidate integrand} ) &\equiv \sum_{{\cal G}_{\rm cut,\,UO} } \frac{n(g) c(g)}{d(g)}\,,\\
{\rm cut_{\rm CO} } ( \text{candidate integrand}  ) &\equiv \sum_{{\cal G}_{\rm  cut,\,O} } \frac{n(g)}{d(g)}\,,\\
{\rm cut_{\rm GR} }(  \text{candidate integrand}) &\equiv \sum_{{\cal G}_{\rm cut,\,UO} } \frac{n_{\rm GR}(g)}{d(g)}\,.
\end{align}

\begin{example}  Planar 2-loop cut.

Consider the functional graph-organized integrand representation of two-loop $\cN=4$ super-Yang-Mills theory (sYM) from section~\ref{ex2l}.  The graphs are given as follows:
\begin{align}
\label{planar2LoopDefn}
{\rm planar} &= \left\{
\begin{array}{c}
 \left(k_1,l_{11}, -l_9\right) \\
 \left(k_2, -l_{10},-l_{11} \right) \\
 \left(k_3,-l_8,l_7\right) \\
 \left(k_4,-l_5,l_8\right) \\
 \left(l_5,l_9,-l_6\right) \\
 \left(l_6,l_{10},-l_7\right) \\
\end{array}
\right\}\\
&= \vcenter{\includegraphics[width=2.75in]{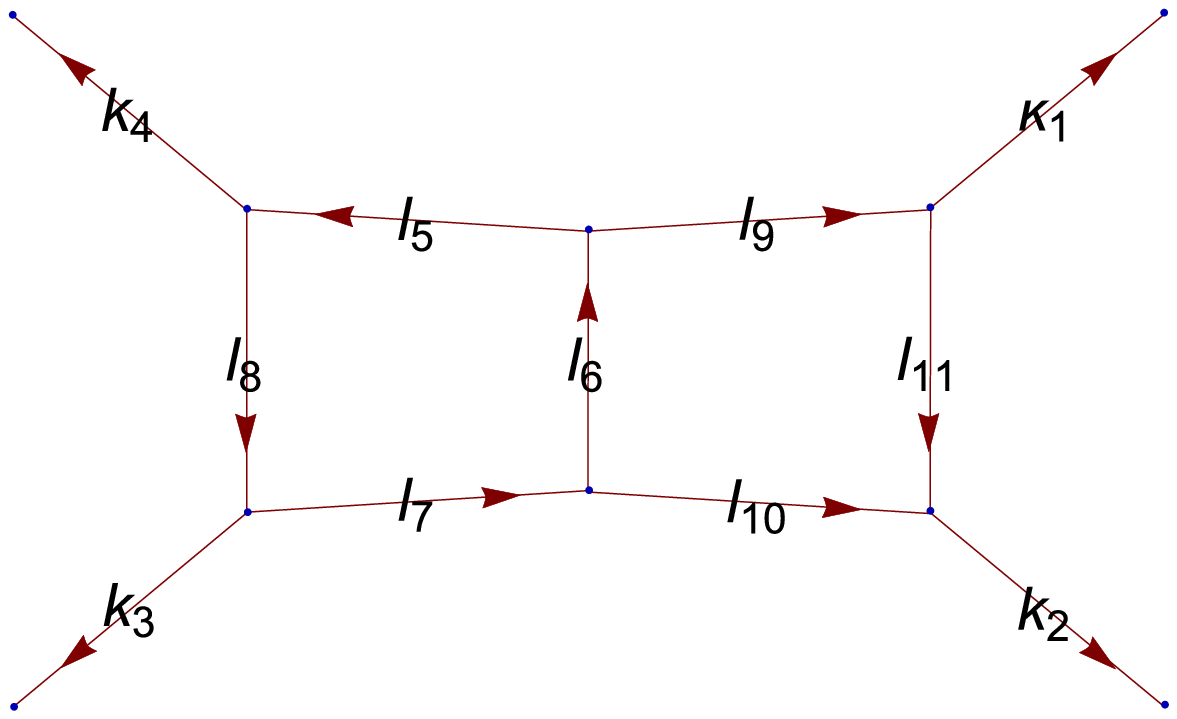}}   \!\!\!\!  \!\!\!\!  \!\!\!\!  \!\!\!\!  \!\!\!\!  \!\!\!\! 
 \!\!\!\!  \!\!\!\!  \!\!\!\!  \!\!\!\!  \!\!\!\!  \!\!\!\!. \nonumber
\end{align}
and
\begin{align}
\label{nonplanar2LoopDefn}
{\rm nonplanar } &= \left\{
\begin{array}{c}
 \left(k_ 1,l_{11},-l_9\right) \\
 \left(k_ 2,-l_{10},-l_{11}\right) \\
 \left(k_ 3,-l_8,l_ 7\right) \\
 \left(k_ 4,-l_5,-l_6\right) \\
 \left(l_ 5,l_ 8,l_ 9\right) \\
 \left(l_ 6,-l_7,l_{10}\right) \\
\end{array}
\right\}\\
&=  \!\!\!\!  \!\!\!\!  \!\!\!\!  \vcenter{\includegraphics[width=2.75in]{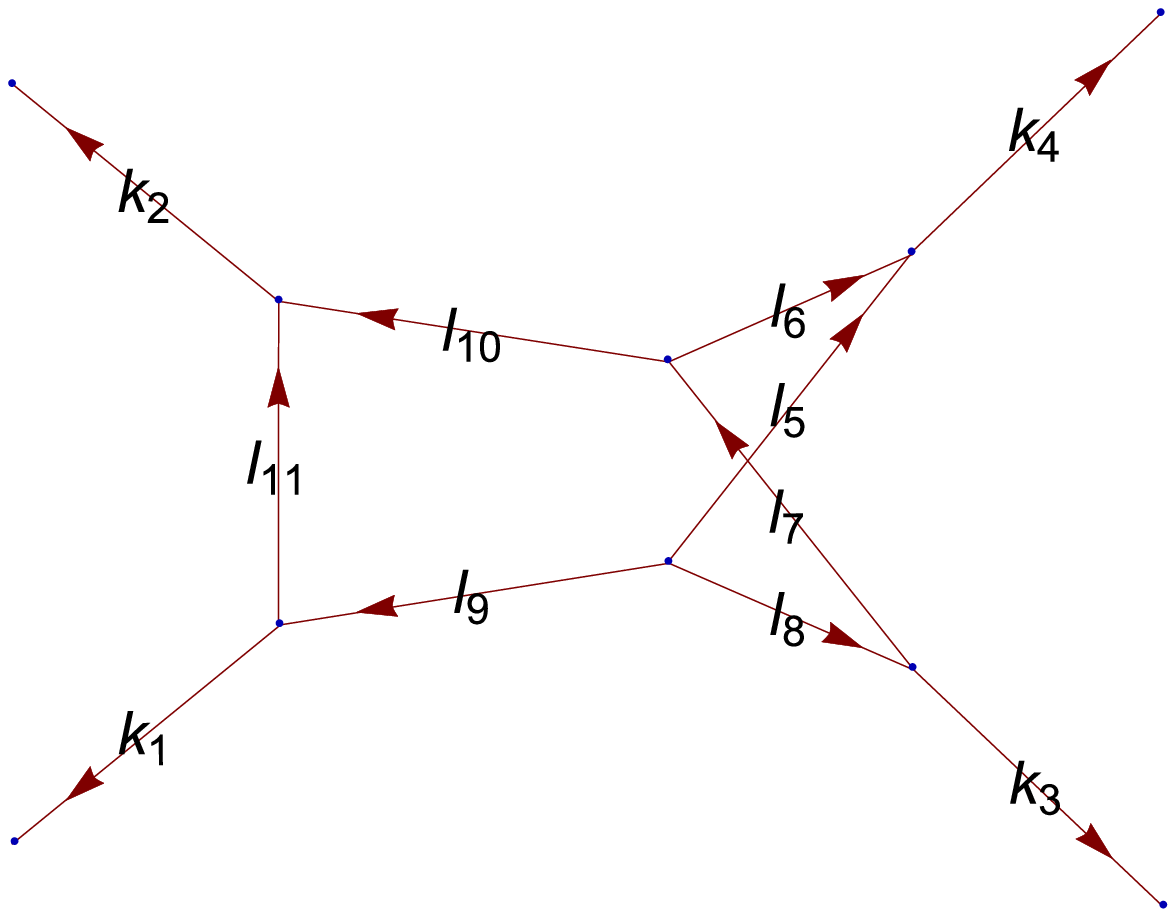}}  \!\!\!\!  \!\!\!\!  \!\!\!\!  \!\!\!\!  \!\!\!\!  \!\!\!\! 
 \!\!\!\!  \!\!\!\!  \!\!\!\!  \!\!\!\!  \!\!\!\!  \!\!\!\!.  \nonumber
\end{align}
There are other two-loop four-point graphs, but they do not contribute to \(\cN=4\) super Yang-Mills, so we define their numerator factors to be zero for this theory.

As discussed in section~\ref{ex2l}, the color factors are obtained by dressing every vertex with $f^{abc}$, as at tree level. The denominators will simply be the product of the square of the momentum of the internal edges. We use conservation of momenta to express them in terms of independent loop momenta.%
 For the displayed vertex order the numerator factors are given as:
 \begin{align}
 n({\rm planar}) &= s_{12}^2 s_{23} \AtreeCO{4}(1234)\,, \\
 n({\rm nonplanar}) &= s_{12}^2 s_{23} \AtreeCO{4}(1234).
 \end{align}
 To know this is correct we must check all cuts.  This will be our candidate functional integrand (this collection of $n$, $c$, and $d$ mappings).
 
Let us consider the following color-ordered cut:
\be 
\label{cutpDefn}
\text{cut}_p=\sum_{\rm states} \AtreeCO{5}(K_1, K_2, l_a, l_b, l_c) \AtreeCO{5}(- l_c, -l_b,-l_a, K_3, K_4). 
\ee
For this cut I choose to use external momenta labels $K_i$ and cut momenta labels $l_j$ to emphasize the functional nature of the kinematic mapping.  

The result of carrying out the cut from ``Feynman graphs'' is:
\begin{multline}
\label{cutpSoln}
\text{cut}_p = s_{K_1,K_2} s_{K_2,K_3} \AtreeCO{4}(K_1,K_2,K_3,K_4)\\
\times \Biggl( \frac{\left(K_1+K_2\right){}^2}{\left(l_a+l_b\right){}^2
   \left(-l_a-K_2\right){}^2 \left(l_b+l_c\right){}^2
   \left(K_4-l_c\right){}^2}\\+\frac{\left(K_1+K_2\right){}^2}{\left(l_a+l_
   b\right){}^2 \left(l_b+l_c\right){}^2 \left(-l_a-l_b-K_2\right){}^2
   \left(l_b+l_c-K_4\right){}^2}\\
   +\frac{\left(K_1+K_4\right){}^2}{\left(-l
   _a-K_2\right){}^2 \left(K_4-l_c\right){}^2
   \left(-l_a-l_b-K_2\right){}^2 \left(l_b+l_c-K_4\right){}^2} \Biggr) \,.
 \end{multline}
The evaluation of the state sum over the product of trees is described in the appendix, \sect{appendix}, and results in expressions like \eqn{cutSoln}, which can be reduced to the above form.

Note the ubiquitous prefactor of $s t \AtreeCO{4}(1234)$  that is associated (appropriately labeled) 
with every cut of every four-point amplitude in the maximally supersymmetric theory.  This factor is critical; if nothing else it encodes the external states of the particles, but one needn't write it all over the place. Its ubiquity means that it is often elided to concentrate on the novel structure that shows up, but for pedagogy I am bucking convention and explicitly calling it out.

If we extract the prediction for the cut from our candidate integrand and land on the same answer as \eqn{cutpSoln}, then we will have verified the candidate integrand on at least this cut (and any cuts that this cut {\em spans} but I will explain that terminology in a bit). If, however, we get a different result, then it means our candidate is wrong --- there is missing information the theory requires that our integrand is failing to provide.

Let me carefully walk through extracting the cut information from the candidate integrand.  Recall the process is to first identify the set of contributing graphs.  Since both trees in the cut are color-ordered 5-point trees, there will be the same five topologies contributing from each tree amplitude, but they will be labeled differently. I generate and label all graphs from each tree in \fig{twoLoopPlanarCutGraphs}. The set of left tree graphs labeled $Ax$ contributes to the amplitude ${\rm A}^{tree}(K_1, K_2, l_a, l_b, l_c) $ and the  set of right tree graphs labeled $By$ contributes to the amplitude ${\rm A}^{tree}(- l_c, -l_b,-l_a, K_3, K_4)$.  The set of all possible two-loop graphs contributing to the cut is the outer product 
\be
{{\cal G}_{\rm  cut,\,O} } \equiv A \otimes B = \left\{AaBa, AaBb,\ldots, AeBd, AeBe\right\} \,.
\ee   Notice that each multi-loop graph we get out of gluing any graph of $A$ with any of $B$ is indeed a four-point two-loop graph. However, only three of these graphs correspond to a graph from our candidate integrand.  Every combination except for $AaBa, AbBb, AcBc$ contains triangles or worse.  In fact, all three of $AaBa, AbBb, AcBc$ are differently-labeled versions of the planar graph defined in \eqn{planar2LoopDefn}.

\begin{figure}
\centerline{\includegraphics[width=3in,bb= 0 0 330 771]{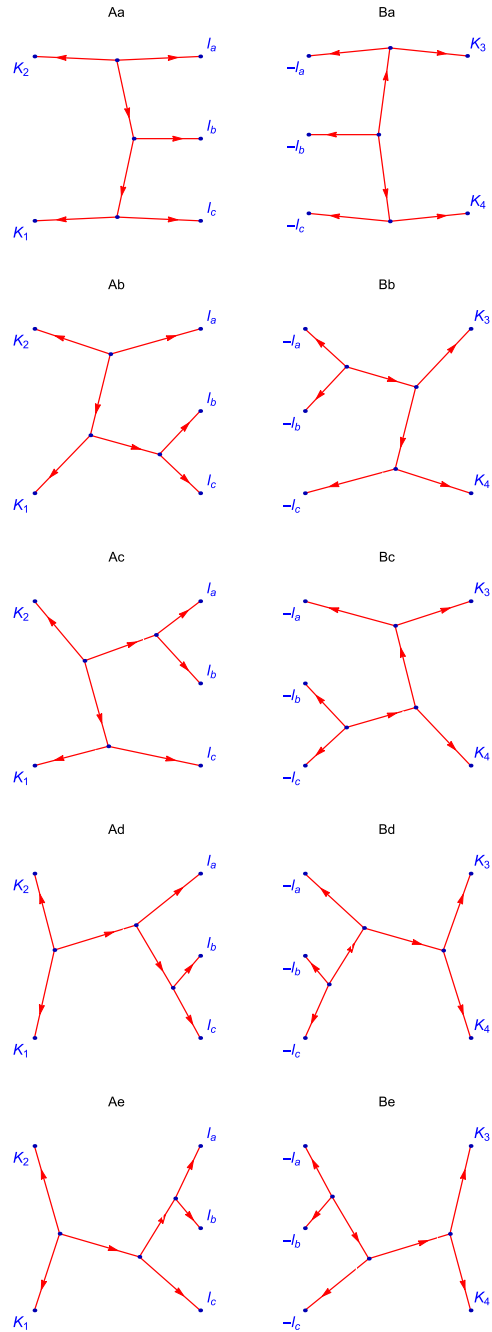}}
\caption{The tree graphs contributing from each of the trees in the planar 2-loop cut \eqn{cutpDefn}. } \label{twoLoopPlanarCutGraphs}
\end{figure}

To be very explicit I'm going to write out the vertex-organized graph representations of tree-graphs $Aa$ and $Ba$, and show what it means to glue them to make the multi-loop cut-graph $AaBa$.  
\begin{align}
Aa &= \left\{\begin{array}{c}
 \left(i_ 1,K_ 2,l_a \right) \\
 \left(-i_1,l_b,i_ 2\right) \\
 \left(-i_2,l_c,K_ 1\right) \\
\end{array}\right\} \,, \\
Ba &= \left\{
\begin{array}{c}
 \left(i_ 4,-l_b,i_ 3\right) \\
 \left(-i_3,-l_a,K_ 3\right) \\
 \left(-i_4,K_ 4,-l_c\right) \\
\end{array}
\right\} \,.
\end{align}
We glue them together simply by collecting all their vertices into the same graph (after first verifying that none of the internal labels $i_x$ conflict --- if they had, we would simply relabel any conflicting internal vertices before gluing),
\begin{equation}
AaBa = \left\{\begin{array}{c}
 \left(i_ 1,K_ 2,l_a \right) \\
 \left(-i_1,l_b,i_ 2\right) \\
 \left(-i_2,l_c,K_ 1\right) \\
 \left(i_ 4,-l_b,i_ 3\right) \\
 \left(-i_3,-l_a,K_ 3\right) \\
 \left(-i_4,K_ 4,-l_c\right) \\
\end{array}
\right\}.
\end{equation}
This amounts to literally connecting $-l_x$ and $l_x$ from each of the tree graphs to form a multi-loop graph.  

How do we know that this is a relabeled version of the planar two-loop graph in our candidate integrand? Mathematically the answer is settled once one finds an isomorphism between the two graphs --- the mapping of labels in the planar graph of \eqn{planar2LoopDefn} to $AaBa$ --- or else one shows that no such isomorphism exists (by ruling out all possible mappings). In this case we do have multiple isomorphisms, one of which is the mapping 
$\{k_1 \to K_1, k_2 \to K_4 , k_3 \to K_3,
 l_{11}\to l_c, l_{8}\to l_a \} $. Only those edges need be specified, all other edges follow from momentum conservation.  We can verify that there is an even number of odd-vertex permutations, so that we can just apply this isomorphism to $n({\rm planar})$ to get 
\begin{multline}
n(AaBa) = ( (K_1+K_4)^2)^2 (K_4+K_3)^2  \AtreeCO{4}(K_1,K_4,K_3,K_2) \\
= (K_1+K_4)^2 (K_1+K_2)^2 (K_2+K_3)^4  \AtreeCO{4}(K_1,K_2,K_3,K_4)\,
\end{multline}
 where in the second line we used the fact that $(a+b)^2 (b+c)^2  \AtreeCO{4}(a,b,c,d)$ is invariant under permutations of $\{a,b,c,d\}$.
Similarly we can find isomorphisms from the planar graph of \eqn{planar2LoopDefn} to both  $AbBb$ and $AcBc$, yielding the 
numerators 
\begin{align}
n(AbBb)  &= ( (K_1+K_2)^2)^2 (K_2+K_3)^2 \AtreeCO{4}(K_1,K_2,K_3,K_4)\,, \\
n(AcBc)  &= ( (K_1+K_2)^2)^2 (K_2+K_3)^2 \AtreeCO{4}(K_1,K_2,K_3,K_4)\,.
\end{align}
I will leave it as an exercise for the reader to verify that the denominators work out for these three graphs to exactly reproduce \eqn{cutpSoln}, thus verifying our graph-organized integrand on this cut.

\begin{problem}
One important thing to note is that, when generating the denominators associated with cut graphs, one must never write down the propagator associated with cut edges.  Why?
\end{problem}

\begin{problem}
\label{isoProblemPlanarGraph}
How many isomorphisms are there between AaBa and the planar graph of  \eqn{planar2LoopDefn}?  Show that they each reproduce the same $n(AaBa)$ and $d(AaBa)$ under momentum conservation. 
\end{problem}

\begin{problem}
If instead we consider the cut where $-l_c$ and $-l_b$ are swapped in the second tree we will have non-planar contributions.  The same tricks apply, but in this case we will also pick up negative signs for the numerators for some of our graphs, because there will be an odd number of odd-vertex permutations between the candidate graphs and some of the cut graphs.  Verify that you get the following result: 
\begin{multline}
\label{cutnpSoln}
\text{\rm cut}({\rm np})
 = -\frac{\left(K_1+K_2\right){}^2}{\left(l_a+l_b\right){}^2 \left(l_a+l_c\right){}^2
   \left(K_4-l_b\right){}^2
   \left(-l_a-l_b-K_2\right){}^2}\\
   -\frac{\left(K_1+K_2\right){}^2}{\left(l_a+l_c\right){}^2 \left(-l_a-K_2\right){}^2 \left(l_b+l_c\right){}^2
   \left(K_4-l_b\right){}^2}\\
   -\frac{\left(K_1+K_2\right){}^2}{\left(l_a+l_b\right){
   }^2 \left(K_4-l_b\right){}^2 \left(-l_a-l_b-K_2\right){}^2
   \left(l_b+l_c-K_4\right){}^2}\\
   -\frac{\left(K_1+K_2\right){}^2}{\left(l_a+l_b\right){}^2 \left(l_b+l_c\right){}^2 \left(-l_a-l_b-K_2\right){}^2
   \left(l_b+l_c-K_4\right){}^2}\\
   -\frac{\left(K_3+K_4\right){}^2}{\left(l_a+l_c\right){}^2 \left(-l_a-K_2\right){}^2 \left(K_4-l_b\right){}^2
   \left(-l_a-l_b-K_2\right){}^2}\\
   +\frac{\left(K_2+K_3\right){}^2}{\left(-l_a-K_2\right){}^2 \left(K_4-l_b\right){}^2 \left(-l_a-l_b-K_2\right){}^2
   \left(l_b+l_c-K_4\right){}^2} \,.
 \end{multline}
 
 Hint: You can just swap $-l_b$ and $-l_c$ labels in the second column graphs of \fig{twoLoopPlanarCutGraphs}.  But you still need to work out what combination of $A\otimes B$ are isomorphic to our two dressed graphs in our candidate integrand.  You can simplify your life by realizing that you can always neglect any graph that involves $Ad$, $Ae$, $Bd$, and $Be$.  (Why?)  This means you just have to see which of the 9 cut loop-graphs are isomorphic to our two integrand graphs --- only 6 will be.
 
 \end{problem}

\end{example}

\subsubsection{Do we really have to do every cut?}
After this example you might be thinking to yourself, well,  great.  But if the only way to know if we have a good answer is to do every cut, and that means pick up every pole in every combination, well, that sure is a tremendous amount of work.  You would not be wrong.  There is a saving grace, however.  You might not have noticed, but I qualified the coverage of what we verified in the previous example using the word ``spans'', and this is where the magic of verification happens. 

 If a cut is satisfied by a graph-organized integrand, then, provably, any cut of that cut is satisfied.  We say that a cut {\em spans} every cut that has the same or additional on-shell constraints.  It turns out that verification requires very few cuts indeed.  Of course a minimal spanning set is the set of color-dressed $L$-particle cuts for gauge theories (or simply the set of $L$-particle cuts for gravity theories) which sews a $2L + m$ tree amplitude to itself to probe every channel in every possible way.  But this is done in a highly redundant manner representing a very computationally expensive cut to perform as the multiplicity and loop order increases. As an alternative, one can start with all edges of all graphs cut (a set of cuts equal to the number of cubic graphs), and aggregate the information obtained by removing on-shell conditions (i.e.~taking on-shell internal edges off shell) until one has saturated the known (or maximal candidate) power-counting\footnote{Power-counting refers to the behavior of the integrand when all the loop momenta become large.  In gauge theory, generically, there is at most one power of the loop momentum at each vertex, and this leads to a power-counting of $m$ powers of loop momentum in all numerator factors for one-loop $m$-point integrands.  In supersymmetric theories the power-counting behavior can be considerably better.  It is $m-4$, not $m$, for one-loop $m$-point processes, and it is expected to be $2(L-2)$ for $L$-loop 4-point processes. See also the discussion in section~\ref{pcsubsubsection} and in Problem~\ref{pcproblem}.}
of the theory  --- thereby establishing a spanning set that is sufficient for a given theory where each individual cut is hopefully still manageable.

\subsubsection{What is really going on here?}
One can ask what an amplitude's job is. Arguably, at least at the integrand level, its job,  its sacred duty even, is to encode the residues of all the poles.  How do we ensure that all the correct residues are present?  We verify that every possible pole  has the appropriate associated information by comparing against (hopefully) compact on-shell quantities.  When we have our integrand organized graphically we can reach inside our integral, with all its redundancy and over-counting, and precisely pick out what contributions coincide with the sewing together of physical on-shell tree amplitudes.  The power of easy verification that can precisely target an integrand, pole by pole, leads to a natural method of construction which we will consider now.

\subsection{Construction}

Americans play a spoken game called ``20 questions''. The goal of the game is for a questioner to figure out within 20 Yes/No questions whatever subject an answerer has in mind.  The game is won if the questioners can correctly guess the object\footnote{A popular variant is called ``Animal/Vegetable/Mineral'' which constrains the answerer from choosing something abstract like the number 3,411,323,423.}.  Given how frequently the game is won, it suggests that many people rarely have more than $2^{20}$ objects to be thinking about.  Clearly not enough people are concerned with which cubic graphs contribute to the 14 graviton tree-level scattering amplitude, or how to fully parametrize a local four-point five-loop integrand. If you have ever played a similar game, you know how valuable having an Oracle telling you ``yes/no'' can be towards finding solutions if you ask questions efficiently.

It turns out that by asking the correct questions, by considering unitarity cuts, we have a natural means of construction of integrands of scattering amplitudes.  The way we play this construction game, when we have an idea of the the rough form of the functional dependence of the numerators $n(g)$ (using locality, power-counting, etc.), will often involve the use of Lorentz-invariant ans\"atze --- parameterized guesses whose parameters will be fixed by evaluating a clever choice of cuts.  Strictly speaking such guesses are not actually necessary --- one can put the data from cuts together like piecing together puzzle pieces --- and this is fantastic.  Still, let me spend a second advocating for ans\"atze.  At the time of writing this lecture, the state sums in four-dimensional cuts are much easier to evaluate than higher-dimensional cuts --- the book-keeping and expressions are far more compact.  One can write down $D$-dimensional ans\"atze, and first impose all of the four-dimensional cuts.  This procedure will leave unconstrained at most terms in the integrand that vanish when the loop-momenta are all four-dimensional.  These terms can be constrained by relevant ``surgical'' higher-dimensional cuts. 

\subsubsection{Sidebar on locality and power-counting}
\label{pcsubsubsection}

We have been discussing {\em local} quantum field theories, whose amplitudes could (in principle) have been calculated using local Feynman rules.  This means that, expressed in terms of polarization tensors and momentum invariants, we should expect to find representations where the only poles associated with graphs are a result of graph-propagators vanishing.  This means that graph numerators can be expressed as polynomials in momentum invariants, and in Lorentz products between momenta and polarization tensors.  In other words, for  {\it local representations}, the numerator functions $n(g)$ will have no denominators.

What degree should the polynomial $n(g)$ have for a given theory?  If $p$ stands for either external momenta or internal loop momenta, then the scaling with $p$ can be determined by dimensional analysis.   An $m$-particle scattering amplitude in gauge theory in four dimensions (where the gauge coupling is dimensionless) has dimension $4-m$.  The measure at $L$ loops has dimension $4L$, and the propagators account for $-6L-2m +6$, for $m\ge4$.  Thus after accounting for the dimensions of the measure and the propagators, each numerator has dimension $m + 2L -2$.  For gravity theories, the double-copy representation tells us that numerators for loop amplitudes should contain twice the power of momenta; the extra $2L$ per loop compensates for the dimension of the gravitational coupling factor $\kappa^2 \propto 1/M_{\rm Planck}^2$.

The notion of {\em power-counting} involves tracking how many {\em loop} momenta one expects to see in the contribution of a graph.  It answers the question, after consideration of the propagators and the dimension-dependent measure of integration, how many of the numerator momenta $p$ can be internal loop momenta $l$?  This is intimately related to the ultraviolet behavior of a theory.  Specifically if one has too many loop-momenta in numerators for a given dimension, without a symmetry forcing cancellations, one can expect ultraviolet divergences.  We will discuss the power-counting for the maximally supersymmetric gauge theory in Problem~\ref{pcproblem}.

\begin{problem}
 For a theory to be finite in $D$ dimensions, what is the maximum power of loop momenta that can be present graph by graph in a local representation?
 \end{problem}

\begin{problem}
The answer to the previous problem was $[l]^\infty$ --- as long as the extraneous loop-momentum dependence cancels between graphs.  Now consider  the more useful question: what's the most loop momenta that can appear so that each graph is manifestly finite without requiring cancellation between graphs?
 \end{problem}
 
\subsubsection{Method of a minimal cut}
 Let us consider first a definitive, but ultimately impractical, way of getting the integrand from on-shell information.  This method is to basically write down anything the answer possibly could be, also called an {\it ansatz}, and ask the Oracle one (very big) question. By evaluating the answer numerically multiple times, we can then tease out what the solution is.

\begin{enumerate}
\item Write down all cubic graphs relevant to the $m$-point amplitude at $L$ loops.
\item For each graph, write an ansatz for the numerator as a polynomial in the available Lorentz products, with a free parameter for each term and an overall degree in the loop momenta that is consistent with the known power-counting of the theory. 
\item Calculate the unordered $L$-particle cut by sewing the unordered $m+2L$ tree to itself.
\item Calculate the prediction of your ansatz on this cut.
\item Evaluate the equality between the two numerically using random momenta for external legs and cut-momenta, until you have a sufficient number of relations to solve for all parameters.
\item Solve for the parameters.
\end{enumerate}

This method, while fine in principle, requires a tremendous number of parameters as multiplicity and loop level increases.  The ``on-shell'' cut grows incredibly unwieldy  at even modest loop orders and multiplicity.  While precise, the redundancy in this object is absolutely overkill\footnote{But perhaps not overkill when thinking about trying to establish loop-level on-shell recursion.  See ref.~\cite{planarRecursion} and consider how to apply such methods to non-planar theories.} for cut-construction. Rather than confronting each pole once, you confront it a myriad of times in many largely irrelevant ways.  We can do better.

\subsubsection{Method of maximal cuts~\cite{FiveLoop,CompactThree,Neq44np}}

\begin{figure}
\centerline{\includegraphics[width=4.5in]{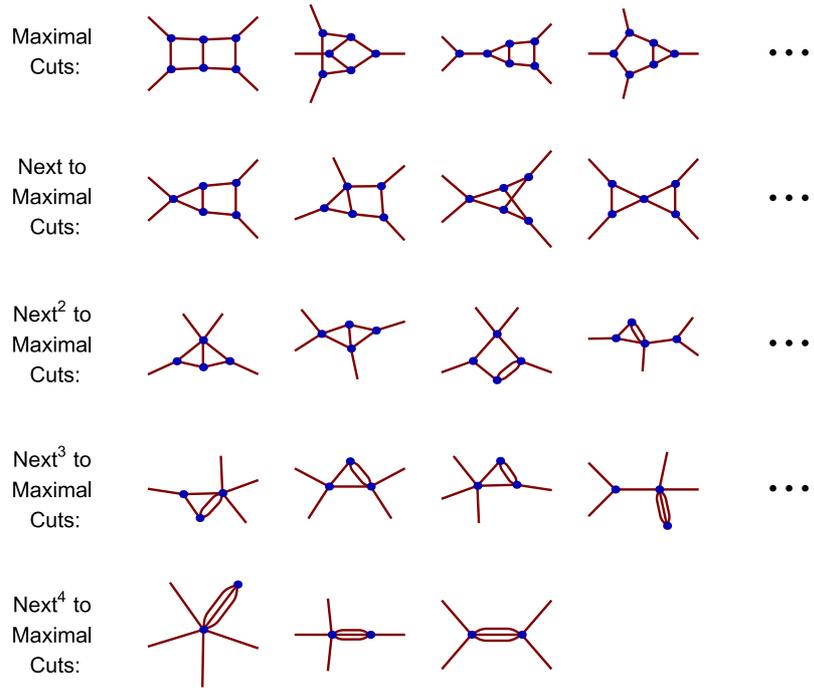}}
\caption{The hierarchy of cuts considered in the method of maximal cuts for a two-loop four-point amplitude.  Every exposed internal leg is taken to be cut, every blob is meant to represent a tree.  First one considers the maximal cuts, then the next-to-maximal cuts and so on.  Note that the next${}^n$-to-maximal cuts involve higher-point trees.   } \label{fig:maxCuts}
\end{figure}

The approach here is to consider a hierarchy of cuts.  First do the simplest cuts that access the least information --- targeting individual graphs --- information that must be pegged to those isolated graphs in any representation.  The next level of hierarchy of cuts allows for information that may be shared between a pair of graphs (for ordered cuts) or three graphs  (unordered cuts).  Such information goes by the name: ``contact terms.''  Once these contact terms have been unambiguously and consistently assigned to parent graphs, then the next level of hierarchy (joining previous clusters pairwise, or in threes) is considered.  So on and so forth, until all the information in the system of cuts is encoded in numerators associated with the cubic parent graphs.  The process is finished when the result is demonstrated to satisfy a spanning set of cuts.   See \fig{fig:maxCuts} for the relevant hierarchy of cuts at two-loops.

In a bit more detail, the method of maximal cuts for the $L$-loop $m$-point amplitude has the following steps :
\begin{enumerate}
\item Generate the set of all maximal $m$-point $L$-loop cuts.  How?  First write down all cubic graphs relevant to $m$ points and $L$ loops\footnote{How? Write a set consisting of one graph with the correct properties, then close it under $\widehat{t}$ and $\widehat{u}$ on all edges of all generated graphs, up to isomorphisms.}. Turn each graph into a cut by taking all vertices and simply regarding them\footnote{I could be pedantic here and introduce an operator to turn graphs into sets of trees, one for each vertex; if this isn't obvious treat it as an exercise.} as three-point trees.  Each graph, so regarded, lists the product of trees whose states need to be summed over to evaluate the cut.  These are called maximal cuts --- they target individual graphs with every propagator put on shell. These cuts will tell us inarguably what local information the theory insists comes along with every graph. Will it be all the information the theory wants?  Absolutely not --- it will be missing information proportional to inverse propagators (things that vanish when propagators are cut).
\item Assign to the numerators of the cubic parent graphs this maximal cut information --- there is no ambiguity. Each cut targets specifically one and only one parent cubic graph.  One must only take care so as to assign the information in a way that is consistent with the automorphic symmetries of the parent graphs.
\item Next consider the set of cuts with one fewer cut condition than the maximal cuts --- i.e.~with one propagator off shell.  These are called ``next-to-maximal cuts''.  To get all of the next-to-maximal cuts associated with a graph, list every internal edge of that graph.  For each internal edge $e$ generate a new graph by applying the collapse operator ${\cal C}_e$ to the original graph (see \eqn{collapseOperator}). You will have generated a set of graphs that have one quartic vertex and the rest cubic.  This process is also referred to as releasing the cut condition associated with edge $e$. If you turn the vertices into trees, then you have the cut that spans all the maximal cuts one can find by expanding out the quartic vertex in all possible ways.  But in  addition, this cut gives us new information: any data proportional to the square of the momentum running along the parent's edge $e$ that was collapsed to make this cut. 
\item Assign any newly identified missing information from the next-to-maximal cuts to any relevant parent cubic graphs.  This new information can be assigned in whole or in part to any of the parent cubic graphs relevant to the cut providing the missing data.  No way of assigning this information can spoil the already satisfied maximal cuts.  However, care must be taken to assign this data so as to simultaneously satisfy the symmetries of the graphs and not disturb other next-to-maximal cuts relevant to this level.  One way of doing this, that does not necessarily maintain manifest power-counting, is to simply tag the missing information with the explicit inverse-propagator when assigning it to the parent graphs. Schematically,
\be
n(\text{parent})=\text{max-cut} + e^2 (\text{missing near-max-cut info})\,,
\ee
where $e^2$ is the inverse propagator associated with the edge $e$. 
\item Repeat the previous two steps --- releasing cut conditions, and assigning any missing information to cubic parent graphs --- until one reaches a stage in the next$^n$-to-maximal hierarchy where there is no missing information in any such cut; i.e.~all these cuts are successfully reproduced by the candidate integrand, expressed in terms of cubic graphs and their associated numerators.
\item Verify the candidate integrand on a spanning set of cuts, or at least those cuts that can detect all possible numerator polynomials consistent with the maximum possible power-counting of the theory.
\end{enumerate}

Despite the large number of cuts performed, this method is remarkably practical.  While the number of cuts obviously grows  faster than the number of graphs, at every level of the hierarchy each cut targets the maximally local missing information. This approach is efficient because it identifies the smallest amount of additional information that needs to be included in the solution.

As an example, consider the maximal cut of the planar graph in the two-loop four-point $\cN=4$ super-Yang-Mills scattering amplitude given in \eqn{planar2LoopDefn}.  It is quite simply:\footnote{At least it is simple once you understand how to manipulate spinor-helicity variables for complex massless 3-point kinematics.  A thorough review is beyond the scope of my current lectures but the ideas are quite straightforward.  I'll direct you to ref.~\cite{Carrasco:2011hw} for a pedagogical review and references to original source material.} $s^2 t \AtreeCO{4}(1234)$. This is the result of sewing together all the 3-point trees that represent cutting all of its propagators. We assign it to the numerator of the planar graph, since the planar graph is the only graph that contributes to that cut. We do the maximal cut of the non-planar graph given in \eqn{nonplanar2LoopDefn}, and again find $s^2 t \AtreeCO{4}(1234)$, and similarly assign that contribution to the non-planar graph.  Checking all other maximal cuts in the theory, the top line of \fig{fig:maxCuts}, we find that all other cubic graphs have vanishing contributions.  At this point we actually have the entire solution. Any cuts other than the maximal ones would require additional information pinned in such a way so as to violate the power-counting of the theory, i.e.~actual dependence of the planar and non-planar numerators on the loop momentum. (See Problem~\ref{pcproblem}.)  Still, to verify the answer, we can release cut conditions.  We first consider the next-to-maximal (NMax) cuts --- the second line of \fig{fig:maxCuts}.  As an example of a NMax cut, consider collapsing the propagator called $l_{11}$ in the planar double box in \eqn{planar2LoopDefn} to get the following color-ordered cut:
\begin{align}
{\rm NMax}_{\,p,\, l_{11}} &= \sum_{\rm states} \Bigl(\AtreeCO{3}\left( l_5,l_9,-l_6 \right)
   \AtreeCO{3}\left( l_6,l_{10},-l_7 \right)\\
&~  \times \AtreeCO{3}\left( k_3,-l_8,l_7\right)
   \AtreeCO{3}\left(k_4,-l_5,l_8\right)\nonumber\\
&~ \times \AtreeCO{4}\left( k_1,k_2,-l_{10},-l_9 \right)  \Bigr)\nonumber\\
&= \frac{s_{12}^2 s_{23} \AtreeCO{4}(1234)}{(l_{10}-k_2)^2}\,.
\end{align}
This result verifies the integrand generated by the maximal cuts --- no missing data is revealed by this cut.  (Note that $l_{10}-k_2 = l_{11}$).  In fact, there will not be any missing information for any other cuts since (as mentioned above) for the maximally supersymmetric theory, the maximal cuts for the two-loop four-point amplitude contain all the information --- we already have the answer.  After verifying the answer on all NMax cuts, you really are done, since you will have verified it on a spanning set above the power-counting of the theory.
 
Let us say we were dealing with some theory deformed from $\cN=4$ sYM, such that all maximal cuts were identical, but in our new toy theory a contact term is hiding\footnote{I'm not saying such a theory would be consistent, but for this toy example it is beside the point.},
\be
{\rm NMax}_{\,p,\, l_{11}, {\rm deformed}} 
= s_{12} s_{23} \AtreeCO{4}(1234)\times \left(\frac{s_{12}}{(l_{10}-k_2)^2} + 1 \right) \,.
\ee
Here I chose the missing information associated with this cut to be $ s t \AtreeCO{4}(1234)$.  We need to assign this contact term to either the planar double-box or to a triangle-box graph.  Apply $\widehat{t}$ to $l_{11}$ of the planar-box and you'll get the triangle-box graph, which is isomorphic to the third maximal-cut graph of \fig{fig:maxCuts}.  We can assign it to the planar-double box with a factor of $l_{11}$ to pin the contribution to this cut:
\be
n({\rm planar}) \to s t \AtreeCO{4}(1234) ( s + l_{11}^2 )
\ee
Or, instead, we could give it to the triangle-box:
\be
n(\text{triangle-box}) \to s^2 t \AtreeCO{4}(1234)\,.
\ee
Notice how for the triangle-box the collapsed propagator consists only of external momentum, whereas for the planar box, the contact term has loop-momenta dependence.

\begin{problem}
Explain why ``$s$'' is the correct factor for the triangle-box to inherit the counterterm present in our toy deformed theory.  Is this consistent with the maximal cut of the triangle-box graph vanishing? If this is equivalent to adding an $l_{11}^2$ to the planar-double box, how is the triangle-box contact term consistent with the correct power-counting? Hint: consider the loop-momentum dependence of each graph's denominator as well.
\end{problem}

\begin{problem}
As stated exactly above the vast majority of cuts one performs at higher and higher N${}^{k}$Max cuts will be trivially satisfied because most of the cuts are redundant with each other.  Identical N${}^{k}$Max cuts (albeit with potentially different labels) will come from collapsing a propagator on multiple N${}^{k-1}$Max cuts.  Describe a way to remove this ambiguity procedurally.  Hint: graph isomorphism.
\end{problem}

\begin{problem}
How far away from maximal cuts was the 3-particle cut verified in the verification example~(\ref{cutpSoln}) above?
\end{problem}

\begin{problem}
{\em Method of Intermediate Cuts.}  
\label{probMethodInt}
Imagine a case where you understand all of the functional dependence of the numerator factors for a theory, modulo some free parameters that will take on some friendly rational numbers like $1,0,-1$.  Is there a happy medium between the methods described above, which lets you fix those free parameters without doing countless trivial cuts, and without doing one big monster cut?  Describe such an approach, and how you would  choose which cuts to evaluate. 

This is very much the case that confronts one in the maximally supersymmetric gauge theory in the limit where the number of colors $N_c\to\infty$.  This is the  limit where only planar graphs  contribute. Planar graphs are those that, after joining all external edges to the same external point, can be drawn on a plane with no crossed edges.  In the planar limit dual conformal symmetry governs all contributions, so one should simply write down all relevant local dual-conformal numerators.  Dual conformal symmetry alone does not fix the coefficients of these integrals, but both the method of maximal cuts and the method of a minimal cut are overkill.
\end{problem}

The construction methods described in this section represent fantastically useful approaches.  As mentioned earlier, we can apply them even when we have no idea what type of functional dependence could show up for a given scattering amplitude (abandoning locality, or expected power-counting) --- as long as we can carry out the cuts and manipulate the expressions into a Lorentz-invariant symmetric form, we can assign the information to parent graphs.

 There are many sophisticated approaches in this spirit that are currently being brought to bear to tackle some of the very serious multi-loop challenges confronting an understanding of QCD backgrounds that are relevan for LHC physics.  There is a problem, however: the number of graphs increases factorially with multiplicity and loop order, so at some point, no matter what, the computational power of any graduate student, or any available software, on any available hardware, can be completely saturated.  We will see in the next section how the imposing of kinematic-Jacobi identities between graphs has the power to potentially fight this growing complexity, and I will outline some of the current problems that confront us.  First I should tell you how to get the actual expression you may want to integrate after understanding all of these $n(g)$ mappings.

\subsection{Yeah, but what do we integrate?}

After we have constructed a graph-organized set of numerator mappings that satisfies all cuts, we still want an algebraic expression that can be integrated.  We write it as:
\begin{align}
\text{alg. YM integrand} &= \sum_{\text{external permutations} }  \sum_{g \in {\cal G}}  \frac{1}{S(g)} \frac{n(g) c(g)}{d(g)}\,, \\
\text{alg. GR integrand} &= \sum_{\text{external permutations} }  \sum_{g \in {\cal G}}  \frac{1}{S(g)} \frac{n_{\rm GR}(g)}{d(g)}\,,
\end{align}
where the sum is over all permutations of external labels. The symmetry factor $S(g)$ is equal to the number of automorphisms of $g$; it accounts for any over-counting due to either the external permutation sum or the eventual integration.

\section{Exploiting Color-Kinematics Duality}
\label{ckexploit}

Now that you have some familiarity with how multi-loop integrands can be organized graphically for full non-planar theories, and the verification criteria they must satisfy, we can discuss the primary difference between applying Jacobi relations at tree level and at loop level.  For space and time restrictions, I will not work out any multi-loop examples here, but fortunately the literature has some very pedagogical discussions of this at three loops --- please see refs.~\cite{BCJLoop, BCJ4Loop, Carrasco:2011hw}.

\begin{figure}
\centerline{\includegraphics[width=4
in]{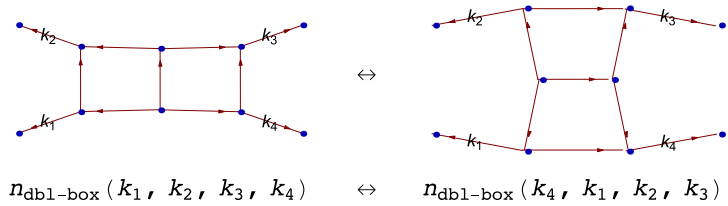}}
\caption{ When a multi-loop graph has an automorphism, the numerator must satisfy automorphism symmetry at least up to the redundancy of the theory.   For ${\cal N}=4$ super-Yang-Mills the above automorphism holds for all external states -- i.e.~the same function applies to every labeling of the same topology that contributes to the integrand.   For pure Yang-Mills the above relation only holds when all four external gluons have the same helicity. } \label{fig:automorphism}
\end{figure}

The numerators at loop level must satisfy automorphism symmetry, at least up to the redundancy allowed by the theory.  For example, for maximally supersymmetric $\NeqFour$ super-Yang-Mills theory at the four-point level, there are supersymmetry Ward identities that imply that all external state dependence can be encoded in the prefactor of a permutation-invariant function of the tree-level amplitude: $ s t \Atree{4}(1234)$.  The entire expression must be invariant under automorphisms (see e.g.~\fig{fig:automorphism}).

However, for pure-YM, I will have different numerators based upon what helicity gluon is on what external leg.  I can encode this graphically by adding extra structure to differentiate plus helicity external legs from minus helicity external legs. (e.g.~by putting a `dot', or two-vertex, on plus-helicity external legs).  So we break what might be automorphic symmetries of all external legs to allow the theory the freedom to behave drastically differently given different external states --- yet still the numerators must obey all remaining  external and (especially!) internal automorphic symmetries.

What we did at tree level suffered no such requirement.  In fact at the level of graphs we might as well have been treating every distinctly labeled external edge as something entirely graphically distinct (give each edge as many `dots' as the label: 1 for $k_1$, 2 for $k_2$, etc.).  With no automorphic symmetry requirement, the Jacobi relations were taken in our tree-level section purely as linear constraints.  Furthermore, because gauge-invariant objects depend on these numerator factors, we could pseudo-invert, defining the Jacobi-satisfying numerators in terms of functions of tree-level color-ordered partial amplitudes.

At loop level, however, Jacobi relations impose functional constraints.  Look at what happens when you apply $\widehat{u}$ to $l_{11}$ of the planar double-box in the previous section: you get back the planar double box but with external legs 1 and 2 exchanged.  Even if you set the triangle-box contribution for $\widehat{t}\circ l_{11}\to0$, this means the following functional constraint from the Jacobi relation on that leg:
\be
\label{JacConstraintFunctional}
n_{\text{planar}}(k_1,k_2,k_3,l_a,l_b) = n_{\text{planar}}(k_2,k_1,k_3,l_a,l_b) \,.
\ee

This is quite different from what occurs at tree level.  If you recall the single four-point tree-level Jacobi looks like:
\be
n_s = n_t + n_u \,.
\ee
We are absolutely free to take $n_u\to 0$, giving us $n_s=n_t$.  But this does not mean that there is some $f(a,b,c,d)$ that dresses all four-point half-ladder topologies such that  $n_s = f(1,2,3,4) = n_t = f(4,1,2,3)$  and $n_u = f(3,1,4,2)=0$.  Rather we are simply assigning three different functions to the graphs, and then equating two of them --- not placing a functional constraint on any.

To get an idea of the types of functional constraints we have to worry about at loop level it is a useful exercise to carry out a similar program at tree level.  Let us take four-point tree-level seriously, and try to find a numerator function $n$ such that:
\begin{align}
n(c,a,d,b) &=  n(a,b,c,d) - n(d,a,b,c) \,, \\
\Atree{4}(a,b,c,d) &=\frac{ n(a,b,c,d)}{s_{ab}} + \frac{ n(d,a,b,c)}{s_{da}} \, ,\\
n(a,b,c,d) & = -n(b,a,c,d)  = -n(a,b,d,c) = n(b,a,d,c)\,.
\end{align}
The first equation imposes the Jacobi relation, the second gives the definition in terms of the one independent scattering amplitude, and the third imposes antisymmetry.  
To solve these functional relations we need an ansatz.    We are free to choose an ansatz involving two independent amplitudes $\AtreeCO{4}(a,b,c,d)$ and $\AtreeCO{4}(a,b,d,c)$; thanks to the Kleiss-Kuijf relations, they span the space of color-ordered amplitudes without the need to put any momentum invariants in the denominator.    We will need one power of $s_{ij}$ on each to satisfy power counting.  One can express $s_{ac}$ as $-s_{ab}-s_{ad}$ (from $u=-s-t$), so there are only two choices of momentum invariant per scattering amplitude:
\begin{multline}
n(a,b,c,d) =  \alpha ~ s_{ab}\AtreeCO{4}(a,b,c,d) 
+ \beta  ~  s_{ad} \AtreeCO{4}(a,b,c,d)  \\
+ \gamma ~ s_{ab} \AtreeCO{4}(a,b,d,c)
 + \delta  ~s_{ad} \AtreeCO{4}(a,b,d,c) 
 \end{multline}
 This ansatz has more parameters than necessary, because the two color-ordered amplitudes are related by the BCJ relations, but this does not matter.  Plugging in our ansatz and solving in terms of the one independent BCJ amplitude gives us an entirely fixed answer: 
\be
 n(a,b,c,d) = \frac{s_{ab} s_{bc}}{ s_{ab} s_{bc} s_{ac}} \AtreeCO{4}(a,b,c,d)  \frac{1}{3} s_{ab} ( s_{ac} - s_{bc})\,.
\ee
We exploited all the residual gauge freedom of our linear Jacobi solution in order to set up this functional Jacobi solution at the four-point level.  Notice that, strictly-speaking, this automorphism-symmetric numerator is non-local in the external momenta.
\begin{problem}
Do this for the five-point tree.  Bonus points: do this for the six-point tree.  If you can do it generically, at any multiplicity, independent of helicity and dimension send me an email! 
Hint: See ref.~\cite{virtuousTrees}.
\end{problem}

As you'll note in this four-point example, only one function is required to describe the kinematic contributions of all three graphs. This pattern persists. To all multiplicity, all tree-level graphs can be expressed by Jacobi relations in terms of linear combinations of the numerators of half-ladder graphs.  So if an automorphic half-ladder representation can be found at each multiplicity, then for each amplitude only one function is necessary to encode the information of all $(2m-5)!!$ graphs.   Such a reduction to one function would be a tremendous compression with respect to the $(m-3)!$ objects that are treated as independent when one allows topologies to not respect external isomorphisms, as was done in earlier sections.

But in any case this is all to demonstrate that, in order to solve the functional relations imposed by Jacobi relations at the loop level, at present we must resort to the introduction of ans\"atze.  Now we need only provide an ansatz for the master graphs.   It turns out that for 3- and 4-loop four-point amplitudes in the maximally supersymmetric theory, there is only one master graph; all the remaining graphs are related by Jacobi. So we only have to provide an ansatz for one graph.  This feature tremendously reduces the number of parameters needed in the ansatz.  At 5 loops, looking at all cubic graphs without triangles, one seems to require two master graphs. Unfortunately, however, as of yet, no local ansatz has been found that satisfies all Jacobi equations and all cuts at 5 loops.

\begin{problem}
\label{pcproblem}
For the four-point amplitude in maximally supersymmetric Yang-Mills theory, in general, a naive power counting would have at worst\footnote{For a more sophisticated look at the power counting of $\cN=4$ sYM see refs.~\cite{NeqFourFiniteness,MS,twoLoopNeq8GR,HoweStelleRevisited}.}  one additional power of $l^2$ in the numerator of cubic graphs, for every loop order above two loops.  At three loops this goes as $l^2$, at four loops $l^4$, and at five loops $l^6$.   For four-point scattering there will be three independent external momenta.  At $L$ loops there will be $L$ independent loop momenta.  What size ans\"atze do we need for a four-point $L$-loop master graph, in order to fully span such a power-counting?  
\end{problem}

\subsection{A discussion of our original two-loop Jacobi-satisfying solution}

Recall that in my introductory lecture I discussed how the two-loop solution was color-dual. Now that we have a firmer understanding of some of the issues at play I would like to return to the discussion.

For any particular theory, one does not necessarily require the presence of all graphs. Notably, in the maximally supersymmetric theories, we have always been able to encode the multi-loop representations in terms of graphs that have no 1-loop triangle, bubble, or tadpole graphs.  This is a result of the freedom to move contact terms away from triangle graphs, and the fact that there are no contributions forced to triangle graphs by maximal cuts.    In any case, in the maximally supersymmetric two-loop four-point amplitude, all triangle graphs come in with a kinematic weight of 0.  Does this mean that the color factors associated with these graphs vanish?  Not at all. In fact they have a very definite value.

\begin{figure}
\begin{multline*}
c\left(  \!\!\!\!\! \vcenter{ \includegraphics[width=2in, bb=0 0 260 166]{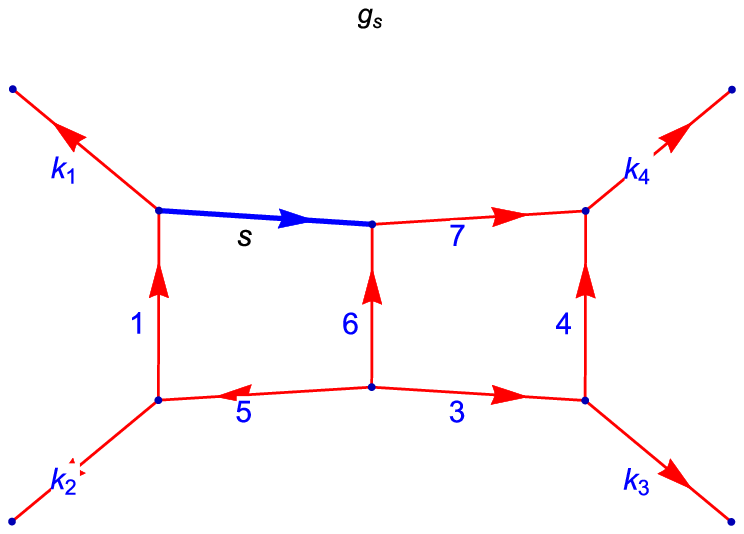}} \!\!\!\!\! \!\!\!\!\! \!\!\!\!\! \!\!\!\!\!  \!\!\!\!\! \!\!\!\!\! \!\!\!\!\! \!\!\!\!\!  \!\!\!\!\! \!\!\!\!\! \!\!\!\!\! \!\!\!\!\!  \!\!\!\!\! \!\!\!\!\!  \!\!\!\!\! \!\!\!\!\! \!\!\!\!\!  \!\!\!\!\! \!\!\!\!\! \!\!\!\!\! \!\!\!\!\! \!\!\!\!\! \right) = 
c\left(  \!\!\!\!\!  \vcenter{ \includegraphics[width=1.8in, bb=0 0 260 212]{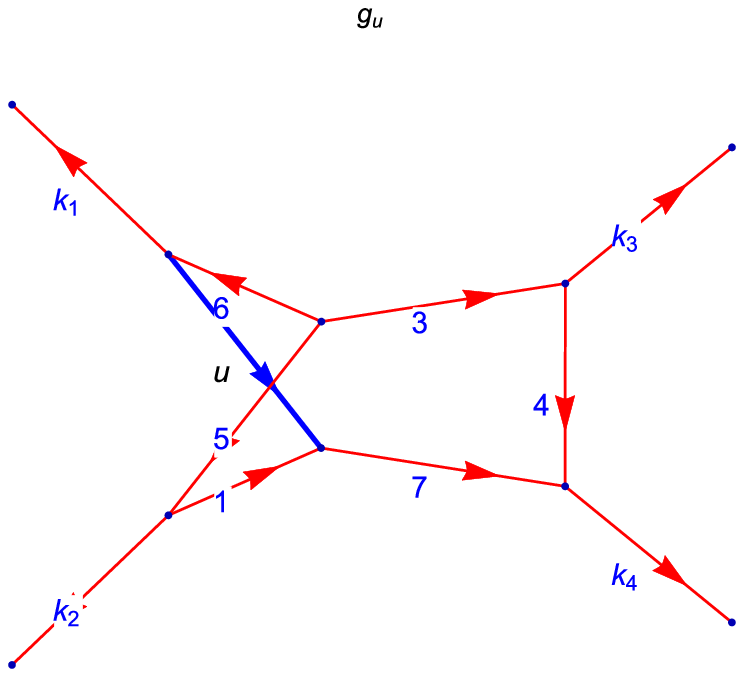}}  \!\!\!\!\! \!\!\!\!\! \!\!\!\!\! \!\!\!\!\!  \!\!\!\!\! \!\!\!\!\! \!\!\!\!\! \!\!\!\!\!  \!\!\!\!\! \!\!\!\!\! \!\!\!\!\! \!\!\!\!\!  \!\!\!\!\! \!\!\!\!\!  \!\!\!\!\! \!\!\!\!\! \!\!\!\!\!  \!\!\!\!\!\!\!\!\!\! \!\!\!\!\! \!\!\!\!\! \!\!\!\!\! \!\!\!\!\! \right) \\
 + c\left( \!\!\!\!\! \!\!\!\!\! \vcenter{ \includegraphics[width=2in, bb=0 0 260 215]{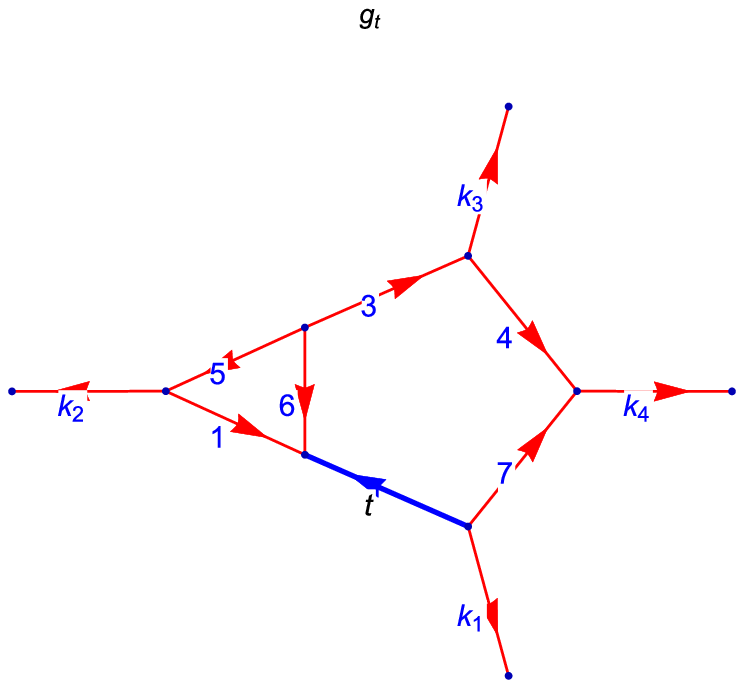}}  \!\!\!\!\! \!\!\!\!\! \!\!\!\!\! \!\!\!\!\!  \!\!\!\!\! \!\!\!\!\! \!\!\!\!\! \!\!\!\!\!  \!\!\!\!\! \!\!\!\!\! \!\!\!\!\! \!\!\!\!\!  \!\!\!\!\! \!\!\!\!\!  \!\!\!\!\! \!\!\!\!\! \!\!\!\!\!  \!\!\!\!\! \!\!\!\!\! \!\!\!\!\! \!\!\!\!\!   \!\!\!\!\!  \!\!\!\!\! \right)
\end{multline*}
\caption{Example of a two-loop color Jacobi relation.  The edge we Jacobi around is labeled $s$, $t$, and $u$ in each graph respectively.  The graph $g_t$ is a triangle graph  which does not contribute to \eqn{ym2Loop}. } \label{twoLoopJacobiFig}
\end{figure}

Consider applying the color-factor Jacobi relations to any of the edges of our two loop planar graph. For every edge you'll find that triangle-containing graphs appear.  See \fig{twoLoopJacobiFig} for an example.  Here we see a Jacobi relation relating the planar graph, the nonplanar graph, and a new triangle-box graph. The color-weight of the triangle-box graph is exactly the difference between the color weights of the other two graphs. You can check for yourself with the structure constants of SU(2) or SU(3) --- it is a great exercise.

In our two-loop amplitude we have a representation that does not require triangles.  
Note that this representation makes the power-counting of the theory manifest.  Triangles have worse power-counting than boxes as they have fewer loop momenta in the denominator.  The introduction of a triangle-containing graph would require loop-momentum dependence in the numerator of one (or both) of our other graphs, in order to cancel the behavior of the triangle-containing graph.  It turns out that if we allow ourselves to violate, graph-by-graph, the naive power-counting of the theory, then it is possible to find non-color-dual representations, but such {\em unfortunate} representations do not have manifest power-counting, and they were not found first (at two loops). 

Instead, setting $n({\rm triangle})=0$ and demanding manifest power-counting (along with the symmetry properties of the numerators) forces color-kinematics duality to hold at two loops:
\be
 n({\rm nonplanar}) = n({\rm planar}).
\ee
Color-kinematics duality then implies the double-copy property; i.e.~these numerators square to get the maximally supersymmetric supergravity two-loop amplitude.

At three loops, manifest power-counting behavior and locality are not enough to restrict to a representation that obeys Jacobi relations for every edge.  For example. the first expression found for the full ${\cal N}=4$ super-Yang-Mills scattering amplitude~\cite{superfinite} was in a representation where many Jacobi relations are satisfied for many edges of many graphs, but not all edges of all graphs.  Consequently, squaring the kinematic weights of that representation does {\em not} result in a corresponding maximally supersymmetric supergravity scattering amplitude.  It was only later, after a representation was found~\cite{BCJLoop} of the gauge theory where all the Jacobi relations are satisfied, that the supergravity amplitude came for free in a double-copy manner. But even in the first, semi-motivated representation, there was a tremendous amount of redundancy. For example, in the first description, which included 9 independent graphs, four of the graphs have a numerator of $s^2 ( s t \AtreeCO{4}(1234))$. Another 3 have a numerator of $s (l+k_4)^2 ( s t\AtreeCO{4}(1234))$. The last 2 have numerator factors that somewhat scramble combinations of the first few. But the number of unique expressions is very small.  Even without imposing the Jacobi relations, there was further evidence for something important:

{\em There is a tremendous amount of redundancy of information in gauge theory scattering amplitudes!}  

The amount of {\em unique} information we need from the theory to specify a given loop amplitude is incredibly small.  When we make manifest the Jacobi relations between kinematic numerators at 3 loops, we need only to specify the contribution of one of the graphs.  That graph only needs to be specified under the most restrictive kinematics that target that particular graph, namely its maximal cut.  Only a very small amount of information is required from the theory, and when the amplitude is written correctly, Jacobi relations propagate that small amount of information to the other graphs to generate the full amplitude.  At four loops there are 83 graphs, but only again one graph's numerator factor is needed~\cite{BCJ4Loop}, once we constrain to a Jacobi-satisfying representation.  The Jacobi relations propagate the information from that one graph to all 83 graphs.  So we see, there is a possibility of fighting the factorial explosion in the number of graphs by exploiting gauge freedom to impose relations that rigidly lock the behavior of graphs to each other.  This is the upshot of the tools we have been exploring: there exists a kernel of information to be grasped at.  Instead of having to look at a sea of 83 graphs we should only need to look at one graph. 

\section{Open Questions and Interesting Paths}

While there is tremendous potential for color-kinematics dual representations to dramatically decrease the complexity of higher-loop and higher-multiplicity calculation, it is clear that solving a new and ever growing ansatz loop order by loop order, multiplicity by multiplicity, cannot be the ultimate revolution.  Sure, it is a great way to collect data  in a variety of theories and lets us move incrementally forward as long as we can guess ans\"atze that are small enough to be manageable but broad enough to be relevant.  But if we really want to rewrite scattering amplitudes to fully exploit the universal structures we are discovering, we may want to look towards an invariant formulation of the duality --- and indeed for the kinematic structure constants, analogous to the color-factor structure constants that make our lives so simple when writing down $c(g)$ simply in terms of $f^{abc}$'s.

What are the barriers?  No known gauge choice allows for off-shell Feynman rules that directly generate cubic Jacobi-satisfying kinematic numerators when calculating on-shell quantities like tree-level and loop-level scattering amplitudes.  At tree level, this does not stop us from writing down all-multiplicity numerators in terms of gauge-invariant color-ordered tree amplitudes.  This is possible because we have an unambiguous algebraic structure we can invert.  So far no such gauge-invariant objects have been written at loop-level. The necessary functional form of loop level numerators, requisite to make manifest the internal automorphic symmetry, presents a non-trivial obstacle.  This often goes under the name ``the label-shifting problem.''   Turning the currently unbounded functional problem of finding Jacobi-satisfying loop-level numerators into a linear problem --- no matter what the resulting complexity bound --- will represent very real progress.

Here I should note the very interesting work looking for explicit kinematic ``structure constants", initially by considering the self-dual Yang-Mills theory in refs.~\cite{Monteiro:2011pc, BjerrumBohr:2012mg}, which has led to some fantastic all-multiplicity one-loop results~\cite{Boels:2013bi,Bjerrum-Bohr:2013iza}.  Speaking of one loop, I should also point the interested student to the  fruitful avenue pioneered by Mafra, Schlotterer, and Steiberger, which has been to consider amplitudes written in pure spinor superspace. This formalism allows for a covariant treatment of ten-dimensional super-Yang-Mills theory, as well as the full-fledged open superstring with manifest supersymmetry~\cite{Berkovits:2000fe}. In this framework, BCJ-satisfying local numerators for all-multiplicity tree-amplitudes  have been constructed~\cite{Mafra:2011kj}, as well as one-loop numerators for up to six external particles~\cite{Mafra:2014gja}. This construction invokes multiparticle superfields, ensuring color-kinematic satisfaction for external tree-level subgraphs at arbitrary loop-order~\cite{Mafra:2014oia} (and whose relation to non-linearities in ten-dimensional SYM was clarified in ref.~\cite{Mafra:2015gia}). An all-multiplicity classification of one-loop kinematic factors with gauge-invariant parity-even parts has recently been given in ref.~\cite{Mafra:2014gsa}.  Arguably one loop is somewhat special --- all graphs are planar.  That being said, these approaches and associated ideas may very well generalize to higher-multiplicity, higher-loop numerators.

On an alternate front, but one deeply related to the quest to solve the label-shifting problem, the fact that there exists a proven double-copy relationship at tree level to all multiplicity suggests that one should rightfully be able to find classical gauge theory solutions that, in the correct gauge, double-copy to classical solutions of general relativity + matter.  The first such explicit construction to my knowledge was for shock-waves by Saotome and Akhoury~\cite{Saotome:2012vy}.  Indeed, since I gave these lectures,  Monteiro, O'Connell and White have published a paper~\cite{KerrSchildBCJ} positing that   {\em Kerr-Schild}  coordinates in pure general relativity may represent  a double-copied gauge, a tantalizing idea as it describes  GR solutions such as Schwarzschild and Kerr Black holes.    Reducing the complexity of problems in classical gravity --- e.g.~astrophysical binary black holes --- to that of classical gauge-theory problems is an attractive prospect, especially now in the advent of direct gravitational wave observation.  If you're interested in understanding double-copy at the level of classical actions it is also worth looking at recent work by Anastasiou, Borsten, Duff, Hughes and Nagy~\cite{Borsten:2013bp,Anastasiou:2013hba,Anastasiou:2014qba}.

Besides the possibility of making tractable tough analytic predictions, understanding the relationship between classical gauge and gravity theories gives a hope of an invariant understanding of what these local perturbative Jacobi relations mean more broadly for gauge theories.   Of course it would be fantastic to be able to formulate these relationships geometrically, generalizing exciting ideas getting active play in the planar maximally supersymmetric theory~\cite{LoopPolytopes}.  Ultimately it will be important to also see the implications these relations have for integrated quantities directly relevant to collider observation (as well as our theoretical understanding of whether all point-like gravitational theories must be effective field theories, or if some can indeed be perturbatively finite in the ultraviolet in four dimensions). 

In any case, as exciting as the developments of these past few years are, there is so much more yet to come. We are very much at the beginning of developing an incredibly playful and intuitive way of looking at some  very serious questions --- a wonderful time to jump aboard!

\section{Appendix: Handy Expressions}
\label{appendix}

I wrote my lectures with the idea that everyone is already familiar with how to write down color-ordered tree-level scattering amplitudes.  It turns out that this is not, yet, a universal human property.  I am including this appendix, not as a replacement for the source material, but as a quick compression of data helpful for theories in four dimensions.  If you do not understand where these formulae come from, please take the time to study  some of the source material referenced.   

\subsection{Spinor Helicity for a Nickel}
In four dimensions there are some pretty spectacular four-dimensional representations of momenta going under the rubric of spinor-helicity notation.  See refs.~\cite{LanceTasi1996, ElvangHuangBook,LanceTasi2013} for all the details you could ever want.  I'm going to just tell you here a consistent representation for spinor products, given {\em massless} 4-momenta $a^\mu$, $b^\nu$:
\begin{align}
\label{helicityProducts}
 \langle a , b \rangle &=\frac{\left(a_1+i a_2\right) \left(b_0+b_3\right)-\left(a_0+a_3\right) \left(b_1+i
   b_2\right)}{\sqrt{a_0+a_3} \sqrt{b_0+b_3}} \,, \\
 \left[ a , b \right] &=\frac{\left(a_0+a_3\right) \left(b_1-i b_2\right)-\left(a_1-i a_2\right)
   \left(b_0+b_3\right)}{\sqrt{a_0+a_3} \sqrt{b_0+b_3}} \,.
\end{align}

\begin{problem}
Look up and read in ref.~\cite{LanceTasi1996} a standard definition of helicity spinors.  Derive the above representation of spinor products.
\end{problem}

Some important identities:
\begin{align}
\langle a , b \rangle \left [ b , a \right ] &= (a+b)^2 = 2 a \cdot b \,,\\
\langle a , a \rangle &= 0 = \left [a \, a \right] \,.
\end{align}
Here, as it is everywhere in these lectures, my dot product is a Minkowski four-product.

\begin{convention}
There is an ambiguity in spinor products for how to handle negative momenta across cuts.  The following convention resolves this ambiguity:
\begin{align}
\label{FermionArrowSign}
  \langle (-a) , b \rangle &\to - \langle a, b\rangle \,,\\
  \langle a , (-b) \rangle &\to -\langle a, b\rangle \,,  \nonumber \\
  \left[  (-a) , b \right  ]  &\to   \left[  a , b \right ]  \,, \nonumber  \\
  \left[  a , (-b) \right  ]  &\to   \left[  a , b \right ] \,. \nonumber
\end{align}
\end{convention}

\subsection{MHV: The equation that launched 1000 ships.}
 Recall that the two gluonic states in 4D can be described as $+$ helicity and $-$ helicity. If, for a given scattering amplitude, only two of the gluons are negative helicity, and the rest are positive, the resulting color-ordered scattering amplitudes go by the name ``maximally helicity violating'' or MHV.  Let us say that the $i$th and $j$th gluon of an $m$-gluon MHV scattering amplitude have negative helicity.  Then the color-ordered-scattering amplitude is given simply by:
\begin{equation}
\AtreeCO{m}(1{}^+,2{}^+,\ldots,i{}^-,\ldots , j{}^-, \ldots , m{}^+) =  \frac{ \langle i  j \rangle^4}{
\langle 1  2 \rangle \, \langle 2  3 \rangle \cdots  \langle m-1 , m \rangle\,\langle m,  1 \rangle } \,.
\label{PTappendixeq}
\end{equation}
Here I'm eliding a potential phase and the relevant powers of the coupling constant.
This is the famous formula, conjectured by Parke and Taylor~\cite{Parke:1986gb}, written in modern spinor-helicity notation; it is almost singlehandedly responsible for the glowingly optimistic introductions of innumerable papers on scattering amplitudes.

Using \eqn{PTappendixeq}, one can verify all the tree-level relations discussed in these lecture notes (at least in the MHV case).  To consider unitarity sums at loop level, however, requires the consideration of amplitudes with less helicity ``violation'': N${}^k$MHV amplitudes have $k$ additional negative helicity gluons.  Each of these cases can be embedded into a supersymmetric generating function indexed by Grassmann variables --- none of which I will describe here, but I encourage readers to look at refs.~\cite{Nair, GGK, FreedmanGenerating}.  That said, in order to get cut-data in 4D I should give you a way of talking about N${}^k$MHV --- fortunately there is such a way that only involves using lower-point MHV tree amplitudes.

\subsection{N$^{k}$MHV Tree Amplitudes}

Cachazo, Svr\v{c}ek, and Witten~\cite{CSW} discovered an MHV-vertex expansion for tree-level gluonic amplitudes --- offering a graph-based approach where N$^{k}$MHV trees are expressed as sums over functions of graphs where each vertex represents an MHV tree.    This can be generalized to superspace approaches for appropriately supersymmetric theories.  The idea is that every N${}^k$MHV tree can be written as a sum over graphs.  There should be $k$ vertices in each graph.  One finds all ways of assigning all external labels to the vertices in a manner consistent with the color-order, such that there is a choice of the helicity of the legs joining the vertices where each vertex is indeed MHV.  One gives the edges in the graph the propagator associated with (off-shell) conserved momenta, dresses the vertices with the representation of the MHV tree and sums over the resulting products.  

For example, consider $A(1^-,2^+,3^-,4^+,5^-,6^+)$.  We can expand it in MHV vertices the following ways:
\be
\begin{array}{c}
 A\left(i_1{}^-,k_2{}^+,k_3{}^-\right)
   A\left(-i_1{}^+,k_4{}^+,k_5{}^-,k_6{}^+,k_1{}^-\right) \\
 A\left(k_1{}^-,k_2{}^+,i_1{}^-\right)
   A\left(-i_1{}^+,k_3{}^-,k_4{}^+,k_5{}^-,k_6{}^+\right) \\
 A\left(-i_1{}^-,k_4{}^+,k_5{}^-,k_6{}^+\right)
   A\left(i_1{}^+,k_1{}^-,k_2{}^+,k_3{}^-\right) \\
 A\left(-i_1{}^+,k_5{}^-,k_6{}^+,k_1{}^-\right)
   A\left(i_1{}^-,k_2{}^+,k_3{}^-,k_4{}^+\right) \\
 A\left(-i_1{}^+,k_3{}^-,k_4{}^+,k_5{}^-\right)
   A\left(k_1{}^-,k_2{}^+,i_1{}^-,k_6{}^+\right) \\
 A\left(-i_1{}^-,k_4{}^+,k_5{}^-\right)
   A\left(i_1{}^+,k_6{}^+,k_1{}^-,k_2{}^+,k_3{}^-\right) \\
 A\left(-i_1{}^-,k_5{}^-,k_6{}^+\right)
   A\left(i_1{}^+,k_1{}^-,k_2{}^+,k_3{}^-,k_4{}^+\right) \\
 A\left(-i_1{}^-,k_6{}^+,k_1{}^-\right)
   A\left(i_1{}^+,k_2{}^+,k_3{}^-,k_4{}^+,k_5{}^-\right) \\
 A\left(-i_1{}^-,k_3{}^-,k_4{}^+\right)
   A\left(k_1{}^-,k_2{}^+,i_1{}^+,k_5{}^-,k_6{}^+\right) \\
\end{array}
\ee
Notice that  each  internal (shared) edge is assigned the correct helicity to support every ``tree'' being MHV.   Each of the above lines represents a graph with two nodes.  For each graph one takes the product of the MHV amplitude associated with each of the two vertices as well as the propagator $\frac{1}{i_1^2}$, where $i_1$ takes on different momenta per graph as per conservation of momenta.  Now you should rightfully ask, what does $\langle i_1 k_2 \rangle$ mean if $i_1$ is off-shell.  I have not defined it, so a very good question!  Now the answer.  What we do here is the following:  everywhere an internal edge $i$ appears in a spinor product, flatten it (convert it into a null-vector) by using an arbitrary null {\it reference} momentum $\xi$:
\begin{align}
\langle i , a \rangle &\to \langle i^\flat , a \rangle \,, \\
i_\mu^\flat &\equiv   i_\mu  -  \xi_\mu  \frac{i^2}{2 \, i \cdot \xi} \,.
\end{align}
The null-momentum $\xi$ can be arbitrary --- but for the $\xi$ dependence to cancel out the same $\xi$ must be used for the entire amplitude.  To make it absolutely clear, the first contribution to our six-point MHV expansion above looks like:
\begin{multline}
\frac{\langle (-k_2 -k_3)^\flat , k_3 \rangle^4}{\langle (-k_2 - k_3)^\flat , k_2 \rangle \, \langle k_2 , k_3 \rangle \, \langle k_3, (-k_2 -k3)^\flat \rangle }\,  \frac{1}{(k_2+k_3)^2}\\
\times\frac{\langle k_5 , k_1 \rangle^4}{  \langle (k_2 +k_3)^\flat , k_4 \rangle \,  \langle k_4 , k_5 \rangle \,  \langle k_5 , k_6 \rangle \,  
\langle k_6 ,  (k_2 +k_3)^\flat  \rangle}\,.
\end{multline} 
As you can see, when the internal edges appear in spinor products their momenta are assigned the flattened values, but they are not flattened when they appear in propagators.  Now you have all you need to quickly write down all pure Yang-Mills tree-level amplitudes.  This generalizes quite simply to a superspace encoding where the MHV expressions are simply replaced by super-MHV expressions --- Grassmann-encoding the various states.

\begin{problem}
Verify on some examples, say through 7 point, that this works all the way to all-negative but two positive-helicity gluons --- the so called $\overline{\rm MHV}$ amplitudes (read ``MHV - bar'').  Now you will arrive at the ugliest, longest expressions for $\overline{\rm MHV}$ color-ordered amplitudes with this approach, but you can easily verify numerically that you are doing things correctly.  How?  Using parity, the ${\overline{\rm MHV}}$ amplitudes can alternatively be written in the same form~(\ref{PTappendixeq}) as the MHV amplitudes, except that everywhere we would write $\langle a \, b \rangle$ for MHV we instead write $[ b \, a]$ for $\overline{\rm MHV}$, and obviously the preferred legs in the numerator are now the two positive-helicity gluons.
\end{problem}

\subsection{But what about cuts?}

It turns out that the MHV-vertex expansion just involving gluons works as a beautiful template for cut summation for pure SUSY gauge theories in 4D.  (Pure SUSY gauge theories are those for which all the states are connected by supersymmetry to gluons.)  You will see that by just tracking gluons, but dressing vertices of graphs appropriately, we will recover the correct sums over the entire multiplet of pure SUSY states crossing the cuts.

Here is the expression~\cite{SuperSum} for some cut with external gluons of specified helicity in some supersymmetric theory with $\cN\le4$~:
\be
\sum_{states} {\rm cut} = \sum_{g \in \Gamma_{\rm MHV}({\rm cut})} \frac{N_{\rm MHV}(g)}{D_{\rm MHV}(g)} \,.
\ee
The set $\Gamma_{\rm MHV}$ is all ways of writing this cut as a graph, where each tree explicit in the cut will be MHV-expanded in such a way that every vertex of the graph is a gluonic MHV tree.  By MHV I mean, only two negative-helicity gluons, every other helicity positive, tracking  the helicity of cut legs as well as tree-internal expanded legs as we did for N${}^{k}$MHV above.  Note that for each of these graphs with MHV vertices $A_i$, there can be multiple helicity assignments for internal propagators (including cut legs) consistent with these vertices remaining MHV.  Let us call the set of helicity assignments for each graph ${\cal H}(g)$.

For each such graph $g$, there will be a denominator 
\be 
D_{\rm MHV}(g) = d(g) \times  \prod_{i \in {\rm vertices(g)}} d_{\rm MHV}(A_i)  \,,
\ee  
which is the product of dressing each (non-cut) internal edge of the graph with the off-shell momentum invariant running through it (the same thing we have called $d(g)$ throughout these lectures), and dressing each vertex with the cyclic spinor-product ``MHV denominator'' of its associated tree $A_i$.    i.e.
\be
d_{\rm MHV}({\rm A}(1,2,3,\ldots,n)) = \langle 1\, 2 \rangle \, \langle 2\,  3 \rangle \cdots \langle n \, 1 \rangle\,.
\ee
Every tree-internal momentum that appears in $d_{\rm MHV}$ should be flattened using some reference momentum inside the spinor product.   E.g.~for the cut leg $l_1$, external leg $k_2$, and exposed tree-internal leg $i_1$,
\be
d_{\rm MHV}({\rm A}\left(i_1{}^-,k_2{}^-,l_1{}^+\right) ) =  \langle i_1^\flat , k_2 \rangle \langle  k_2 , l_1 \rangle \langle   l_1  , i_1^\flat \rangle \,.
\ee

There will also be a numerator associated with the graph, $N_{\rm MHV}(g)$, which now depends on the amount of SUSY, 
\begin{multline}
N_{\rm MHV}(g)=\left( \sum_{h \in {\cal H}(g)}  {\rm sig}(  \{  {\rm A}_i \} ,h)\, \prod_i {\rm Sp}{}^-({\rm A}_i,h)  \right)^\cN  \\
\times \left( 
\begin{array}{c}
         \sum_{h \in {\cal H}(g)}  \left(    {\rm sig}( \{  {\rm A}_i \}, h)\, \prod_i {\rm Sp}{}^-({\rm A}_i,h)  \right)^{4-\cN} \mbox{ for $\cN\le4$}\\
                1 \mbox{ for $\cN=4$}
\end{array} \right) \,,
\end{multline}
where ${\rm sig}( \{{\rm A}_i\},h)$ takes the negative-helicity legs of each tree ${\rm A}_i$ under helicity assignment $h$ in the list as ordered, and returns the signature of the permutation needed to place the list in some canonical order (say the lexicographically sorted list of edges that take negative helicity somewhere in the graph). E.g.
\begin{multline}
{\rm sig}\Bigl( \{{\rm A}\left(i_1{}^-,k_2{}^-,l_1{}^+\right), {\rm A}\left(-i_1{}^+,l_2{}^-,l_3{}^+,k_1{}^-\right), \\
{\rm A}\left(k_3{}^+,k_4{}^+,-l_3{}^-,-l_2{}^+,-l_1{}^-\right) \}  \Bigr) \\
={\rm Signature}(\left\{i_1,k_2,l_2,k_1,l_3,l_1\right\} | \left\{i_1,k_1,k_2,l_1,l_2,l_3\right\} ) \\
= 1 \,.
\end{multline}
Note that one can ignore the direction of the momentum when calculating the signature.  These signs arise naturally from the Grassmann-encoding of the state-sum, a discussion I am skipping for lack of space, but please read ref.~\cite{SuperSum}.   The function ${\rm Sp}{}^-({\rm A}_i, h)$ returns the spinor-helicity product of the two negative-helicity legs of the MHV tree ${\rm A}_i$ under helicity assignment $h$ in the order depicted:
\be
{\rm Sp}{}^-\left({\rm A}(1{}^+,2{}^+,\ldots,i{}^-,\ldots , j{}^-, \ldots , m{}^+) \right)=  \langle i j \rangle \,.
\ee
Every tree-internal momentum that appears in Sp${}^-$ should flatten against some reference momenta inside the spinor product.   E.g.
\be
{\rm Sp}{}^-(A\left(i_1{}^-,k_2{}^-,l_1{}^+\right) ) =  \langle i_1^\flat , k_2 \rangle \,.
\ee

\noindent {\bf IMPORTANT.} There are three important points to remember with the entire expression:
\begin{enumerate}
\item  One must remember to apply the spinor-product sign convention described in \eqn{FermionArrowSign} to the resulting cut spinor expressions.
\item The same reference momenta must be used in all flattening operations on the cut.
\item The entire discussion carries over to different (non-gluonic) external states
\end{enumerate}

Here is a complete example (or target, if you like) for the planar $\cN=4$ cut considered earlier:  
\begin{tiny}
\begin{multline}
\label{cutSoln}
\sum_{\rm states} A(k_1^-, k_2^-,a,b,c) A(-c,-b,-a,k_3^+,k_4^+) =
\frac{\langle 1\,2\rangle ^3 \left(\left\langle a\,i_1 ^{\flat }\right\rangle  \langle b\,c\rangle -\langle a\,c\rangle  \left\langle b\, i_1 ^{\flat }\right\rangle
   \right)^4}{\langle a\,b\rangle ^2 \left\langle a\, i_1^{\flat }\right\rangle  \langle a\,3\rangle  \langle b\,c\rangle  \left\langle b\, i_1^{\flat   }\right\rangle  \left\langle c\,i_1^{\flat }\right\rangle  \langle c\,1\rangle  \langle c\,4\rangle  \left\langle i_1^{\flat }\,2\right\rangle  \langle   3\,4\rangle   i_1^2} \\
   -\frac{\left(\langle a\,c\rangle  \left\langle b\,i_2^{\flat }\right\rangle 
   -\langle a\,b\rangle  \left\langle
   c\,i_2^{\flat }\right\rangle \right)^4 \langle 1\,2\rangle ^3}{\langle a\,b\rangle  \left\langle a\,i_2^{\flat }\right\rangle  \langle a\,2\rangle  \langle
   a\,3\rangle  \langle b\,c\rangle ^2 \left\langle b\,i_2^{\flat }\right\rangle  \left\langle c\,i_2^{\flat }\right\rangle  \langle c\,4\rangle  \left\langle
   i_2^{\flat }\,1\right\rangle  \langle 3\,4\rangle  i_2^2}  
   -\frac{\left\langle a\, i_2^{\flat }\right\rangle ^3 \langle b\,c\rangle ^2 \langle 1\,2\rangle
   ^3}{\langle a\,b\rangle  \langle a\,2\rangle  \langle a\,3\rangle  \left\langle b\,i_2^{\flat }\right\rangle  \left\langle c\, i_2^{\flat }\right\rangle  \langle
   c\,1\rangle  \left\langle i_2^{\flat }\,4\right\rangle  \langle 3\,4\rangle i_2^2}  \\
   -\frac{\langle a\,b\rangle ^2 \left\langle c\, i_3^{\flat }\right\rangle ^3
   \langle 1\,2\rangle ^3}{\langle a\,2\rangle  \langle a\,3\rangle  \langle b\,c\rangle  \left\langle b\, i_3^{\flat }\right\rangle  \langle c\,1\rangle  \langle
   c\,4\rangle  \left\langle i_3^{\flat }\,3\right\rangle  \left\langle i_3^{\flat }\,4\right\rangle  i_3^2}   
   +\frac{\left((\langle a\,2\rangle 
   \langle b\,c\rangle -\langle a\,c\rangle  \langle b\,2\rangle ) \left\langle i_4^{\flat }\,1\right\rangle +\langle a\,b\rangle  \langle c\,1\rangle  \left\langle
   i_4^{\flat }\,2\right\rangle \right)^4}{\langle a\,b\rangle ^2 \langle a\,2\rangle  \langle a\,3\rangle  \langle b\,c\rangle  \left\langle b\,i_4^{\flat
   }\right\rangle  \left\langle c\, i_4^{\flat }\right\rangle  \langle c\,1\rangle  \langle c\,4\rangle  \left\langle i_4^{\flat }\,1\right\rangle  \left\langle
   i_4^{\flat }\,2\right\rangle  \langle 3\,4\rangle  i_4^2}   \\
   -\frac{\left(\left\langle a\, i_5^{\flat }\right\rangle  \langle b\,c\rangle -\langle
   a\,c\rangle  \left\langle b\,i_5^{\flat }\right\rangle +\langle a\,b\rangle  \left\langle c\,i_5^{\flat }\right\rangle \right)^4 \langle 1\,2\rangle ^3}{\langle
   a\,b\rangle ^2 \left\langle a\,i_5^{\flat }\right\rangle  \langle a\,3\rangle  \langle b\,c\rangle ^2 \left\langle c\,i_5^{\flat }\right\rangle  \langle
   c\,4\rangle  \left\langle i_5^{\flat }\,1\right\rangle  \left\langle i_5^{\flat }\,2\right\rangle  \langle 3\,4\rangle 
   i_5^2}
    -\frac{\left\langle a\,i_7^{\flat }\right\rangle ^3 \langle b\,c\rangle ^2 \langle 1\,2\rangle ^3}{\langle a\,b\rangle  \langle
   a\,2\rangle  \langle a\,3\rangle  \left\langle b\, i_7^{\flat }\right\rangle  \langle c\,1\rangle  \langle c\,4\rangle  \left\langle i_7^{\flat
   }\,3\right\rangle  \left\langle i_7^{\flat }\,4\right\rangle  i_7^2}  
     \\
     +\frac{\left(\langle a\,2\rangle  \langle b\,c\rangle  \left\langle i_6^{\flat }\,1\right\rangle +(\langle a\,b\rangle  \langle
   c\,1\rangle -\langle a\,c\rangle  \langle b\,1\rangle ) \left\langle i_6^{\flat }\,2\right\rangle \right)^4}{\langle a\,b\rangle  \left\langle a\,i_6^{\flat
   }\right\rangle  \langle a\,2\rangle  \langle a\,3\rangle  \langle b\,c\rangle ^2 \left\langle b\,i_6^{\flat }\right\rangle  \langle c\,1\rangle  \langle
   c\,4\rangle  \left\langle i_6^{\flat }\,1\right\rangle  \left\langle i_6^{\flat }\,2\right\rangle  \langle 3\,4\rangle 
  i_6^2} 
     +\frac{\langle a\,b\rangle ^2 \left\langle c\,i_8^{\flat
   }\right\rangle ^3 \langle 1\,2\rangle ^3}{\left\langle a\, i_8^{\flat }\right\rangle  \langle a\,2\rangle  \langle b\,c\rangle  \left\langle b\,i_8^{\flat
   }\right\rangle  \langle c\,1\rangle  \langle c\,4\rangle  \left\langle i_8^{\flat }\,3\right\rangle  \langle 3\,4\rangle  i_8^2}\,, 
\end{multline}
\end{tiny}
where the internal momenta are given by,
\begin{align}
i_1&= k_3+k_4-c \,, \nonumber \\
i_2&=b+c \, , \nonumber\\
i_3&=k_4-c \,, \nonumber\\
i_4&=k_1+c\,, \nonumber\\
i_5&=k_3+k_4 \, ,\nonumber\\
i_6&=k_1+b+c\,,\nonumber\\
i_7&=b+c-k_4 \,, \nonumber\\
i_8&= k_3+k_4-c \,. \nonumber
\end{align}

Notice how big the expression is --- and I'm telling you that this is  equivalent to  \eqn{cutpSoln}, which I reproduce here:
\begin{multline}
\label{cutpSoln2}
\text{cut}_p = s t \AtreeCO{4}(k_1,k_2,k_3,k_4)\\
\times \Biggl( \frac{\left(k_1+k_2\right){}^2}{\left(l_a+l_b\right){}^2
   \left(-l_a-k_2\right){}^2 \left(l_b+l_c\right){}^2
   \left(k_4-l_c\right){}^2}\\+\frac{\left(k_1+k_2\right){}^2}{\left(l_a+l_
   b\right){}^2 \left(l_b+l_c\right){}^2 \left(-l_a-l_b-k_2\right){}^2
   \left(l_b+l_c-k_4\right){}^2}\\
   +\frac{\left(k_1+k_4\right){}^2}{\left(-l
   _a-k_2\right){}^2 \left(k_4-l_c\right){}^2
   \left(-l_a-l_b-k_2\right){}^2 \left(l_b+l_c-k_4\right){}^2} \Biggr) \,.
 \end{multline}
 Now there are some points I'd like to make.  Since this was a ``MHV"-cut, sewing a MHV tree with a $\overline{\text{MHV}}$ tree, the method I provided is maximally verbose --- writing the MHV-expansion of a  $\overline{\text{MHV}}$ amplitude is one of the most exhausting ways of writing it --- and this verbosity carries through to the entire cut.  The reason I do so is because this method is quite general, and this example gives you a verbose enough example to test your work on and make sure you aren't missing steps.   The second point is to emphasize that these very different looking expressions \eqn{cutSoln}  and \eqn{cutpSoln2} are equivalent, and offer a few tools to help make this clear.  First you should verify that they're equivalent numerically (perhaps up to a phase convention).

\begin{problem}
 Verify numerically, up to a possible phase convention, that \eqn{cutSoln} $=$ \eqn{cutpSoln2} .  How?  Find 4-momenta for $k_1,k_2,k_3,k_4, l_a,l_b,l_c$ such that:
 \begin{align}
 k_1+k_2+l_a+l_b+l_c&=0,\\
 k_1+k_2+k_3+k_4&=0,\\
 k_i^2 &=0, \\
 l_i^2&=0.
 \end{align}
Do you get it?  For any cut, each tree that contributes to a cut imposes the momentum conditions: 
\begin{equation}
\Atree{m}( {\rm labels}) \to \left \{ \begin{matrix} 
 \sum_{p \in {\rm labels}} p =\{0,0,0,0\}\\
  {\rm and }\\
  p^2=0 ~ \forall~ p \in \{{\rm labels}\} \end{matrix} \right .
\end{equation}
Here's a handy trick that works for a tremendous number of cuts --- at least by considering some path through trees\footnote{\begin{problem} Find a cut where this approach will fail to generate useful momenta.  Find your own solution to getting good kinematics for that cut. \end{problem}}.  For all but two of the momenta in any set of labels, just generate any random null momenta you like.  For the last two momenta, you need to satisfy:  $p_{(-2)}^2 =0$ and $ \left( p_{(-2)} + \left(P \equiv \sum_{i = 1}^{|p|-2} p_i \right)\right)^2=0$.   How?  Define some random null momenta $\xi$.   Use:
\begin{align}
p_{(-2)} &=  -  \xi \, \frac{P^2}{2 \,\xi \cdot P} \\
p_{(-1)}&= -P-p_{(-2)}
\end{align}
OK, once you have some set of momenta, it should be trivial for you to validate the equality to any degree of precision you like.
 \end{problem}
 
  Once you've convinced yourself that they're equal numerically you're ready to try to find the map analytically.  Let me be clear --- this exercise is good for building muscles and convincing oneself that these spinor-helicity representations make sense.  That said, I would caution against building too strong a devotion towards exercising these muscles.  Numerics, when you're careful, can take you very very far, and four dimensions is, after all, only four dimensions.  In general, one can miss higher-dimensional data necessary for the complete dimensionally-regularized integrand.  Getting comfortable with these manipulations is an important part of modern calculation, even if dealing with restricted kinematics. 
  
\begin{problem}
Find an analytic path between  \eqn{cutSoln} and \eqn{cutpSoln2}.  Hint:  In these expressions one can without loss of generality, choose the reference momenta such that we can replace, 
\be
\label{spab}
\langle  j, i^\flat  \rangle  \to  \langle j | i | X ]
\ee
where $X$ is an arbitrary (but uniform throughout the cut) massless 4-vector, and
\begin{align}
 \langle a | P | b ] &\equiv  P_\mu  \langle a | \gamma^\mu | b]  \\
  & = \frac{1}{ \sqrt{a_0+a_3} \sqrt{b_0+b_3} }\times \nonumber\\
  &~ P \cdot  \left ( \begin{matrix} 
   \left(a_0+a_3\right) \left(b_0+b_3\right)+\left(a_1+i a_2\right) \left(b_1-i b_2\right)\\
  \left(a_0+a_3\right) \left(b_1-i b_2\right)+\left(a_1+i a_2\right) \left(b_0+b_3\right)\\
   \left(a_0+a_3\right) \left(b_2+i b_1\right)+\left(a_2-i a_1\right) \left(b_0+b_3\right)\\
   \left(a_0+a_3\right) \left(b_0+b_3\right)-\left(a_1+i a_2\right) \left(b_1-i b_2\right) 
   \end{matrix}  \right )\, .
\end{align}
I include this latest expansion for completeness (and see \cite{LanceTasi1996} for more details), but please ignore it unless you really want to try to prove this equivalence using components (or will verify on numerics to line up book-keeping on intermediate steps).  
 
  The replacement in \eqn{spab} holds in these expressions because a common factor cancels between the numerators and the denominators (see, e.g., the CSW section of ref.~\cite{ElvangHuangBook}).   As the sandwiched $i$ are sums of labeled on-shell momentum $p_i$ these expressions will expand out to
  \begin{align}
  \langle j | p_1+ p_2+\cdots | X ]&= \sum_i  \langle j | p_i | X] \\
  &= \sum_i  \langle j  p_i \rangle \, [p_i  X] \,.
  \end{align}  
  (Remember to use the fermion sign ambiguity resolution given in \eqn{FermionArrowSign}.)
  As any $p_i$ sandwiched between the $\langle j |$ and $ | X ]$ will spinor-product to 0 if equivalent to $j$ or $X$,  we are in a happy situation.  A judicious choice of setting $X$ equal to a particular labeled momenta (uniform throughout the expression) will cause many of these expressions to vanish. At this point, through  angelic  consideration of conservation of momentum, Fierz, and Schouten identities, one can now shake these expressions towards equality.  
\end{problem}

\begin{problem}
 Execute the four-dimensional state-sum of the non-planar cut described earlier: $\sum_{\rm states} A(k_1^-, k_2^-,a,b,c) A(-b,-c,-a,k_3^+,k_4^+)$.
\end{problem}


\subsection{From Structure Constants ($f^{abc}$) to Trace Basis}

While almost everything I discussed involved having color-factors $c(g)$ expressed in terms of the structure constants associated with the graph topologies and orientations, it is often handy to be able to express these in terms of a color trace basis.  For the path to group theoretic enlightenment, I cannot recommend highly enough the famous ``Bird-track" monograph by Predrag Cvitanovi\'c~\cite{Cvitanovic:2008zz}. That said, for the case that all particles are in the adjoint representation of SU($N_c$),
here is a quick procedure to go from $f^{abc}$s to a trace basis.  One starts with $c(g)$ given by the product of a string of $f^{abc}$s.  To this expression, repeatedly apply the following rules until you reach a fixed point:
\begin{align}
 f^{abc} &\to  {\rm Tr}\left(\cTi{a} \cTi{b} \cTi{c}\right) - {\rm Tr}\left( \cTi{b} \cTi{a} \cTi{c}\right) \,, \\
 {\rm Tr}\left(\{w\},  \cTi{a}\right){}^2 &\to {\rm Tr}\left(\{w\},\{ w\}\right) - \frac{1}{N_c} {\rm Tr}\left(\{w\}\right)^2\, ,\\
{\rm Tr}\left(\{w \}, \cTi{a} , \{x\} \right) {\rm Tr}\left(\{y\} , \cTi{a},\{z\}\right)  &\to 
  {\rm Tr}\left(\{x\},\{ w\}, \{z\}, \{y\}\right) \\
  &~ - \frac{1}{N_c} {\rm Tr}\left(\{x\},\{ w\}\right){\rm Tr}\left(\{z\},\{ y\}\right)  \, , \nonumber \\ 
 {\rm Tr}\left(\{x\}, \cTi{a}, \{y\}, \cTi{a},\{ z\}\right) &\to {\rm Tr}\left(\{x\},\{ z\}\right) {\rm Tr}\left(\{y\}\right) 
 \\
  &~- \frac{1}{N_c} {\rm Tr}\left(\{x\},\{ y\},\{ z\}\right) \,, \nonumber \\
 {\rm Tr}\left(\cTi{a}, \cTi{a}\right) &\to N_c^2 - 1\,, \\ 
 {\rm Tr}\left(\right) &\to N_c \,, \\ 
 {\rm Tr}\left(\cTi{a}\right) &\to 0 \, ,
\end{align}
 where bracketed labels $\{w\},\{x\},\{y\},\{z\}$ denote variable length sub-lists of SU($N_c$) generator labels (of minimum 0 length), and all other labels are taken to be individual.  Recall that traces are cyclic, so after achieving a fixed point one can rotate the arguments to some canonical lexicographic ordering.

 \section*{Acknowledgments}
I would first very much like to the thank the TASI organizers, Lance Dixon and Frank Petriello, for the invitation to lecture at the TASI school 2014, and the students and  hosts for generating such an exciting atmosphere.   I took my first stab at presenting these ideas and tools pedagogically at the 2012 Arnold Sommerfeld School so I should also like to express my sincere gratitude to its organizers: Michael Haack, Stefan Hofmann, Dieter L\"ust and Stephan Stieberger, as well as its students. My thoughts and approaches to this material have been deeply informed by all of my collaborators but especially Zvi Bern, Johannes Broedel, Marco Chiodaroli, Tristan Dennen, Lance Dixon, Yu-tin Huang, Harald Ita, Henrik Johansson,  Murat G\"unaydin, Renata Kallosh, David Kosower,  Radu Roiban, and Arkady Tseytlin.  I am particularly grateful for the  insightful comments on early drafts by Lance Dixon and Oliver Schlotterer.  I am supported by the John Templeton foundation grant `Quantum Gravity Frontiers' and the Stanford Institute for Theoretical Physics.

\end{document}